\documentclass[a4paper, 11pt]{article}
\usepackage{latexsym,amsmath,amsfonts,amssymb}
\usepackage{tikz}
\usetikzlibrary{decorations.pathmorphing,cd,decorations.markings,calc}
\usepackage{mathrsfs}
\usepackage[american]{babel}
\usepackage{graphicx}
\usepackage{bbm}
\usepackage{cite}
\usepackage{tcolorbox}

\usepackage[colorlinks=true, citecolor=blue, linkcolor=blue, linktocpage=true]{hyperref}

\renewcommand{\baselinestretch}{1.2}
\newcommand{\ds}{\displaystyle}

\newcommand{\fP}{\mathfrak{P}}

\newcommand{\DWA}{\text{DW}(\mathbb{A})}
\newcommand{\TYA}{\text{TY}(\mathbb{A})_{\gamma,\epsilon}}

\newcommand{\eg}{\textit{e.g.}}

\newcommand{\ie}{\textit{i.e.}}

\numberwithin{equation}{section}

\newcommand{\mat}[1]{\begin{pmatrix} #1 \end{pmatrix}}
\newcommand{\smat}[1]{\big( \begin{smallmatrix} #1 \end{smallmatrix} \big)}
\newcommand{\be}{\begin{equation}} \newcommand{\ee}{\end{equation}}
\newcommand{\bea}{\begin{equation} \begin{aligned}} \newcommand{\eea}{\end{aligned} \end{equation}}

\newcommand{\pd}[1]{ #1^{\vee} }

\newcommand{\cA}{\mathcal{A}}
\newcommand{\cB}{\mathcal{B}}
\newcommand{\cC}{\mathcal{C}}
\newcommand{\cD}{\mathcal{D}}

\newcommand{\cK}{\mathcal{K}}

\newcommand{\calL}{\mathscr{L}}
\newcommand{\cM}{\mathcal{M}}

\newcommand{\calN}{\mathscr{N}}

\newcommand{\cN}{\mathcal{N}}

\newcommand{\cQ}{\mathcal{Q}}

\newcommand{\cS}{\mathcal{S}}
\newcommand{\cT}{\mathcal{T}}
\newcommand{\cU}{\mathcal{U}}

\newcommand{\cZ}{\mathcal{Z}}
\newcommand{\bA}{\mathbb{A}}
\newcommand{\bB}{\mathbb{B}}

\newcommand{\bG}{\mathbb{G}}

\newcommand{\bN}{\mathbb{N}}

\newcommand{\bR}{\mathbb{R}}

\newcommand{\bZ}{\mathbb{Z}}

\newcommand{\fg}{\mathfrak{g}}

\newcommand{\rmD}{\mathrm{D}}
\newcommand{\rme}{\mathrm{e}}
\newcommand{\rmm}{\mathrm{m}}
\newcommand{\rmO}{\mathrm{O}}

\newcommand{\unit}{\mathbbm{1}}

\newcommand{\linv}{\calL_D}
\newcommand{\teta}{\tilde{\eta}}

\def\SU{\mathrm{SU}}

\def\repa{\raise4pt\hbox{$\square$}\mkern-14mu\raise-4pt\hbox{$\square$}}
\def\repab{\overline{\raise4pt\hbox{$\square$}\mkern-14mu\raise-4pt\hbox{$\square$}\mkern-1mu}}

\DeclareMathOperator{\sign}{sign}

\DeclareMathOperator{\Arf}{Arf}

\DeclareMathOperator{\Hom}{Hom}
\DeclareMathOperator{\Rad}{Rad}
\newcommand{\abA}{\mathbb{A}}
\newcommand{\abB}{\mathbb{B}}

\begin{document}

\thispagestyle{empty}
\fontsize{12pt}{20pt}
\begin{flushright}
	SISSA  XX/2023/FISI
\end{flushright}
\vspace{13mm}  
\begin{center}
	{\huge Anomalies of non-invertible self-duality symmetries: \\[1em]
fractionalization and gauging}
	\\[13mm]
    	{\large Andrea Antinucci$^{a,\, b}$, \, Francesco Benini$^{a,\, b,\, c}$, \, Christian Copetti$^{a,\, b}$,
	\\[6mm]
	Giovanni Galati$^{a,\, b}$, and Giovanni Rizi$^{a,\, b}$}
	
	\bigskip
	{\it
		$^a$ SISSA, Via Bonomea 265, 34136 Trieste, Italy \\[.0em]
		$^b$ INFN, Sezione di Trieste, Via Valerio 2, 34127 Trieste, Italy \\[.6em]
		$^c$ ICTP, Strada Costiera 11, 34151 Trieste, Italy
	}
\end{center}

\bigskip

\begin{abstract}
We study anomalies of non-invertible duality symmetries in both 2d and 4d, employing the tool of the Symmetry TFT. In the 2d case we rephrase the known obstruction theory for the Tambara-Yamagami fusion category in a way easily generalizable to higher dimensions. In both cases we find two obstructions to gauging duality defects. The first obstruction requires the existence of a duality-invariant Lagrangian algebra in a certain Dijkgraaf-Witten theory in one dimension more. In particular, intrinsically non-invertible (\textit{a.k.a.} group theoretical) duality symmetries are necessarily anomalous.
The second obstruction requires the vanishing of a pure anomaly for the invertible duality symmetry. This however depends on further data. In 2d this is specified by a choice of equivariantization for the duality-invariant Lagrangian algebra. We propose and verify that this is equivalent to a choice of symmetry fractionalization for the invertible duality symmetry. The latter formulation has a natural generalization to 4d and allows us to give a compact characterization of the anomaly. We comment on various possible applications of our results to self-dual theories.

\noindent

\end{abstract}

\newpage
\pagenumbering{arabic}
\setcounter{page}{1}
\setcounter{footnote}{0}
\renewcommand{\thefootnote}{\arabic{footnote}}

{\renewcommand{\baselinestretch}{.88} \parskip=0pt
\setcounter{tocdepth}{2}
\tableofcontents}

\newpage

\section{Introduction}

In recent years a great deal of effort has been devoted to deepen our understanding of the role played by symmetries in quantum field theories (QFTs), following the seminal work \cite{Gaiotto:2014kfa}. 
A growing amount of evidence suggests that the general mathematical framework suited to describe the symmetries of a $d$-dimensional QFT is that of fusion $(d-1)$-categories \cite{douglas2018fusion}. This description includes non-invertible symmetry defects, whose fusion algebra does not have a group-like structure, and are by now known to be ubiquitous in 2d QFTs (see \eg \cite{Verlinde:1988sn, Petkova:2000ip, Fuchs:2002cm, Frohlich:2003hm, Frohlich:2004ef, Frohlich:2006ch, Frohlich:2009gb, Carqueville:2012dk, Bhardwaj:2017xup, Chang:2018iay, Thorngren:2019iar, Komargodski:2020mxz, Thorngren:2021yso, Huang:2021zvu, Burbano:2021loy, Benini:2022hzx, Inamura:2022lun, Lin:2022dhv, Choi:2023xjw}) as well as in higher dimensions (see \eg \cite{Nguyen:2021yld, Roumpedakis:2022aik, Bhardwaj:2022yxj, Choi:2022zal, Kaidi:2022uux, Choi:2022jqy, Cordova:2022ieu, Antinucci:2022eat, Bashmakov:2022jtl, Damia:2022bcd, Choi:2022rfe, Bhardwaj:2022lsg, Bartsch:2022mpm, Apruzzi:2022rei, Bartsch:2022ytj,Mekareeya:2022spm, Damia:2022seq, GarciaEtxebarria:2022vzq, Heckman:2022muc, Niro:2022ctq, Antinucci:2022cdi, Antinucci:2022vyk, Heckman:2022xgu, Bhardwaj:2022kot,Bhardwaj:2022maz, Choi:2022fgx, Bashmakov:2022uek, Apte:2022xtu, Argurio:2023lwl, Damia:2023ses, Copetti:2023mcq, Bartsch:2023pzl,Bartsch:2023wvv, Bhardwaj:2023ayw, Bashmakov:2023kwo, Bhardwaj:2023wzd, Apruzzi:2023uma, Cordova:2023ent, vanBeest:2023dbu, Sun:2023xxv}). While various applications have been already found in the 2d context \cite{Chang:2018iay, Komargodski:2020mxz, Jacobsen:2023isq, Choi:2023xjw} thanks to the large amount of available results on fusion 1-categories \cite{etingof2016tensor} (see \cite{Bhardwaj:2016clt, Chang:2018iay} for reviews aimed at physicists), dynamical consequences of non-invertible symmetries in higher dimensions have been somewhat elusive, requiring a better understanding of the relevant categorical structures. 

A step towards this objective is the development of a concrete characterization of 't Hooft anomalies and of their dynamical consequences for RG flows. While for invertible (higher) symmetries a complete classification of 't Hooft anomalies is given by the appropriate cobordism group \cite{Kapustin:2014dxa, Kapustin:2014tfa, Freed:2016rqq, Yonekura:2018ufj}, for non-invertible symmetries the correct general framework remains unclear.
A standard approach is to define anomalies as obstructions to the gauging of a symmetry $\cC$. Gauging (or condensation) in higher fusion categories is however a subtle procedure, as it requires the specification of a certain type of consistent algebra objects $\cA \in \cC$. While the mathematical theory governing such objects has been developed for 1-categories \cite{Fuchs:2002cm, Kong:2013aya} and recently for 2-categories \cite{Decoppet:2022dnz}, a complete characterization of the required consistency conditions is to this day still missing.
A more modern perspective would be to characterize 't~Hooft anomalies as obstructions to the existence of a trivially-gapped $\cC$-symmetric phase.
As pointed out in \cite{Choi:2023xjw}, for non-invertible symmetries this latter definition of 't Hooft anomalies implies also that there is an obstruction to gauging, while the converse is generically not true. This observation has been recently reformulated as the existence of certain weakly (respectively, strongly) symmetric boundary conditions in \cite{Choi:2023xjw}.%
\footnote{The two notions coincide for invertible symmetries and the obstruction to define them is equivalent to an 't~Hooft anomaly \cite{Jensen:2017eof, Thorngren:2020yht}. In the non-invertible case the two notions bifurcate.}

In this work we focus on self-duality symmetries,  which appear when a $d=2n$ dimensional QFT $\cT$ is mapped back to itself after gauging a discrete $(n-1)$-form symmetry $\bA$ \cite{Choi:2021kmx, Kaidi:2021xfk, Choi:2022zal, Kaidi:2022uux}:
\be
\cT/ \bA \,\cong\, \cT \;,
\ee
possibly with a choice of discrete torsion which we leave implicit. Above $\cong$ means equivalence up to a change of duality frame. The corresponding symmetry category $\cC$ is best described as a graded category, graded by the group $G$ of self-dualities:
\be
\cC = \bigoplus_{g \in G} \cC_g \;, \qquad\qquad \cC_0 = n\text{Vec}_\bA \;, \qquad\qquad \cC_g = \{ \cN_g \} \;.
\ee
Here $n\text{Vec}_\bA$ is the category describing an anomaly-free $(n-1)$-form symmetry $\bA$, and in the last equality we meant that the connected component%
\footnote{\label{footnote: connected components}%
Given a (higher) category $\cC$, $\pi_0(\cC)$ denotes the set of simple objects of $\cC$ modded out by the equivalence relation $x \sim y$ if $\Hom(x,\, y)$ is nontrivial \cite{Johnson-Freyd:2021chu}. Physically, the modding procedure corresponds to the condensation of symmetries localized on the defects.}
$\pi_0(\cC_g)$ has a single simple object $\cN_g$ for $g \neq 0$. The fusion rules of the $\cN_g$'s respect the $G$-grading up to condensates $C_\bA$ of the symmetry $\bA$ \cite{Roumpedakis:2022aik}. In particular
\be
\cN_g \times \overline{\cN}_g = C_\bA \;.
\ee
We will consider the cases of $G=\bZ_2$ in $d=2$ and $G=\bZ_4, \, \bZ_3$ in $d=4$. Our analysis however can in principle be extended to more general cases.%
\footnote{In $d=4$, theories of class $\cS$ \cite{Gaiotto:2009we} can have self-duality defects with non-Abelian $G$ \cite{Antinucci:2022cdi, Bashmakov:2022uek}. Moreover, it has been recently pointed out that there exist duality defects in 6d SCFTs \cite{Lawrie:2023tdz}.}
Examples of theories with duality symmetries are the Ising CFT and the $c=1$ boson in 2d \cite{Chang:2018iay,Thorngren:2021yso}, and $\cN=4$ SYM and pure Maxwell theory (for specific values of the complexified gauge coupling) in 4d \cite{Choi:2021kmx}.%
\footnote{See also \cite{Damia:2023ses} for an extension to theories with lower amounts of SUSY, and \cite{Decoppet:2023bay} for the mathematical treatment of self-duality categories in 3d.}

In this work will define an 't Hooft anomaly for a non-invertible duality defect as the obstruction to constructing a condensable algebra $\cA$ containing all the $\cN_g$'s.
It is believed (although not proven) that compatibility with a trivially-gapped phase is equivalent to the existence of such an algebra which furthermore contains the full category $\cC$. In two dimensions for Tambara-Yamagami (TY) categories \cite{tambara1998tensor}, this viewpoint has been examined in \cite{Thorngren:2019iar} by exploiting the concept of \emph{fiber functor}. In this case, condensable algebras are of the form $\cA = \bB \oplus n_\nu \, \cN$ with $\bB \subset \bA$ a subgroup and $n_\nu$ an integer. The symmetry admits a trivially-gapped realization only if $\bB = \bA$. If instead $\bB \subsetneq \bA$, the symmetry only admits a duality-invariant TQFT. We will regard $\cN$ to be anomaly-free in both cases.

The aim of this work is to give a unified treatment of anomalies for duality symmetries which can be generalized to higher dimensions. A fundamental tool to this purpose is the Symmetry TFT $\cZ(\cC)$ \cite{Freed:2022qnc}.%
\footnote{See \eg{} \cite{Kaidi:2022cpf, Antinucci:2022vyk, Antinucci:2022cdi, Kaidi:2023maf, Bhardwaj:2023ayw, Apruzzi:2023uma} and references therein for recent applications. Notice that in the mathematics literature the notation $\cZ(\cC)$ is used to denote the Drinfeld center \cite{etingof2016tensor} of a Fusion category $\cC$, while here it denotes the Symmetry TFT for the symmetry $\cC$. While the two concepts are equivalent in 2d, one must be careful when extending the correspondence to higher categories. See for example \cite{Johnson-Freyd:2021chu}.} 
Given a fusion $(d-1)$-category $\cC$, $\cZ(\cC)$ is a $(d+1)$-dimensional TQFT which encodes the full categorical data, and in particular the anomalies, of the symmetry $\cC$. Topological manipulations (generalized gaugings) in the QFT are belived to be in one-to-one correspondence with topological boundary conditions of $\cZ(\cC)$. Hence an 't Hooft anomaly corresponds to the absence of certain ``magnetic'' topological boundary conditions in $\cZ(\cC)$ which would trivialize $\cN$ on the boundary. A similar perspective has been considered recently in \cite{Kaidi:2023maf, Zhang:2023wlu}. 

The Symmetry TFT for duality defects has been identified as a $(2n + 1)$-dimensional Dijkgraaf-Witten (DW) theory, further gauged by a 0-form symmetry \cite{Kaidi:2022cpf, Antinucci:2022vyk}. In Section~\ref{sec:proposal} we use this fact to lay out a general approach to identify obstructions to the existence of those magnetic boundary conditions.
The resulting obstruction theory consist of two conditions:
\begin{enumerate}

\item The first one is the existence of a $G$-invariant Lagrangian algebra $\linv$ in the ungauged DW theory. In the language of \cite{Kaidi:2022cpf}, the duality symmetry is ``non-intrinsic".
Hence intrinsically non-invertible symmetries are necessarily anomalous.

\item When that condition is satisfied, the second obstruction corresponds to the cancellation of a pure 't Hooft anomaly for an invertible symmetry. Given a $G$-invariant Lagrangian algebra $\linv$, the group $G$ generally acts on it through a nontrivial automorphism. This action is not unique and corresponds to a choice of \emph{equivariantization} $\teta$ of $\linv$ \cite{Bischoff:2018juy}. We propose and check in several examples that such a choice encodes symmetry fractionalization data for the boundary symmetry, which can sometimes be used to cancel the cubic 't Hooft anomaly (see \cite{Delmastro:2022pfo} for other examples of this phenomenon). This allows to overcome the difficulty of generalizing the equivariantization procedure to higher categories.
\end{enumerate}

In Section \ref{sec: 3d} we examine, in two dimensions, duality defects that are described by Tambara-Yamagami categories $\TYA$. The full classification of their anomalies is known in both the mathematical \cite{tambara2000representations, gelaki2009centers, meir2012module} and physical literature \cite{Thorngren:2019iar}. We show how our prescription precisely reproduces the known results. We then generalize this strategy to analyze the anomalies of non-invertible duality defects in 4d \cite{Choi:2021kmx, Kaidi:2022uux, Choi:2022zal} in Section \ref{sec: 5d}. Notice that even though a complete definition of ``gauging" for higher categories is still absent, the obstructions are nevertheless accessible from the Symmetry TFT. 
These defects are present, for instance, in  $\cN=4$ SYM and the analysis of their anomalies is a crucial first step in understanding the dynamics of duality-preserving RG flows \cite{Damia:2022bcd}. Gauging these symmetries in $\cN=4$ SYM also leads to certain $\cN=3$ SCFTs \cite{Bourton:2018jwb}. Our findings shed light on their consistency. In Section \ref{sec: dimred} we consider the compactification of the 4d/5d setup on a torus, obtaining a 2d/3d system with a Tambara-Yamagami symmetry associated with a finite group $\widetilde{\bA}=\bA\times \bA$, where $\bA$ is the 4d 1-form symmetry group. We show how in this case the 4d obstruction theory correctly descends to the 2d one. We conclude in Section~\ref{sec: conclusions} with several applications of our results and further directions. Technical material and general proofs are gathered in various appendices.

\paragraph{Notation} We use additive notation for the Abelian groups $\bA$, $\bB$, \textit{etc}.
We indicate the Pontryagin dual to $\bA$ as $\bA^\vee = \operatorname{Hom}\bigl(\bA, U(1) \bigr)$.
For simplicity, we indicate group-cohomology groups as $H^p\bigl( \bA, U(1) \bigr)$ as opposed to the lengthier (though equivalent) notation $H^p\bigl( B\bA, U(1) \bigr)$ for the cohomology groups of the classifying space $B\bA$ of $\bA$. For group-cohomology classes, and more generally for cochains, we use multiplicative notation with values in $U(1)$, with the exception of integrals, examples or where otherwise stated where we use additive notation with values in $\bR/\bZ$. In order to limit confusion, we sometimes use the notation $\underline{0}$ and $\underline{1}$ for the trivial element and the generator of $\bZ_n$, respectively.

\emph{Note added: while this work was being completed we became aware of \cite{CPZ} whose results overlap with ours. We are grateful to the authors of \cite{CPZ} for coordinating submission.}

\section{A proposal from the Symmetry TFT}
\label{sec:proposal}

A promising approach to analyze anomalies of categorical symmetries is the Symmetry TFT \cite{Freed:2022qnc,Apruzzi:2021nmk}. Given a symmetry category $\cC$ in $d$ dimensions the associated Symmetry TFT is a $d+1$-dimensional TQFT $\cZ(\cC)$ admitting a gapped boundary condition $\calL_\cC$, which we call electric, giving rise to the symmetry $\cC$ on the boundary. Formally this means that the category $\text{Mod}_{\calL_\cC}\left(\cZ(\cC)\right)$ of $\calL_\cC$-modules, which describes topological operators confined to the gapped boundary, coincides with $\cC$: 
\be
\text{Mod}_{\calL_\cC}\left( \cZ(\cC) \right) = \cC \, .
\ee
General gapped boundary conditions are in one-to-one correspondence with Lagrangian algebra objects $\calL$ of the bulk category (see \eg \cite{Kapustin:2010hk}). This correspondence is realized by noting that $\cZ(\cC)$ is trivialized after condensing $\calL$, and topological operators charged under $\calL$ are confined on the boundary where they form the category $\text{Mod}_{\calL}\left(\cZ(\cC)\right)$. 

This setup can be coupled to a dynamical theory $\cT$ via a ``sandwiching" procedure:
\bea
\label{fig: freedsandiwch}
	\begin{tikzpicture}[scale=1.75]
		\filldraw[fill=white!70!blue, opacity=0.5] (0,0) -- (1,0.25) -- (1,1.25) -- (0,1) -- cycle;
		\node at (0.5, 0.625) {$\cT$}; 
		\node at (1.5,0.625) {$\simeq$};
		\begin{scope}[shift={(2,0)}]
				\draw (0,0) --(2,0); \draw (1,0.25) -- (3,0.25); \draw (1,1.25) -- (3,1.25); \draw (0,1) -- (2,1);
		\filldraw[fill=white!70!blue, opacity=0.5] (2,0) -- (3,0.25) -- (3,1.25) -- (2,1) -- cycle;	
		
			\filldraw[fill=white!70!red, opacity=0.5] (0,0) -- (1,0.25) -- (1,1.25) -- (0,1) -- cycle;	
		\node at (1.5,0.625) {\small $ \cZ(\cC)$};
			\node at (0.5, 0.625) {$\cC$};
   \node at (0.5, 1.35) {$\calL_{\cC}$};
		\end{scope}
	\end{tikzpicture}
\eea
where the $d+1$-dimensional manifold is a slab with two boundaries, one supporting a dynamical theory (with free boundary conditions) while on the other we impose the gapped boundary condition $\calL_{\cC}$. Different choices of $\calL$ give rise to symmetry categories belonging to the same Morita equivalence class $\cM(\cC)$ of the symmetry $\cC$. Physically, elements in the same class are related by (generalized) discrete gauging operations.
The reason why the Symmetry TFT is a useful tool for detecting anomalies is that, at least in a large class of examples, the correspondence between gapped boundary conditions and elements of the Morita equivalence class $\cM(\cC)$ is one-to-one:\footnote{
This is certainly true for maximally degenerate categories, such as those considered in this work. Physically this means that there is no charged object which is also topological. In these cases the Symmetry TFT is constructed by a state-sum and is a generalization of the Turaev-Viro theory \cite{turaev1992state} (see also \cite{Crane:1994ji, Levin:2004mi, Walker:2011mda, Walker:2021esr} for concrete generalizations in both condensed matter and mathematics).} all the allowed topological manipulations can be realized by condensing the appropriate Lagrangian algebra in the bulk.
In QFT language we define an 't Hooft anomaly as an obstruction to gauging the symmetry $\cC$. In the Symmetry TFT language this translates to the lack  of a dual boundary condition $\calN_{\calL}$, which would implement the topological manipulation of ``gauging $\cC$". 

In this work we focus on duality defects \cite{Thorngren:2019iar,Choi:2021kmx,Choi:2022zal,Kaidi:2021xfk,Kaidi:2022uux}, for which $\cZ(\cC)$ is known \cite{Kaidi:2022cpf} (see also \cite{Antinucci:2022vyk}). The Symmetry TFT description of $\cC$ is closely related to the one for $\text{nVec}(\abA)$. The latter corresponding to a generalized untwisted Dijkgraaf-Witten theory
\be
\cZ(\text{nVec}(\abA)) = \DWA \, ,
\ee
with $\text{nVec}(\abA)$ associated to the canonical Dirichlet boundary condition in $\DWA$. Concretely, in $d=2n$ dimensions $\DWA$ is a pure $(n-1)$-form $\bA$ gauge theory in $d+1$ dimensions with action
\begin{equation}
    S=2\pi i \int _{X_{d+1}}A\cup dB \ , \ \ \ \ \ A\in C^n(X_{d+1},\bA) \ , \ \ B\in C^n(X_{d+1},\bA^\vee)
\end{equation}
which has an $n-$form symmetry $\bA\times \bA^\vee$ generated by the Wilson surface operators of $B$ and $A$ respectively. The canonical Dirichlet boundary conditions simply sets the pull back of $A$ to the boundary to zero. The duality symmetry $G$ is a subgroup of the 0-form symmetry of the Dijkgraaf-Witten theory,%
\footnote{Depending on $\bA$ there could be other zero-form symmetries in the theory, in this work we only consider those that are present for any $\bA$.}
which acts by exchanging electric and magnetic operators by specifying an isomorphism
\be
    \phi : \bA\rightarrow \bA^\vee \;.
\ee
We will consider the cases in which $G$ is isomorphic to $\bZ_2$ for $n=1$ and to either $\bZ_3$ or $\bZ_4$ for $n=2$, corresponding respectively to duality and triality symmetries. Generically $G$ acts non-trivially on boundary conditions through its action on the associated Lagrangian algebras $\calL$. Gauging the $G$ symmetry, possibly with discrete torsion $\epsilon$, leads to the Symmetry TFT $\cZ(\cC)$:
\bea
\begin{tikzcd}
    \DWA \arrow[rr, dashed, shift left, bend left=30, "G^\epsilon"]& & \arrow[ll, dashed, shift left, bend left=30, "\text{Rep}(G)"] \cZ(\cC) 
\end{tikzcd}
\eea
the upper arrow indicating gauging with discrete torsion and the lower one the ``inverse" operation of gauging the dual symmetry $\text{Rep}(G)$.\footnote{To be precise the gauging of $\text{Rep}(G)$ should be accompanied by the stacking of the inverse SPT $\epsilon^{-1}$.} We will argue that the choice of $\epsilon$ acts as a kind of pure $G$ anomaly for the duality defects.
From the bulk perspective the duality defects $\cN_g$ are related to the liberated twist defects $\Sigma_g$ of the 0-form symmetry $G$ in $\DWA$ \cite{Antinucci:2022vyk, Kaidi:2022cpf} which are the objects carrying charge under the quantum $\text{Rep}(G)$ symmetry.

Gapped boundary conditions $\calL_0$ in $\DWA$ are in correspondence with maximal (Lagrangian) lattices $\calL_0$ of mutually local charges. Condensing such objects in the bulk leads to a trivial theory whose unique state is the gapped boundary condition \cite{Kaidi:2021gbs}. 
From the previous observation we can always induce a gapped boundary condition $\calL$ for $\cZ(\cC)$ from a gapped boundary $\calL_0$ in the DW theory by first condensing $\text{Rep}(G)$ and then $\calL_0$:
\bea \label{eq: seqgauging}
\begin{tikzcd}
    \DWA \arrow[dd, dashed, "\calL_0"] \\
    \\
    \text{nVec} 
\end{tikzcd}
\ \ \ \ \ \ \ \ \ \ \ \ \ \ \ \ \ \ \ \ \ \ \ \ \ \ \ \ \ \ \ \ \ \ \ \ \ \ \ \ \ 
\begin{tikzcd}
   \DWA  \arrow[dd, dashed, "\calL_0"] & &  \arrow[ll, dashed, swap, "\text{Rep}(G)"] \arrow[lldd, color=blue, "\calL"]  \cZ(\cC) \\
    & & \\
    \text{nVec} & &
\end{tikzcd}
\eea
Here with nVec we denote the trivial $d+1$ dimensional theory obtained after condensing $\calL_0$ in $\cZ(\cC_0)$. When $\calL_0=\calL_{\cC_0}$ this two step gauging defines a canonical Dirichlet boundary condition $\calL_\cC$ of the Symmetry TFT $\cZ(\cC)$. Since the twist defects $\Sigma_g$ are charged under the dual $\text{Rep}(G)$ symmetry they are confined to the boundary $\calL$, which thus describes a system with a non-invertible duality symmetry. This construction was implicitly used in \cite{Antinucci:2022vyk}. 

Gauging the non-invertible symmetry $\cN_g$ must correspond to a gapped boundary condition on which the twist defects $\Sigma_g$ are trivialized.
Thus to quantify a duality anomaly we must construct a different set of boundary conditions $\calN_{\calL_0}$ of $\cZ(\cC)$ associated with $\calL_0$ whose symmetry $\text{Mod}_{\calN_{\calL_0}}(\cZ(\cC))$ is trivially charged under $\text{Rep}(G)$. We will refer to this as a Neumann boundary condition, since the $G$ gauge field will remain dynamical on the boundary.

The crucial insight comes from considering a $G$-invariant Lagrangian algebra $\calL_0 = \linv$ in $\DWA$. This ensures that gauging $\linv$ leads, in the bulk, to an SPT phase for $G$, rather than to a completely trivial theory.
The SPT is completely specified by an element $Y$ living in the appropriate cobordism group.\footnote{For $d=2$ this is just a bosonic SPT $\in \, H^3(G, \, U(1))$. For $d=4$ instead we will work on spin manifolds and the correct group to consider is either $\text{Tors}(\Omega_5^{\text{spin}-G}(pt))$ or $\text{Tors}(\Omega_5^{\text{spin}}(BG))$ depending on whether $(-)^F$ sits inside the duality group or not. The same observations apply to the dicrete torsion $\epsilon$.}
It turns out that the datum $Y$ cannot be fixed by the choice of $\linv$ alone, but requires by further data, which we dub $\teta$, describing how the $G$ symmetry acts on the algebra morphisms of $\linv$. This is called an \emph{equivariantization} of $\linv$ \cite{etingof2016tensor,Bischoff:2018juy}. We denote the equivariantized algebra by a pair $(\linv, \teta)$.
The SPT $Y$ also contains a nonempty $G$ twisted sector with a unique simple object $M_g$ for each $g \in G$. In the 3d setting these can be formally described as twisted local modules for $\linv$, see Appendix \ref{app:Frobenius}.
Since $\linv$ is $G$-invariant the operation of gauging $G$ commutes with the condensation of $\left( \linv, \, \teta\right)$ and composing the two operations we end up with a bulk $G$ Dijkgraaf-Witten theory with twist $Y \epsilon$:
\bea \label{eq: introtwist}
\begin{tikzcd}
    \DWA \arrow[dd, dashed, "{\left(\linv , \, \teta\right)}"] \\
     \\
     \text{SPT}(G)_Y 
\end{tikzcd}
\ \ \ \ \ \ \ \ \ \ \ \ \ \ \ \ \ \ \ \ \ \ \ \ \ \ \ \ \ \
\begin{tikzcd}
   \DWA \arrow[dd, dashed, "{\left(\linv, \, \teta\right)}"] & & \arrow[ll, dashed, swap, "\text{Rep}(G)"] \cZ(\cC)   \\
    & & \\
    \text{SPT}(G)_Y \arrow[rr, dashed , swap, "G^\epsilon"] & & \text{DW}(G)^{Y \, \epsilon}
\end{tikzcd}
\eea
Its magnetic operators are the former twist defects $M_g$. In $3d$ their spin is determined by the total twist $Y \epsilon$.\footnote{In the 5d case the twist determines a triple linking between magnetic defects \cite{Kaidi:2023maf} instead.}

The $\text{DW}(G)^{Y \epsilon}$ theory always admits a canonical Dirichlet boundary condition, which gives rise to an invertible $G$-symmetry on the boundary (such occurrences have been dubbed \emph{non-intrinsic} in \cite{Kaidi:2022cpf}). This also coincides with the algebra $\calL$ we have previously introduced, in the special case in which $\calL_0 = \linv$ is $G$-invariant. If however $Y \epsilon = 1$ magnetic defects $M_g$ are also condensable. This allows to define the Neumann boundary condition $\calN_\calL$ we are looking for:
\bea
\begin{tikzcd}
   \DWA \arrow[dd, dashed, "{\left(\linv, \, \teta\right)}"] & & \arrow[ll, dashed, swap, "\text{Rep}(G)"] \arrow[dddd, color=blue, bend left=60, "\calN_{\calL}"] \cZ(\cC)   \\
    & & \\
    \text{SPT}(G)_Y \arrow[rr, dashed , swap, "G^\epsilon"] & & \text{DW}(G)^{Y \, \epsilon} \arrow[dd, dashed, "\calN"] \\
    & & \\
    & & \text{nVec} 
\end{tikzcd}
\eea
Thus the existence of $(\linv, \, \teta)$ and the triviality of $Y \epsilon$ are sufficient conditions for the duality symmetry to be anomaly-free.

Let us also argue that they are also necessary.\footnote{See also \cite{Sun:2023xxv} for an other argument, in the case $\bA=\bZ_n$, in favour of the necessity of the existence of a duality invariant algebra.} Suppose that an algebra $\calN_{\calL_0}$ of $\cZ(\cC)$ containing the twist defects $\Sigma_g$ exists, \ie{} $\Hom(\calN_{\calL_0}, \, \Sigma_g) \neq \emptyset$. $\calN_{\calL_0}$ has a natural grading given by the charge under $\text{Rep}(G)$ of its elements:
\be
\calN_{\calL} = \bigoplus_{g \, \in \, G} \calN^g \, ,
\ee
The algebra product  $\times_{\calN_{\calL_0}}$ mapping $\calN_g \times_{\calN_{\calL_0}} \calN_h \ \to \ \calN_{gh}$.
The consistency conditions for $\calN_{\calL_0}$ are also graded over $G$, and in particular imply that $\calN^0$ must itself be an algebra, although not a maximal one. On the other hand $\calN^g$, $g \neq 1$ are (local) modules over $\calN^0$, \ie{} they must survive the condensation of $\calN^0$. Physically this corresponds to gauging $\calN_{\calL_0}$ sequentially.
Since $\calN^0$ has trivial grading condensing it must also preserve the $\text{Rep}(G)$ symmetry. Due to the algebra product, furthermore, the $\calN^g$ become invertible $G$ defects in the gauged theory. Lastly, since the $\calN^g$ participate in the $\calN_{\calL_0}$ algebra, the symmetry $G$ must be anomaly-free. 

Assuming that $\calN^g, \, g \,  \in G$, and $\text{Rep}(G)$ are the only defects surviving the $\calN^0$
condensation it follows that the gauged theory is the $G$-Dijkgraaf-Witten theory with trivial twist $\text{DW}(G)$.\footnote{This can be proven for 3d MTCs. The fact that $\calN^g$ are invertible as bimodules and the fact that $\calN_{\calL_0}$ is Lagrangian implies that $\dim(\calN^g)=\dim(\calN^0)=|\abA|$. After gauging $\calN^0$ the resulting MTC has dimension $\cD=|G|$, which is saturated by the $G$ invertible $\calN^g$ and the $G$ elements of $\text{Rep}(G)$. The fact that $\calN^g$ is charged under $\text{Rep}(G)$ gives the canonical braiding of the DW theory.} Gauging the $\text{Rep}(G)$ symmetry (and stacking with a discrete torsion $\epsilon^{-1}$) now leads us to a $G$ SPT $Y=\epsilon^{-1}$. 

Chasing the diagram below shows that we can induce a $G$-invariant algebra $(\linv , \, \teta)$ in $\DWA$ by sequentially gauging $G^\epsilon - \calN^0 - \text{Rep}(G)$:
\bea
\label{eq: necesscond}
\begin{tikzcd}
    \cZ(\cC) \arrow[dd, dashed, "\calN_{\calL_0}"] \\
     \\
    \text{nVec}
\end{tikzcd}
\ \ \ \ \ \ \ \ \ \ \ \ \ \ \ \ \ \ \ \ \ \ \ \ \ \ \ \ \ \
\begin{tikzcd}
   \DWA \arrow[dd, color=blue, "{(\linv, \, \teta)}"] \arrow[rr, dashed, "G^\epsilon"]& &  \cZ(\cC) \arrow[dd, dashed, "\calN^0"]   \\
    & & \\
    \text{SPT}(G)_{\epsilon^{-1}}  & & \text{DW}(G) \arrow[ll, dashed , "\text{Rep}(G)"]
\end{tikzcd}
\eea
Intuitively one should think of $\calN^0$ as the image of $\linv$ after gauging the $G$ symmetry.
Strictly speaking our reasoning is rigorous only in the case of 3d TFTs, where the concepts above can be explicitly implemented. We however expect the same ideas also to apply to the higher categorical setting, once a complete definition of the higher Symmetry TFT is given.
We thus arrive at the following proposal for the anomalies of duality defects:

\paragraph{First obstruction.}
There exist, in $\DWA$, a $G$-invariant boundary condition $(\linv, \, \teta)$. In the language of \cite{Kaidi:2022cpf} the duality symmetry is non-intrinsic. We further explain in Appendix \ref{app: invTFT} that this condition is equivalent to the existence of a duality-invariant TQFT \cite{Thorngren:2021yso, Apte:2022xtu}. A similar analysis connecting these two concepts when $\abA$ is cyclic has recently appeared \cite{Sun:2023xxv}. Our results coincide with theirs when they overlap.

\paragraph{Second obstruction.}
The (invertible) duality symmetry is anomaly-free. This is equivalent to the vanishing of the total Dijkgraaf-Witten twist, which depends on the equivariantization data $\tilde\eta$. In practice the invertible duality symmetry suffers from a mixed anomaly with (a subgroup of) the symmetry $\cS$ on the non-intrinsic boundary which can be computed explicitly. We conjecture (and show in examples) that the equivariantization data encode how the 0-form symmetry fractionalizes with the $(n-1)$-form symmetry $\cS$. Following \cite{Delmastro:2022pfo} this can be used to shift the value of the cubic anomaly $\epsilon$, \ie{} to change the SPT phase $Y$ in the bulk.

\section{Anomalies of duality symmetries in 1+1 dimensions}
\label{sec: 3d}

This section is devoted to the discussion of anomalies in two-dimensional QFTs whose symmetries are described by Tambara-Yamagami (TY) categories \cite{tambara1998tensor}. We indicate such categories as $\TYA$ and we review their definition in Section~\ref{sec: algebras TY}.
The results are already known both in the physics and mathematics literature \cite{meir2012module, Thorngren:2019iar} and here we present their derivation from the point of view of the Symmetry TFT which can be generalized to higher dimensions. Physically $\TYA$ is the symmetry of a 2d theory which is self-dual under the gauging of an Abelian symmetry $\bA$. Examples, especially in the realm of 2d CFTs, are ubiquitous, the most famous one being the realization of Kramers-Wannier duality in the Ising CFT \cite{Chang:2018iay} for $\bA = \bZ_2$. Many other examples come from either considering free theories%
\footnote{Examples include the $c=1$ boson at squared radius $R^2= 2k$, its $\bZ_2$ orbifold, or multiple bosons at special points on the Narain moduli space.}
as described for instance in \cite{Thorngren:2021yso}, or other minimal models which can flow to the Ising CFT, such as the $c=7/10$ tricritical Ising CFT.  

As a concrete example of anomalous versus non-anomalous theories, the $\text{TY}(\bZ_2)_{\gamma, 1}$ symmetry of the Ising CFT is anomalous on non-spin manifolds,%
\footnote{On spin manifolds it can be fermionized to the $(-1)^{F_L}$ symmetry of the Majorana CFT.}
while the diagonal $\text{TY}(\bZ_2\times \bZ_2)_{\gamma, 1}$ symmetry of $(\text{Ising})^2$ is anomaly-free. The question of which TY categories are anomalous is not a purely academic one, as it can imply strong constraints on duality-preserving RG flows \cite{Chang:2018iay}. For instance, the aforementioned tricritical Ising model has an \emph{anomalous} non-invertible symmetry as well as a duality-preserving relevant deformation.
As a direct consequence of the anomaly, the resulting RG flows cannot end in a trivially gapped theory. Depending on the sign of the deformation, the theory either flows to the gapless Ising model or to a gapped theory with three degenerate vacua.

Our Symmetry TFT analysis gives a simple characterization of the known obstruction theory in two steps, as already pointed out in Section \ref{sec:proposal}. The first obstruction is equivalent to the absence of a duality-invariant Lagrangian algebra $\linv$, which otherwise gives rise to a global variant with invertible symmetry $\cS \rtimes_\rho \bZ_2$. The second obstruction is instead the standard 't~Hooft anomaly for $\bZ_2$ subgroups of $\cS \rtimes_\rho \bZ_2$ in that global variant. When there exists such an anomaly-free $\bZ_2$ subgroup, it can be gauged. The combined sequential gauging of $\linv$ and of $\bZ_2$ corresponds, in the original theory, to a gauging that involves the non-invertible symmetry defect.

\subsection{Algebras in TY categories and anomalies}
\label{sec: algebras TY}

We start by reviewing the basic properties of Tambara-Yamagami categories $\TYA$ \cite{tambara1998tensor}. We assume that the reader has some familiarity with the theory of Fusion Categories, for which excellent reviews are \cite{Fuchs:2002cm, Kitaev:2005hzj, Barkeshli:2014cna, etingof2016tensor, Bhardwaj:2017xup, Chang:2018iay}. 
Given an Abelian group $\bA$, the Tambara-Yamagami symmetry is a $\bZ_2$-graded fusion category
\be
\cC = \cC_{\underline{0}} \oplus \cC_{\underline{1}} \;,
\ee
where $\cC_{\underline{0}} = \text{Vec}_\bA$ (the category of $\bA$-graded vector spaces) describes an Abelian 0-form symmetry $\bA$ with trivial anomaly,%
\footnote{Such an anomaly would be represented by a trivial class $[0] \in H^3 \bigl( \bA, U(1) \bigr)$.}
while $\cC_{\underline{1}}$ has a single simple object $\cN$. The fusion rules consistent with the grading are uniquely fixed and read:
\be
\label{fusion rules of TY}
a \times b = (a + b) \;,\qquad\qquad a \times \cN = \cN \times a = \cN \;,\qquad\qquad \cN \times \cN = \bigoplus\nolimits_{a \in \bA} a \;.
\ee
Here $a,b \in \bA$ are the simple objects in $\cC_{\underline{0}}$, and $+$ in the first equation is the binary group operation in $\bA$.
The category is uniquely determined by a triplet $(\bA, \gamma, \epsilon )$ where $\gamma: \bA \times \bA \to U(1)$ is a non-degenerate symmetric bicharacter, whilst $[\epsilon] \in H^3 \bigl( \bZ_2, U(1) \bigr) \cong \bZ_2$ is the Frobenius-Schur indicator of the self-dual defect $\cN$. This data enters in the associator, or $F$-symbol, of the TY category:
\be
\Bigl[ F^{a,\, \cN ,\, b}_\cN \Bigr]_{\cN,\, \cN} = \Bigl[ F^{\cN,\, a ,\, \cN}_b \Bigr]_{\cN,\, \cN} = \gamma(a,b) \;,\qquad\qquad
\Bigl[F^{\cN,\, \cN,\, \cN}_\cN \Bigr]_{a,\, b} = \frac{\epsilon}{\sqrt{\lvert \bA \rvert}} \, \gamma(a, b)^{-1} \;,
\ee
where $\epsilon = \pm1$, while all other non-vanishing associators are equal to 1. The bicharacter $\gamma$ has a nice physical interpretation (see, \eg,  \cite{Thorngren:2019iar}). Since a theory $\cT$ with TY symmetry is mapped to itself under the gauging of $\bA$, symbolically $\cT/\bA \cong \cT$, we can consider constructing the defect $\cN$ as a topological domain wall between $\cT$ and $\cT/\bA$:
\bea
\begin{tikzpicture}
    \filldraw[color=white, left color=white!50!red, right color= white] (0,0) -- (2,0) -- (2, 1.5) -- (0, 1.5) -- cycle; 
    \draw[color= blue, line width = 1.5] (0,0) -- (0, 1.5);
    \node at (1, 0.75) {$\cT/\bA$}; \node at (-1, 0.75) {$\cT$}; \node[below] at (0,0) {$\cN$};
\end{tikzpicture}
\eea
On the two sides of the wall the 0-form symmetries are $\bA$ and $\bA^\vee$, respectively. 
To identify them we must specify a group isomorphism $\phi: \bA \to \bA^\vee$ such that the associated bicharacter $\gamma$ defined as
\be
\gamma(a_1, a_2) = \phi(a_1) \, a_2 \in U(1)
\ee
is symmetric.%
\footnote{The requirement that $\gamma$ be symmetric is equivalent to the canonical pairing between $\bA^\vee$ and $\bA$ being invariant under $\phi$: $\alpha(a) = \phi(a) \bigl( \phi^{-1}(\alpha) \bigr)$ where $a$ and $\alpha$ are elements of $\bA$ and $\bA^\vee$, respectively. Physically this translates into the fact that the electric and magnetic frames are equivalent due to self-duality.}
We can then consider lines $a \in \bA$ and $\alpha \in \bA^\vee$ ending on the defect $\cN$ from the two sides. The endpoints of these objects are mutually nonlocal, and pick up a canonical phase $\alpha(a)$ as they pass through each other. Indeed, from the point of view of the left side, $a$ is a topological symmetry defect for the symmetry $\bA$ and the endpoint of $\alpha$ is a charged operator (while vice versa from the point of view of the right side).
Using the isomorphism $\phi$ this is converted into the symmetric bicharacter $\gamma$:
\bea
\begin{tikzpicture}
\filldraw[color=white, left color=white!50!red, right color= white] (0,0) -- (2,0) -- (2, 1.5) -- (0, 1.5) -- cycle; 
    \draw[color= blue, line width = 1.5] (0,0) -- (0, 1.5);
    \draw[dashed] (-1.5, 0.5) -- (0, 0.5); \draw[dashed] (0, 1) -- (2, 1); \draw[fill=black] (0, 0.5) circle (0.05); \draw[fill=black] (0, 1) circle (0.05); \node[above] at (-1, 0.5) {$a$}; \node[below] at (1.5, 1) {$\phi(b)$};
    \node[right] at (2.5, 0.75) {$ = \quad \gamma(a, b)$};
    \begin{scope}[shift={(6.5, 0)}]
    \filldraw[color=white, left color=white!50!red, right color= white] (0,0) -- (2,0) -- (2, 1.5) -- (0, 1.5) -- cycle;
         \draw[color= blue, line width = 1.5] (0,0) -- (0, 1.5);
    \draw[dashed] (-1.5, 1) -- (0,1); \draw[dashed] (0, 0.5) -- (2, 0.5); \draw[fill=black] (0, 0.5) circle (0.05); \draw[fill=black] (0, 1) circle (0.05); \node[below] at (-1, 1) {$a$}; \node[above] at (1.5, 0.5) {$\phi(b)$};
    \end{scope}
\end{tikzpicture}
\eea

We now review the classification of \emph{bosonic} gaugeable symmetries $\cA$ in $\TYA$, which are described by symmetric Frobenius algebras in $\cC$ \cite{Fuchs:2002cm}. These are defined by an object $\cA \in \cC$ together with a choice of three-valent junction $m: \cA \times \cA \to \cA$ which is strictly associative:
\bea
\begin{tikzpicture}
    \coordinate (g_hk) at (1.5, 1.2); \coordinate (h_k) at (2.25, 0.6); \coordinate (g_h) at (0.75, 0.6); \coordinate (gh_k) at (1.5, 1.2); \coordinate (g) at (0, 0); \coordinate (h) at (1.5, 0); \coordinate (k) at (3, 0); \coordinate (ghk) at (1.5, 1.8);
    \draw[color=red!70!black, line width=1] (g) node[below] {$\cA$} to (g_h) to (gh_k) to (ghk) node[above] {$\cA$};
    \draw[color=red!70!black, line width=1] (h) node[below] {$\cA$} to (g_h);
    \draw[color=red!70!black, line width=1] (k) node[below] {$\cA$} to (gh_k);
    \draw[fill=red] (g_h) circle (0.1) node[above left] {$m$};
    \draw[fill=red] (gh_k) circle (0.1) node[above left] {$m$};
    \node at (4.5, 0.9) {$=$};
\begin{scope}[shift={(6,0)}]
    \coordinate (g_hk) at (1.5, 1.2); \coordinate (h_k) at (2.25, 0.6); \coordinate (g_h) at (0.75, 0.6); \coordinate (gh_k) at (1.5, 1.2); \coordinate (g) at (0, 0); \coordinate (h) at (1.5, 0); \coordinate (k) at (3, 0); \coordinate (ghk) at (1.5, 1.8);
    \draw[color=red!70!black, line width=1] (k) node[below] {$\cA$} to (h_k) to (gh_k) to (ghk) node[above] {$\cA$};
    \draw[color=red!70!black, line width=1] (h) node[below] {$\cA$} to (h_k);
    \draw[color=red!70!black, line width=1] (g) node[below] {$\cA$} to (g_hk);
    \draw[fill=red] (h_k) circle (0.1) node[above right] {$m$};
    \draw[fill=red] (g_hk) circle (0.1) node[above right] {$m$};
\end{scope}
\end{tikzpicture}
\eea
This diagram encodes the cancellation of 't~Hooft anomalies for the symmetry $\cA$. We give a brief review of the relevant concepts in Appendix \ref{app:Frobenius}.
The classification problem has been solved in the mathematics literature in \cite{meir2012module} and given a physical interpretation from the viewpoint of TQFTs in \cite{Thorngren:2019iar}. 

Such algebras for the TY category can be divided into two types depending on whether $\cA$ also contains the element $\cN$ or not. If not, the gaugeable algebras correspond to gauging a subgroup $\bB$ of $\bA$ with discrete torsion $[\nu] \in H^2 \bigl( \bB, U(1) \bigr)$. 
From the latter one constructs its commutator%
\footnote{Notice that $\chi_\nu$ is well defined (it is independent from the choice of representative $\nu$), alternating (meaning that $\chi_\nu(a,a)=1$) and antisymmetric (meaning that  $\chi_\nu(a,b) = \chi_\nu(b,a)^{-1}$). One can prove that  $\chi_\nu : \bB \times \bB \rightarrow U(1)$ is bilinear (in the multiplicative sense), see for instance \cite{mooregroup}. Then alternating implies antisymmetric, while the opposite is false and in fact dropping the alternating property one can describe fermionic Lagrangian algebras, see also after (\ref{vanishing spin in 3d}).}
\be
\label{def chi_nu}
\chi_\nu(b_1,b_2) = \frac{\nu(b_1,b_2)}{\nu(b_2,b_1)} \;.
\ee
This defines a map $[\nu]\rightarrow \chi_\nu$ from $H^2\bigl( \bB, U(1) \bigr)$ to the group of alternating bicharacters which turns out to be an isomorphism \cite{tambara2000representations}. The Frobenius algebra corresponding to the pair $\bigl( \bB, [\nu] \bigr)$ is:
\be
\cA \equiv \cA_{\bB, \, \nu} = \bigoplus\nolimits_{b \in \bB} b \;,\qquad\qquad\qquad m_{b,\, b'}^{b + b'} = \nu(b, b') \;.
\ee
On the other hand, when including also $\cN$ the most general algebra reads:%
\footnote{Notice that, if we restrict to spin manifolds, there are more candidate algebras because it is possible to gauge discrete symmetries with a nontrivial Arf twist (\ie, a discrete torsion constructed out of the spin connection, see \eg{} \cite{Karch:2019lnn, Gaiotto:2020iye}). This difference will become apparent in the Symmetry TFT.}
\be
\label{2d algebra gauging TY}
\cA \equiv \cA_{\underline{0}} \oplus \cA_{\underline{1}} = \cA_{\bB,\, \nu} \oplus n_\nu \, \cN \;,  \qquad\qquad n_\nu \geq 1 \;.
\ee
The choices of $\bB$, $\nu$ and $n_\nu$ for which such an object can be consistently defined on orientable 2-manifolds are severely restricted by the following two obstructions \cite{tambara2000representations, meir2012module, Thorngren:2019iar}.

\paragraph{First obstruction.} 
We introduce the subgroup $N(\bB) \subset \bA^\vee$ of characters annihilating $\bB$:
\be
\label{def N(B)}
N(\bB) = \Bigl\{ \beta \in \bA^\vee \Bigm| \beta(b)=1 \;,\; \forall b \in \bB \Bigr\} \;.
\ee
This group is canonically isomorphic to $(\bA/\bB)^\vee$, while the quotient $\bA^\vee / N(\bB)$ is canonically isomorphic to $\bB^\vee$. We also define the \emph{radical} $\text{Rad}(\nu) \subset \bB$ of the class $[\nu]$:
\be
\label{def Rad(nu)}
\text{Rad}(\nu) = \Bigl\{ b \in \bB \Bigm| \chi_\nu(b,b') = 1 \;,\; \forall b' \in \bB  \Bigr\} \;.
\ee
The alternating bicharacter $\chi_\nu$ is non-degenerate on $\bB/\text{Rad}(\nu)$.

For the first obstruction to vanish these subgroups must be related as
\be
\label{eq:firstob1}
\phi \bigl( \text{Rad}(\nu) \bigr) = N(\bB) \;.
\ee
Besides, there must exist an involutive automorphism
\be
\label{eq:firstob2}
\sigma: \bB/\text{Rad}(\nu) \,\to\, \bB/ \text{Rad}(\nu) \qquad\text{with}\qquad \sigma^2 = 1
\ee
such that the symmetric and alternating bicharacters, when restricted to $\bB$ and projected to $\bB/\text{Rad}(\nu)$, satisfy
\be
\label{eq:firstob3}
\gamma \bigl( \sigma(a), b \bigr) = \chi_\nu(a,b) \qquad\qquad\text{for}\quad a, b \in \bB/\text{Rad}(\nu) \;.
\ee
Note that the projections from $\bB$ to $\bB/\text{Rad}(\nu)$ are well defined.
One can prove, using the equation above, that $\nu(a,b)$ and $\nu \bigl( \sigma(b), \sigma(a) \bigr)$, when projected to $\bB/\text{Rad}(\nu)$, define equivalent cohomology classes in $H^2 \bigl( \bB/\text{Rad}(\nu), U(1) \bigr)$.%
\footnote{To prove it, one checks that the 2-cochains $\nu(a,b)$ and $\nu\bigl( \sigma(b), \sigma(a) \bigr)$ produce the same bicharacter $\chi_\nu$ in (\ref{def chi_nu}) and so, by isomorphism, must be different representatives of the same cohomology class. \label{foo: exactness}}
Thus there exists a 1-cochain $\teta \in H^1\bigl( \bB/\Rad(\nu), U(1) \bigr)$ such that%
\footnote{It is always possible to choose $\teta$ such that it satisfies both relations. Consider the case $\bB=\bA$ and define the two subgroups $\bA^{\sigma} = \{a \in \bA\, | \, a = \sigma(a)\}$ as well as $\bA_{\sigma} = \{a + \sigma(a) \, | \, a \in \bA\}$, clearly $\bA_{\sigma}\subset \bA^{\sigma} \subset \bA$. It is easy to see (see \eg \cite{tambara2000representations}) that $\gamma$ can be consistently reduced to a bicharacter $\bar{\gamma}$ on the quotient $\bA^{\sigma}/\bA_{\sigma}$, \ie{} $\gamma(a+ b+\sigma(b), a'+b'+\sigma(b'))= \gamma(a,a')$ for any $a,a' \in \bA^{\sigma}$. Now take the first equation and restrict it to $\bA^{\sigma}$, since $\gamma$ is well defined on $\bA^{\sigma}/\bA_{\sigma}$ we rewrite it as
\be
d \teta(a,b) = \frac{{\nu}(a,b) }{{\nu}(b,a)} = \chi_{\nu}(a,b) = \gamma(a,b) = \bar{\gamma}(\pi(a), \pi(b)) \qquad\Rightarrow\qquad d\teta(a,b) = d \mu (\pi(a), \pi(b))
\ee
where we used \eqref{eq:firstob3}, $\pi$ is the projection $\bA^{\sigma}\rightarrow \bA^{\sigma}/\bA_{\sigma}$ and $\mu: \bA^{\sigma}/\bA_{\sigma} \rightarrow U(1) $  is a quadratic refinement of $\bar{\gamma}$.
It follows that there is always a solution $\teta$ that, restricted on $A^{\sigma}$, takes the form $\teta(a) = \mu(\pi(a))$ and hence obeys $\teta|_{\bA_{\sigma}} =\teta(a) \teta(\sigma(a)) = 1 $. The general case for $\bB \subset \bA$ is a straightforward generalization.}
\be
\label{eq:firstob4}
\frac{{\nu}(a,b) }{{\nu} \bigl( \sigma(b), \sigma(a) \bigr)} = d \teta(a,b) \;, \qquad\qquad \teta(a) \:  \teta\bigl( \sigma(a) \bigr) = 1 \;.
\ee
From (\ref{def N(B)})--(\ref{eq:firstob1}) it easily follows that
\be
\label{little n}
|\bA| \, n_\nu^2 = |\bB|^2 \;,
\ee
where%
\footnote{Since $\chi_\nu$ is a non-degenerate alternating bicharacter on $\bB/\Rad(\nu)$ with values in $U(1)$, there exists an isotropic subgroup $\bG$ such that $\bB/\Rad(\nu) = \bG \times \bG^\vee$ and in particular $\bigl\lvert \bB/\Rad(\nu) \bigr\rvert = |\bG|^2 = n_\nu^2$ is a perfect square, where $n_\nu = |\bG|$. See \eg{} Lemma 5.2 in \cite{Davydov:2007}.}
$n_\nu^2 = \bigl\lvert \bB/\Rad(\nu) \bigr\rvert$. The positive integer $n_\nu$ appearing here turns out to be the same as the one in (\ref{2d algebra gauging TY}). Notice in particular that a necessary condition to satisfy the first obstruction is that $|\bA|$ is a perfect square. Since, as it follows from \eqref{fusion rules of TY}, the quantum dimension of $\cN$ is $|\bA|^{\frac{1}{2}}$, this reproduces the known fact that gauging is not possible in presence of non-integer quantum dimensions \cite{Chang:2018iay}.

The rough idea that leads to these formulas is the following. Decomposing the defining equation of a Frobenius algebra into its graded components, one finds that $\cA_{\underline{1}}$ must be an invertible $\cA_{\underline{0}}$-bimodule: $\cA_{\underline{1}} \times_{\cA_{\underline{0}}} \cA_{\underline{1}} = \unit_{\cA_{\underline{0}}}$, where $\times_{\cA_{\underline{0}}}$ is the tensor product in the category of $\cA_{\underline{0}}$-bimodules \cite{Fuchs:2002cm, Fuchs:2004dz, etingof2016tensor}.
Physically this means that we can gauge $\cA_{\underline{0}}$, and then $\cA_{\underline{1}}$ will become an invertible $\bZ_2$ global symmetry of the gauged theory.
Eqns.~\eqref{eq:firstob1}--\eqref{eq:firstob3} are necessary in order to endow $\cA_{\underline{1}}$ with a bimodule structure, and in particular \eqref{eq:firstob4} ensures that the bimodule is invertible. This furthermore implies that $\text{dim}(\cA_{\underline{1}}) = \text{dim}(\cA_{\underline{0}})$ which reproduces (\ref{little n}) in terms of the integer $n_\nu$ in (\ref{2d algebra gauging TY}).

A beautiful alternative perspective on this condition has been given in \cite{Thorngren:2019iar}. Suppose we want to construct a TQFT in which the duality symmetry generated by $\cN$ is preserved. As the TQFT must have symmetry $\bA$, it can be labelled by a doublet $(\bB, \nu)$ where $\bB$ denotes the preserved (as opposed to spontaneously broken) subgroup while $\nu$ is an SPT phase for $\bB$.
The partition function is
\be
\label{Z of 2d TQFT}
Z[B] = \begin{cases}
\exp \bigl( 2 \pi i \int\! B^* \nu \bigr) \quad &\text{if } \pi(B) = 0 \;, \\
0 &\text{otherwise} \;.
\end{cases}
\ee
Here $B$ is a background field coupled to $\bA$, whilst $\pi$ is projection map of the short exact sequence $1 \to \bB \xrightarrow{i\,} \bA \xrightarrow{\pi\,} \bA/\bB \to 1$.
The class $B^*\nu \in H^2\bigl( X_2, U(1) \bigr)$ (using now additive notation) is integrated over the spacetime 2-manifold $X_2$.%
\footnote{The class $B^*\nu$ can be thought of in two ways. Abstractly, $B$ is a homotopy class of maps from $X_2$ to the classifying space $B\bA$, while $\nu$ is a 2-form in $H^2\bigl( B\bA, U(1)\bigr)$, so that the pull back $B^*\nu$ is a 2-form on $X_2$ that can be integrated. More concretely, $B \in H^1(X_2, \bA)$ is a 1-cochain in simplicial cohomology that to each edge $(ij)$ of a triangulation of $X_2$ associates an element of $\bA$, while $\nu:\bA \times \bA \to U(1)$ is an element of group cohomology $H^2\bigl( \bA, U(1) \bigr)$, so that the 2-cochain $(B^*\nu)_{ijk} = \nu(b_{ij}, b_{jk}) \in H^2\bigl( X_2, U(1) \bigr)$ associates to each face $(ijk)$ a value in $U(1)$ and can be integrated.
\label{foo: pull back}}
Imposing that the duality symmetry be unbroken means that
\be
\label{eq:invTQFT1}
\cN \cdot Z[B] \,\equiv\, \frac{1}{\sqrt{ \rvert H^1(X, \bA) \lvert }} \, \sum_{a \, \in \, H^1(X, \bA)} \exp \biggl( 2\pi i \int_X a \cup \phi(B) \biggr) \, Z[a] \,\stackrel{!}{=}\, Z[B] \;,
\ee
where we used the symmetric bicharacter to identify $\bA^\vee$ with $\bA$. It can be checked that \eqref{eq:invTQFT1} reproduces exactly the first obstruction condition. We report the detailed manipulation and its extension to the four-dimensional case in Appendix~\ref{app: invTFT}.

\paragraph{Second obstruction.}
Having discussed the structure of $\cA_{\underline{1}}$, we can simply gauge $\cA$ sequentially by first gauging $\cA_{\underline{0}}$ and then $\cA_{\underline{1}}$. After the first step, $\cA_{\underline{0}}$ becomes the identity defect while $\cA_{\underline{1}}$ becomes an invertible $\bZ_2$ symmetry: $\cA_{\underline{1}}^2 = 1$. In order to be able to gauge the full algebra, it must happen that $\cA_{\underline{1}}$ has a trivial self-anomaly $\epsilon_\text{tot}$. This comes in two parts: a ``bare" contribution from the original Frobenius-Schur indicator $\epsilon$ of the duality defect, and a further contribution $Y$ coming from the bimodule morphism $\cA_{\underline{1}} \times \cA_{\underline{1}} \to \unit$. The latter turns out to be given by the Arf invariant of $\teta$ restricted to the elements of $\bB/\text{Rad}(\nu)$ invariant under the involution $\sigma$:
\be
\label{eq: secondob}
Y = \sign \Biggl( \;\; \raisebox{0.5em}{$\ds\sum_{\substack{ b \,\in\, \bB/\text{Rad}(\nu) \\ \sigma(b) \,=\, b}} $} \, \teta(b) \; \Biggr) = \operatorname{Arf}(\teta) \;.
\ee
We stress that here we are using multiplicative notation for $\teta$, so that $Y$ is the sign of a sum of phases (alternatively, in (\ref{eq:spin arf}) we indicate the correct normalization).
The second obstruction then vanishes if and only if
\be
\epsilon_\text{tot} = \epsilon \, Y = 1 \;.
\ee

Later on, around eqn.~(\ref{new formula for Y 2d}), we will find an alternative formula for the spectrum of values that $Y$ can take as we explore the possible consistent choices of $\tilde\eta$ --- the so-called fractionalization classes.

\subsubsection{A note on quadratic refinements}
\label{sec: quadratic refs}

At various points in this work we use the existence and properties of quadratic refinements.

A function $q:\bA \to U(1)$ (with $\bA$ a finite Abelian group) is called a quadratic function if $q(a) = q(-a)$ and (using multiplicative notation)
\be
\label{def associated symm bichar}
\zeta(a,b) \,\equiv\, \frac{ q(a+b) }{ q(a) \, q(b) }
\ee
is a symmetric bicharacter. One easily derives that $q(0) = 1$, $q(ta) = q(a)^{t^2}$ for any $t \in \bZ$, and
\be
\zeta(a,a) = q(a)^2 \;.
\ee
Any quadratic function $q$, by definition, comes equipped with an associated symmetric bicharacter $\zeta$ as in (\ref{def associated symm bichar}).
However also the converse is true: any symmetric bicharacter $\zeta$ arises from a (not necessarily unique) quadratic function $q$, which is called a quadratic refinement of $\zeta$. The set of quadratic refinements forms a torsor over $\Hom(\bA, \bZ_2)$, indeed one easily proves that the ratio of two quadratic refinements is a $\bZ_2$-valued character on $\bA$.%
\footnote{The set of quadratic functions $q:\bA \to U(1)$ is an extension of the group of symmetric bicharacters $\zeta:\bA \times \bA \to U(1)$ by $\Hom(\bA, \bZ_2)$. For each bicharacter, a quadratic function is easily constructed. For $\bA = \bZ_n$ the bicharacters are $\zeta_r(a,b) = \exp\bigl( \frac{2\pi i r}n ab \bigr)$ with $r \in \bZ_n$. Given one of them, a quadratic refinement is $q_r(a) = \exp \bigl( \frac{\pi i r (n+1)}n a^2 \bigr)$. If $n$ is odd then $r \in \bZ_n$ and the quadratic function is unique. If $n$ is even then $r \in \bZ_{2n}$ and the quadratic functions for $r$ and $r+n$ produce the same bicharacter $\bZ_r$. The case that $\bA$ is a product of cyclic factors is similar.}

A closely related statement is that any symmetric 2-cocycle $\nu \in Z^2\bigl( \bA, U(1) \bigr)$ is exact. This follows from the isomorphism between altenating bicharacters $\chi_\nu$ (\ref{def chi_nu}) and $H^2\bigl( \bA, U(1) \bigr)$. In the special case that the symmetric 2-cocycle $\nu(a,b)$ is bilinear, exactness is equivalent to the existence of a quadratic refinement.

\subsection{Symmetry TFT description}

It is possible to reformulate the properties of Tambara-Yamagami categories $\TYA$ in terms of their 3d Symmetry TFT, \ie, using the language of modular tensor categories (MTCs).%
\footnote{Given a fusion category $\cC$ as the symmetry of some 2d theory, the corresponding 3d Symmetry TFT is given via the Turaev-Viro construction \cite{turaev1992state} by the TQFT whose MTC is the \emph{Drinfeld center} of $\cC$ denoted $\cZ(\cC)$.}
Let us review this fact, that will be useful in order to discuss anomalies in the following sections.
In particular let us describe how the data $(\bA, \gamma, \epsilon)$ appears from the bulk viewpoint.

One starts from a pure 3d gauge theory for $\bA$ (\ie, a Dijkgraaf-Witten theory for $\bA$ with no torsion), which is the Symmetry TFT describing the invertible symmetry $\text{Vec}_\bA$.%
\footnote{Mathematically this corresponds to the fact that the Drinfeld center of $\text{Vec}_\bA$ is $\bA \times \bA^\vee$.}
The spectrum of lines of the $\bA$ gauge theory is $\bA \times \bA^\vee$, the lines being labelled 
by pairs $(a, \alpha) \in \bA \times \bA^{\vee}$. All the $F$-symbols are trivial while the braiding is canonically determined by the pairing between $\bA$ and $\bA^{\vee}$:
\be
    \cB_{(a_1, \alpha_1), (a_2, \alpha_2)} = \alpha_1(a_2) \, \alpha_2(a_1) \;.
\ee
It follows that the topological spins are 
\be
    \theta_{(a, \alpha)} = \alpha(a) \;.
\ee
Crucially, the theory enjoys \textit{electric-magnetic} (EM) duality due to $\bA$ and $\bA^{\vee}$ being isomorphic. More precisely, the choice of an isomorphism $\phi$ naturally induces an automorphism of the Drinfeld center
\be
\begin{array}{cccc}
\Phi : & \bA\times \bA^{\vee} & \rightarrow  & \bA\times \bA^{\vee} \\
& (a,\alpha) & \mapsto & \bigl( \phi ^{-1}(\alpha), \phi(a) \bigr) \;.
\end{array}
\ee
However not all choices of isomorphism are consistent EM dualities since $\Phi$ needs to preserve the braiding. This condition is equivalent to the bicharacter $\gamma(a,b) = \phi(a)\, b$ associated with $\phi$ being symmetric. Note that $\Phi$ squares to 1, so that the duality group is $G \cong \bZ_2$.

If the boundary theory is self-dual under gauging, we can construct the full Symmetry TFT that includes the duality defect by gauging the duality symmetry $G$ \cite{Kaidi:2022cpf}. The gauging operation comes with a choice of discrete torsion $\epsilon \in H^3 \bigl( G, U(1) \bigr) \cong \bZ_2$ which translates to the Frobenius-Schur indicator of the duality defect $\cN$ on the boundary.
To summarize, the data $(\bA, \gamma, \epsilon)$ of the boundary Tambara-Yamagami category appear from the bulk viewpoint as the choice of a duality symmetry $G \cong \bZ_2$ of the $\bA$ Dijkgraaf-Witten theory and of a discrete torsion for the gauging.

To properly discuss the gauged theory, we first describe the data of the 3d Dijkgraaf-Witten theory enriched by the 0-form symmetry (a $G$-crossed category in the language of \cite{Barkeshli:2014cna}). This includes data describing the topological twist defects for the $G \cong \bZ_2$ symmetry.
The full tensor category is graded:
\be
\cZ(\text{Vec}_\bA)_{\bZ_2} = \cZ(\text{Vec}_\bA) \oplus \cZ(\text{Vec}_\bA)_\Phi \;,
\ee
where $\cZ(\text{Vec}_\bA)_\Phi$ describes the twisted sector for the $\bZ_2$ symmetry. The number of simple components of $\cZ(\text{Vec}_\bA)_\Phi$ is the same as the number of $\Phi$-invariant anyons \cite{Barkeshli:2014cna}. The latter are all of the form $\bigl(a , \phi(a) \bigr)$ with $a \in \bA$. Thus there are $|\bA|$ simple objects in $\cZ(\text{Vec}_\bA)_\Phi$ which we denote as $\sigma_a$, $a \in \bA$ (not to be confused with the involution $\sigma$).
The fusion and braiding data for the $\bZ_2$ extension have been computed in \cite{gelaki2009centers}, although we use here a slightly different notation similar to the one employed in \cite{Kaidi:2022cpf}. We find
\bea
\label{eq: fusiontft}
(a, \alpha) \times (b, \beta) &= (a + b ,\, \alpha + \beta) \;,\qquad\qquad\qquad (a, \alpha) \times \sigma_b = \sigma_{b + a + \phi^{-1}(\alpha)} \\
\sigma_a \times \sigma_b &= \bigoplus_{c \,\in\, \bA} \, \bigl( a + b + c ,\, \phi(-c) \bigr) \;.
\eea
These fusion rules are derived by realizing the $G$ symmetry defects as 2d condensates \cite{Roumpedakis:2022aik} of the anti-diagonal lines $\bigl( a, \phi(-a) \bigr)$ (see \eg{} \cite{Kaidi:2022cpf} for the case $\bA = \bZ_n$).%
\footnote{Indeed, the anti-diagonal lines are absorbed by the $\sigma_b$'s, and $\sigma_b \times \sigma_{-b}$ is a 1d condensate of anti-diagonal lines.}
Since the quantum dimension of $(a,\alpha)$ is 1, we also have
\be
\dim( \sigma_a ) = \sqrt{|\bA|} \;.
\ee
The non-vanishing $R$-matrices, in a gauge, are given by \cite{gelaki2009centers}:
\bea
\label{choice of R-matrices}
R_{(a_1 ,\, \alpha_1) ,\; (a_2 ,\, \alpha_2)}^{(a_1 + a_2 ,\; \alpha_1 + \alpha_2)} &= \alpha_2(a_1) \;, \qquad\qquad& R_{(a_1 ,\, \alpha_1) ,\; \sigma_{a_2}}^{ \sigma_{a_1 + a_2 + \phi^{-1}(\alpha_1)}} &= f_{a_2}(a_1) \;, \\[.5em]
R_{\sigma_{a_1} ,\; (a_2 ,\, \alpha_2)}^{\sigma_{a_1 + a_2 + \phi^{-1}(\alpha_2)}} &= 1 \;, \qquad\qquad& R_{\sigma_{a_1} ,\; \sigma_{a_2}}^{(a_3, \alpha_3)} &= f_{a_1}(-a_3)^{-1} \;.
\eea
In the last entry, $(a_3, \alpha_3)$ must be a fusion channel of $\sigma_{a_1} \times \sigma_{a_2}$. Besides, $f_a : \bA \to U(1)$ is a collection (for $a \in \bA$) of functions given by $f_a = \phi(a) \cdot f_0$, or more explicitly $f_a(b) = \gamma(a,b) \, f_0(b)$, required to satisfy
\be
\label{def torsor f_a}
f_a(b) \; f_a(b') = \gamma(b, b') \; f_a(b + b') \;.
\ee
Notice that the equations for different values of $a$ are all equivalent. In these equations, distinct choices for $f_0$ differ by an $\bA$-character and only reshuffle the $f_a$'s, therefore the set of $f_a$'s forms a torsor over $\bA^\vee$. However $f_0$ should be chosen such that $f_0(b) = f_0(-b)$, in other words $f_0$ is a quadratic refinement of $\gamma^{-1}$, which alwasys exists (see Section~\ref{sec: quadratic refs}). Possible different choices are related by $\Hom(\bA, \bZ_2)$ and correspond to different gauge choices. The (gauge dependent) spins of the twisted sector lines are \cite{gelaki2009centers}:
\be
\label{spins in twisted sector}
\theta(\sigma_a) = \sqrt{\frac{1}{|\bA|^{1/2}} \sum_{b \, \in \, \bA} \; f_a(b)^{-1} } \;,
\ee
where the choice of sign for the square root is gauge.

We can now discuss the gauging of the symmetry $G \cong \bZ_2$ with a twist $\epsilon \in H^3 \bigl( G,  U(1) \bigr) \cong \bZ_2$. The gauged theory $\cZ(\text{Vec}_{\bA})_{\bZ_2} / \bZ_2$ is isomorphic to $\cZ \bigl( \text{TY}(\bA)_{\gamma, \epsilon} \bigr)$ and is graded by the quantum $\bZ_2$ 1-form symmetry whose charged objects are the liberated twisted sectors $\sigma_a$. There are three types of objects, whose properties are summarized in Table~\ref{tab: objs in 3d Symm TFT}.
\begin{table}[t]
$$
\begin{array}{c|c|c|c|c}
\text{Object}  & \text{Definition} & \text{Dim} & \# \ \text{of Objects} & \text{Spin } \theta \\[0.5em] \hline \hline \rule[-1em]{0pt}{2.8em}
    L_{(a, \, x)}  & \eta^x \times \bigl( a ,\, \phi(a) \bigr) & 1 & 2 \, \lvert \bA \rvert & \gamma(a,a)  \\
\hline \rule[-1em]{0pt}{2.8em}
    X_{(a, \, b)} & \bigl( a ,\, \phi(b) \bigr) \oplus \bigl( b ,\, \phi(a) \bigr) & 2 & \lvert \bA \rvert \, \bigl( \lvert \bA \rvert -1 \bigr)/2  & \gamma(a,b) \\
\hline \rule[-1em]{0pt}{3.2em}
    \Sigma_{(a, \, x)} & \eta^x \times \sigma_a & \sqrt{\lvert \bA \rvert} & 2 \lvert \bA \rvert &  \ds (-1)^x \sqrt{\frac{\epsilon}{|\bA|^{1/2}} \sum_{b \,\in\, \bA} \, f_a(b)^{-1} }
\end{array}
$$
\caption{Objects (lines) of the 3d Symmetry TFT $\cZ\bigl( \TYA \bigr)$.
\label{tab: objs in 3d Symm TFT}}
\end{table}
In the first line, $L_{(a, \, x)}$ arise from the $\Phi$-invariant elements $\bigl(a , \phi(a) \bigr)$ in the ungauged theory. The label $x \in \{ \underline0, \underline1 \}$ specifies the dressing by the $\bZ_2$ line $\eta \equiv L_{(0, \underline1)}$ generating the dual 1-form symmetry $\text{Rep}(\bZ_2)$. The lines $X_{(a, \, b)}$ with $a\neq b$ arise from long orbits of generic invertible objects and absorb the $\bZ_2$ line $\eta$. Finally, $\Sigma_{(a, \, x)}$ are the liberated twisted sectors, which are the charged objects under the dual $\text{Rep}(\bZ_2)$ symmetry and thus span the non-trivially graded component.
The total dimension of the category is
\be
\text{dim}\Bigl( \cZ \bigl( \TYA \bigr) \Bigr) = \Biggl( \; \sum_{\ell \text{ simple} } \dim(\ell)^2  \Biggr)\rule{0pt}{1.6em}^{1/2} = 2 \, |\bA| \;.
\ee

The topological manipulations of the theory with TY symmetry correspond to Lagrangian algebras of this Symmetry TFT. 
By definition of Drinfeld center, there should exist a Lagrangian algebra corresponding to the global variant with full TY symmetry. As an object, this is given by
\be
\calL_\text{TY} = \unit \oplus \eta \oplus \bigoplus_{b \,\neq\, 0 } X_{(0 ,\, b)} \;,
\ee
and indeed:
\be
\text{dim} \bigl( \calL_\text{TY} \bigr) = 2 \, \lvert\bA\rvert = \text{dim}\Bigl( \cZ \bigl( \TYA \bigr) \Bigr) \;.
\ee
This is the algebra induced by the electric Lagrangian subgroup $\calL_\rme = \bigoplus_{\alpha \,\in\, \bA^\vee} (0, \alpha)$ in the pure $\bA$ gauge theory, following our discussion in Section \ref{sec:proposal}.
While $\calL_\rme $ is clearly not duality invariant, it can be uplifted to an algebra in $\cZ \bigl( \TYA \bigr)$ by adding to it its images under $\Phi$.%
\footnote{If a line $\bigl(a ,\, \phi(a) \bigr)$ (like the identity in this case) is duality invariant, we must add $L_{(a, \underline0 )} \oplus L_{(a, \underline1 )}$ to the algebra.}
The resulting object is well defined in $\cZ \bigl( \TYA \bigr)$, it has vanishing spin (see Table~\ref{tab: objs in 3d Symm TFT}) and has dimension $2 \,|\bA|$ so it is Lagrangian.
This provides an explicit realization of the sequential gauging procedure outlined in \eqref{eq: seqgauging}.
The symmetry on the corresponding boundary can be computed using the sequential gauging prescription. In the trivially-graded sector $\cC_0$ the simple objects are the elements of the quotient $(\bA \times \bA^\vee) / \bA^\vee \simeq \bA$. They generate the 0-form symmetry and we label them simply by $a$. On the other hand, in the $\cC_\Phi$ sector all of the twist defects fall into a single orbit, without fixed points under fusion with $\calL_\rme$ as can be checked from \eqref{eq: fusiontft}. Let us denote this object by $\cN$. The bulk fusion rules imply (\ref{fusion rules of TY}), giving back the $\TYA$ symmetry.

\subsection{First obstruction and Lagrangian algebras}
\label{sec: invalgebra}

Our first goal is to describe how the first obstruction appears from the Symmetry TFT perspective. We have already mentioned in Section~\ref{sec:proposal} that the first obstruction precludes the existence of a discrete gauging $(\bB, \nu)$ which renders the duality symmetry $\cN$ invertible. Since, from the Symmetry TFT perspective, discrete gauging operations correspond to different choices of gapped boundary condition $\calL$, it is natural to rephrase the first obstruction in the language of Lagrangian algebras of the DW theory.
A similar logic has been followed recently in \cite{Zhang:2023wlu}, where the obstructions to gauge the entire symmetry category (\ie{} the case $\bB = \bA$ in our notation) when $|\bA|$ is odd have been found counting the number of bulk lines with trivial spin. 
However such method is hard to generalize to higher dimensions (which is the main aim in our work) since it relies on the notion of topological spin which has no known analog in higher categories. In this and the next two sections, instead, we provide a complete bulk classification of the obstruction theory for $\TYA$ and besides we develop methods that allow us to extend the results to higher-dimensional cases. 

The crucial point which makes this problem accessible is that the Symmetry TFT for the TY category is a $G\cong\bZ_2$ gauging of the Dijkgraaf-Witten theory $\text{DW}(\bA)$ \cite{Kaidi:2022cpf}. By gauging $G$ back and forth, we can rephrase the problem in terms of gauging Lagrangian algebras of a bulk theory that only consists of invertible symmetries. As already argued in Section \ref{sec:proposal}, a sufficient condition for $\cN$ to be anomalous is the absence of $G$-invariant Lagrangian algebras in $\text{DW}(\bA)$, namely, of Lagrangian algebras $\linv$  satisfying
\be
\label{Phi-invariance of L_D}
\Phi(\linv) = \linv \;.
\ee
A duality-invariant Lagrangian algebra of $\DWA$ also gives rise to a duality-invariant boundary condition, where the duality symmetry becomes invertible. Hence we realize that, in the terminology of \cite{Kaidi:2022cpf}, \emph{intrinsic non-invertible symmetries are anomalous}.

Notice that the obstruction we are discussing here is a priori distinct from the first obstruction we discussed in Section~\ref{sec: algebras TY}. However the main result of this section is to show that the two obstructions are equivalent. In order to do this, we make the first obstruction more explicit by classifying all Lagrangian algebras of $\text{DW}(\abA)$ and providing explicit equivalent conditions for their duality invariance in terms of the data $(\bA, \gamma)$.

The 3d theory $\text{DW}(\bA)$ can be thought of as the Symmetry TFT of any theory with a non-anomalous 0-form symmetry $\bA$, and as such the correspondence between topological manipulations and bulk Lagrangian algebras is particularly explicit, but yet non-trivial. The (bosonic) topological manipulations of the boundary are determined by two pieces of data \cite{Gaiotto:2020iye}:
\begin{itemize}
    \item The choice of a subgroup $\bB \subset \bA$ to be gauged.
    \item The choice of a class $[\nu] \in H^2 \bigl( \bB, U(1) \bigr)$ which plays the role of the discrete torsion.
\end{itemize}
The resulting symmetry is an extension of $\bA/\bB$ by the quantum symmetry $\bB^\vee$ \cite{Tachikawa:2017gyf} (see Appendix~\ref{sec:app B} for details). 

On the other hand, global variants of the boundary theory correspond to different interfaces between $\text{DW}(\bA)$ and the trivial 3d theory (\ie, to gapped boundaries), and thus are specified by gauging a subgroup $\calL \subset \bA\times \bA^\vee$, Lagrangian with respect to the braiding. Correspondingly, the lines of $\calL$ can end on the boundary, and the topological lines of the boundary theory generating the 0-form symmetry are labelled by the quotient group
\be
\label{eq:boundary symm}
\cS = \bA \times \bA^\vee / \calL \;.
\ee
Thus we expect a correspondence between pairs $\bigl( \bB, [\nu] \bigr)$ and Lagrangian algebras $\calL$ such that \eqref{eq:boundary symm} coincides with the symmetry after gauging $\abB$ with discrete torsion $[\nu]$. Notice that the braiding induces a canonical isomorphism%
\footnote{This can be seen as follows. The braiding is a bilinear non-degenerate pairing on $\abA \times \abA^\vee$ and thus induces an isomorphism $\abA \times \abA^\vee\rightarrow (\abA \times \abA^\vee)^\vee$. Saying that $\calL$ is Lagrangian is equivalent to saying that its image under this isomorphism is the subgroup of linear functions on $\abA \times \abA^\vee$ which vanish over $\calL$. The latter is canonically isomorphic to the Pontryagin dual of $\abA \times \abA^\vee / \calL = \cS$.}
\be
\calL \cong \cS^\vee \;.
\ee

The simplest case is when $H^2 \bigl( \bA, U(1) \bigr) = 0$ (\eg, if $\bA = \bZ_n$) so that the topological manipulations are simply labelled by the gauged subgroup $\bB \subset \bA$.%
\footnote{Indeed if $H^2 \bigl( \bA, U(1) \bigr) = 0$ then $H^2 \bigl( \bB, U(1) \bigr) = 0$ for every subgroup $\bB$ of $\bA$.}
Then we consider
\be
    \calL_{\bB} \,\equiv\, \bB \times N(\bB) \,\subset\, \bA \times \bA^\vee
\ee
which has cardinality $|\bA|$ and is made of lines of vanishing spin (in particular it trivializes the braiding, see (\ref{def N(B)})), hence it is Lagrangian. Moreover
\be
    \cS_{\bB} = \bA \times \bA^\vee / \calL_\bB = (\bA/\bB) \times \bB^\vee
\ee
is precisely the symmetry after gauging $\bB$.

In the general case we define the linear map $\psi_\nu: \bB \rightarrow \bB^\vee$ associated to $\chi_\nu$:
\be
    \psi_\nu(b_1) \, b_2 = \chi_\nu(b_1, b_2) \;.
\ee
Given the pair $\bigl( \bB, [\nu] \bigr)$ we construct the subgroup $\calL_{\bB,[\nu]} \subset \bA \times \bA^\vee$ as follows. Since $\bB^\vee = \bA^\vee / N(\bB)$, any element of $\bA^\vee$ can be presented as a pair $(\beta,\eta) \in N(\bB)\times \bB^\vee$ (even though the sum is different from the one in $\bA^\vee$) and we denote this element simply as $\beta \eta \in \bA^\vee$. The association is not canonical, however different choices agree on $\eta$ (which is the projection from $\bA^\vee$ to $\bB^\vee$) while may differ on $\beta$.
We denote by $\tilde{c} \in H^2 \bigl( \bB^\vee, N(\bB) \bigr)$ the cocycle which makes $\bA^\vee$ an extension of $\bB^\vee$ by $N(\bB)$.  Then we construct
\be
    \calL_{\bB,[\nu]} = \Bigl\{ \bigl(b ,\, \beta \psi_\nu(b) \bigr) \in \bA \times \bA^\vee \Bigm| b \in \bB ,\quad \beta \in N(\bB) \Bigr\} \;.
\ee
This contains $N(\bB)$ as a subgroup (for $b = 0$), while its quotient by $N(\bB)$ is isomorphic to $\bB$, hence $\calL_{\bB,[\nu]}$ is a group extension
\be
\label{eq:L as an extension}
1 \;\rightarrow\; N(\bB) \;\rightarrow\; \calL_{\bB,[\nu]} \;\rightarrow\; \bB \;\rightarrow\; 1
\ee
whose corresponding cocycle is $\tilde{c} \circ \psi _\nu \in H^2\bigl( \bB, N(\bB) \bigr)$ (see Appendix~\ref{sec:app B} for details).
Moreover $\calL_{\bB,[\nu]}$ has cardinality $|\bA|$, and since $\chi_\nu$ is alternating the spin of the lines is trivial:
\be
\label{vanishing spin in 3d}
\theta_{(b,\,\beta)} = \chi_\nu(b,b) = 1 \;.
\ee
Here $(b,\beta)$ is a shorthand for $\bigl( b,\, \beta \psi_\nu(b) \bigr)$, and $\beta$ does not contribute because it belongs to $N(\bB)$.
One could weaken the alternating condition and just ask $\chi_\nu$ to be antisymmetric. In that case the spins would be $\pm 1$ and one would allow for fermionic Lagrangian algebras, which correspond to fermionizations of the boundary symmetry. We will not discuss such cases here, but note that they are a natural candidate to explain why certain duality symmetries --- such as $\text{TY}(\bZ_2)_{\gamma, 1}$ --- can be gauged on spin manifolds.

We have thus shown that $\calL_{\bB,[\nu]}$ is a Lagrangian algebra with respect to the braiding. In Appendix~\ref{sec:app B} we prove that any Lagrangian algebra of $\abA \times \abA^\vee$ arises in this way. This classification of boundary conditions of the Dijkgraaf-Witten theory coincides with previously known results from category theory \cite{Ostrik:2002}. The boundary condition corresponding to $\calL_{\bB,[\nu]}$ is obtained from the original one by gauging $\bB$ with discrete torsion $[\nu]$. Indeed the symmetry on that boundary is
\be
    \cS = \bA \times \bA^\vee / \calL_{\bB,[\nu]} \,\cong\, \bigl( \calL_{\bB,[\nu]} \bigr)^\vee \;,
\ee
which is the group extension dual to \eqref{eq:L as an extension}, namely
\be
\label{eq:S as extension}
    1 \;\rightarrow\; \bB^\vee \;\rightarrow\; \cS \;\rightarrow\; \bA / \bB \;\rightarrow\;  1 \;.
\ee
The cocycle is $\psi _\nu \circ c \in H^2(\bA/\bB, \bB^\vee)$, where $c\in H^2(\bA/\bB, \bB)$ determines $\bA$ as an extension of $\bA/\bB$ by $\bB$. One can show that this is indeed the symmetry after gauging $\bB$ with discrete torsion $[\nu]$ (see Appendix~\ref{sec:app B} for the proof). 

We should now determine whether $\DWA$ admits duality-invariant Lagrangian algebras $\calL_{\bB,[\nu]}$. In the simplest case that $[\nu]=0$ and hence $\calL_\bB = \bB \times N(\bB)$, duality invariance is simply equivalent to $\phi(\bB) = N(\bB)$. Since $\bigl\lvert N(\bB) \bigr\rvert = |\bA|/|\bB|$ this requires $|\bB|^2 = |\bA|$ and in particular the cardinality of $\bA$ must be a perfect square ($n_\nu = 1$ in (\ref{2d algebra gauging TY}) in this case). However this is in general not sufficient: $\phi(b) \in \bA^\vee$ must vanish on $\bB$, so that $\abB$ must be a Lagrangian subgroup of $\bA$ with respect to the symmetric bicharacter $\gamma$ associated with $\phi$. 

In the cases with discrete torsion, we observe that
\be
    \Phi \bigl( \calL_{\bB,[\nu]} \bigr) = \Bigl\{ \Bigl( \phi^{-1} \bigl( \beta\psi_\nu(b) \bigr) ,\, \phi(b) \Bigr) \in \bA \times \bA^\vee \Bigm| b \in \bB ,\quad \beta \in N(\bB) \Bigr\}
\ee
is equal to $\calL_{\bB,[\nu]}$ if and only if for all $b \in \bB$ and $\beta \in N(\bB)$ there exist $b' \in \bB$ and $\beta' \in N(\bB)$ such that 
\be
    b' = \phi^{-1} \bigl( \beta \psi _\nu(b) \bigr) \;,\qquad\qquad b = \phi^{-1} \bigl( \beta' \psi_\nu(b') \bigr) \;.
\ee
Before stating the general condition under which these equations can be solved, consider the simpler case $\bB = \bA$ for which $N(\bB)=0$. Define the group homomorphism $\sigma \,=\, \phi^{-1} \circ \psi _\nu : \bA \rightarrow \bA$ in terms of which the two conditions become $b'=\sigma(b)$, $b=\sigma(b')$. They have a solution if and only if $\sigma^2=1$.
In particular both $\sigma$ and $\psi_\nu$ must be automorphisms.

When $\bB \subsetneq \bA$ is a proper subgroup, there are further conditions for duality invariance. The proof is technical and we report it in Appendix~\ref{sec:general duality invariance}. Let us remind that the \emph{radical} of the class $[\nu]$ is
\be
    \operatorname{Rad}(\nu) = \operatorname{Ker}(\psi_\nu) \,\subset\, \bB \;.
\ee
Besides, the projection of $\chi_\nu$ to $\bB/\Rad(\nu)$ being non-degenerate gives an isomorphism
\be
\psi_\nu: \bB/\Rad(\nu) \to \bigl( \bB/\Rad(\nu) \bigr){}^\vee \;.
\ee
Then duality invariance of $\calL_{\bB,[\nu]}$ is equivalent to the following conditions:
\begin{enumerate}

\item $\phi \bigl( \text{Rad}(\nu) \bigr) = N(\bB)$. In particular $N(\bB) \subset \phi(\bB)$, and $|\bB| = n_\nu \, |\bA|^{1/2} \geq |\bA|^{1/2}$. In other words, $\bB$ cannot be smaller than Lagrangian and $|\bA|$ must be a perfect square, hence reproducing the known obstruction induced by non-integer quantum dimensions \cite{Chang:2018iay}.

\item Assuming condition 1., also $\phi$ projects to an isomorphism $\phi: \bB/\Rad(\nu) \to \bigl( \bB/\Rad(\nu) \bigr){}^\vee$. Then we can define an automorphism
\be
\label{eq:def sigma}
    \sigma \,\equiv\, \phi^{-1} \circ \psi_\nu: \bB/\text{Rad}(\nu) \rightarrow \bB/\text{Rad}(\nu)
\ee
which, by construction, satisfies $\gamma\bigl( \sigma(a), b\bigr) = \chi_\nu(a,b)$.
The second condition is that
\be
\label{eq: condition duality invariance}
    \sigma^2 = 1 \;.
\ee
\end{enumerate}

Notice that the conditions we obtained for $\calL_{\bB,[\nu]}$ to be duality invariant are equivalent to the first obstruction we reviewed in Section~\ref{sec: algebras TY}. We thus arrive to the punchline of this section: the  first obstruction is equivalent to the absence of duality-invariant Lagrangian algebras in $\DWA$, or in other words, to the non-invertible duality symmetry being intrinsic.

A straightforward consequence of the conditions above concerns the action of the duality symmetry $G$ on the symmetry $\cS = \bA \times \bA^\vee / \calL_{\bB,[\nu]}$ of the invariant boundary. To this purpose, it is convenient to present $\cS$ as a group extension \eqref{eq:S as extension} and further view $\bB^\vee$ as an extension of $\text{Rad}(\nu)^\vee$ by $\bigl( \bB/\text{Rad}(\nu) \bigr)^\vee$, hence presenting the elements of $\cS$ as triplets $(\beta, \eta, \tilde{a})$ with $\beta \in \bigl( \bB/\text{Rad}(\nu) \bigr)^\vee$, $\eta \in \text{Rad}(\nu)^\vee$ and $\tilde{a} \in \bA/\bB$. Using that $\cS = \calL_{\bB,[\nu]}^\vee$ we find that the duality exchanges $\text{Rad}(\nu)^\vee$ with $\bA/\bB$, while it acts on $\bigl( \bB/\text{Rad}(\nu) \bigr)^\vee$ as the automorphism $\sigma ^\vee$:
\be
    \Phi:\, (\beta, \eta, \tilde{a}) \,\rightarrow\, \Bigl( \sigma^\vee(\beta) ,\: \phi(\tilde{a}) ,\: \phi^{-1}(\eta) \Bigr) \;.
\ee

When the data $\bigl( \bB, [\nu] \bigr)$ defines a duality-invariant Lagrangian subgroup, using the definition of $\sigma$ in \eqref{eq:def sigma} and $\sigma^2=1$ we can relate the symmetric and the antisymmetric bicharacters as
\be
\label{eq:relsymmantisymm}
\chi_\nu(b_1, b_2) = \gamma \bigl( \sigma(b_1), b_2 \bigr) \;,\qquad\qquad \gamma(b_1, b_2) = \chi_\nu \bigl( \sigma(b_1), b_2 \bigr) \;.
\ee
This in turn implies a condition for the class $[\nu]$:
\be
\label{eq:G cond on nu}
\nu(b_1,b_2) \, \nu \bigl( \sigma(b_1), \sigma(b_2) \bigr) = d \tilde{\zeta}(b_1,b_2)
\qquad\text{or equivalently}\qquad
\frac{ \nu(b_1,b_2) }{ \nu\bigl( \sigma(b_2), \sigma(b_1) \bigr)} = d \teta(b_1,b_2) \;.
\ee
This is because the l.h.s. of both equations is a symmetric 2-cocycle (see Section~\ref{sec: quadratic refs} or footnote~\ref{foo: exactness}). Those relations coincide with the known relation \eqref{eq:firstob4} (also appearing in the equivariantization of the algebras in TY categories, see Section~\ref{sec: Equivariantization and second obstruction}).

We can neatly express the condition \eqref{eq:G cond on nu} by noticing that the action $\rho : G \rightarrow \text{Aut}(\bA)$ of any group $G$ on a generic Abelian group $\bA$ induces an action on $H^2 \bigl( \bA, U(1) \bigr)$ given by
\be
\label{eq:G action on bicharacters}
(\rho_g \, \xi)(a_1, a_2) = \xi \bigl( \rho_g^{-1}(a_1) ,\, \rho_g^{-1}(a_2) \bigr)
\ee
for each $g\in G$ and $\xi \in H^2\bigl( \bA, U(1) \bigr)$.
Then \eqref{eq:G cond on nu}  can be expressed as
\be
\label{eq:sigma action on bicharacters 3d}
   \sigma \, [\nu] = \rho_{\underline1} \, [\nu] = [\nu^{-1}] \;,
\end{equation}
where $\underline1$ is the generator of $G \cong \bZ_2$. This reformulation will be convenient later on.

\subsubsection*{Examples}

To make concrete the discussion above, we show a few examples. For convenience, here we use additive notation for the phases by thinking of them as elements of $\bR/\bZ$ instead of $U(1)$.

\paragraph{1.}
The simplest example is $\bA=\bZ_n$ for which there is no discrete torsion, and the subgroups are in correspondence with the divisors of $n$. Let $n=pq$, and $\bB = \{ qx \ | \ x=0, \dots, p-1 \} \cong \bZ_p$ so that $N(\bB) = \{ py \ | \ y = 0, \dots, q-1 \} \cong \bZ_q$.

When we gauge $\bB$ on the boundary, the global symmetry is the direct product of the dual symmetry $\bZ_p$ and the quotient $\bZ_q$. From the bulk perspective, the prescription is that this boundary condition is obtained by allowing the lines of the form $(qx,py)$ to terminate on the boundary hence becoming transparent there. On the other hand, the 0-form symmetry is generated by the remaining lines stacked at the boundary, which indeed form the group $\bZ_p \times \bZ_q$.

For what concerns duality invariance, we need $\bB \cong N(\bB)$ and hence $p=q$: this implies that $n=p^2$ must be a perfect square.
Any symmetric bicharacter takes the form $\gamma(a,b) = rab/n \text{ (mod } 1)$ for some $r \in \bZ_n$ ($r$ must be coprime with $n$ for the bicharacter to be non-degenerate), and we notice indeed that $\bZ_p\subset \bZ_{p^2}$ is Lagrangian in all cases:
\be
\gamma(px,py) = 0 \;.
\ee
The integer coefficient introduced in \eqref{2d algebra gauging TY} here is $n_{\nu}=|\bB|/|\bA|^{1/2} = 1$.

\paragraph{2.}
A less trivial example is $\bA = \bZ_n \times \bZ_n$ with $n$ a prime number. There are $n+3$ subgroups: the trivial one, the $n+1$ subgroups isomorphic to $\bZ_n$ generated by $(1,0)$ and $(s,1)$ with $s = 0, \dots , n-1$, and the full $\bA$. Only the last one admits non-trivial discrete torsion $[\nu] \in H^2 \bigl( \bZ_n \times \bZ_n, U(1) \bigr) \cong \bZ_n$ which could be represented as
\be
\nu \bigl( (x_1,x_2) ,\, (y_1,y_2) \bigr) = \frac{r}{n} \, x_1 y_2 \qquad\text{or equivalently as}\qquad  \nu \bigl( (x_1,x_2) ,\, (y_1,y_2) \bigr) = - \frac{r}{n} \, x_2 y_1 \;.
\ee
The corresponding alternating bicharacter is given by the matrix
\be
\label{eq:discrete torsion 3d}
\chi_\nu = \frac{1}{n} \, \biggl( \, \begin{matrix} 0 & r \\ -r & 0 \end{matrix} \, \biggr) \;,\qquad\qquad \text{with}\qquad r \in \bZ_n \;.
\ee
In total there are $2n+2$ boundary theories. One can explicitly see that these are in one-to-one correspondence with the Lagrangian algebras $\calL_{\bB,[\nu]}$ in $\abA \times \abA^\vee$.

Let us show that the induced global symmetry at the boundary is the one obtained by gauging $\bB$ with discrete torsion $\nu$. The cases $\bB = \{0\}$ or $\bB \cong \bZ_n$ are similar to the one discussed above and the corresponding Lagrangian algebras are, respectively:
\bea
\calL_{\bB,[0]} &= \Bigl\{ \bigl( (0,0); (a_1,a_2) \bigr) \Bigm| a_1, a_2 \in \bZ_n \Bigr\} && \text{for } \bB = \{0\} \;, \\
\calL_{\bB,[0]} &= \Bigl\{ \bigl( (a_1,0); (0,a_2) \bigr) \Bigm| a_1,a_2 \in \bZ_n \Bigr\} && \text{for } \bB \cong \bZ_n \text{ generated by } (1,0) \;, \\
\calL_{\bB,[0]} &= \Bigl\{ \bigl( (sa_1,a_1); (a_2, -sa_2) \bigr) \Bigm| a_1,a_2 \in \bZ_n \Bigr\} \quad && \text{for } \bB \cong \bZ_n \text{ generated by } (s,1) \;.
\eea
When $\bB =\bZ_n \times \bZ_n$ the resulting boundary theory has symmetry $\bB^\vee\cong \bZ_n \times \bZ_n$ even for non-trivial discrete torsion. According to our prescription, and using that $N(\bB)$ is trivial and the map $\psi_\nu$ has the same matrix form of $\chi_\nu$ defined in \eqref{eq:discrete torsion 3d}, the corresponding Lagrangian subgroup of $\bA \times \bA^\vee$ is
\be
\label{eq: Lag subgroup in exampl}
    \calL_{\bB, [\nu]} = \Bigl\{ \bigl( (a_1,a_2); (-ra_2, ra_1) \bigr) \Bigm| a_1, a_2 \in \bZ_n \Bigr\} \;, 
\ee
and indeed $(\bA \times \bA^\vee) / \calL = \bZ_n \times \bZ_n$. To see the effect of the discrete torsion we use a Lagrangian description of the DW theory:
\be
\label{Lagrangian description DW example}
    S = \frac{2\pi i}{n} \int_{X_3} \bigl( A_1 \cup dB_1 + A_2 \cup dB_2 \bigr) \;.
\ee
The generic line with charges $\bigl( (a_1,a_2); (b_1,b_2) \bigr)$ is 
\be
\exp \biggl[ \, \frac{2\pi i}{n}\int_\gamma \bigl( a_1A_1 + a_2 A_2 + b_1 B_1 + b_2 B_2 \bigr) \biggr] \;.
\ee
Hence, as a boundary condition, $\calL$ in (\ref{eq: Lag subgroup in exampl}) corresponds to $A_1 + rB_2 =A_2 - rB_1=0$. Changing variables according to $A_1 \to A_1 + r B_2$, $A_2 \to A_2 - r B_1$ we obtain the same bulk Lagrangian as in (\ref{Lagrangian description DW example}) but with an extra boundary term
\be
\delta S_\text{bdry} = \frac{2\pi i r}{n} \int _{\partial X_3} \! B_1\cup B_2 \;,
\ee
which is precisely the discrete torsion for the gauging on the boundary.

Let us discuss which of those algebras are duality invariant, and in particular which symmetric bicharacters admit duality-invariant algebras. There are two natural classes of non-degenerate symmetric bicharacters, diagonal and off-diagonal:
\be
\label{eq:duality 3d ZnxZn}
\gamma^{(\rmD)}_{t_1 ,\, t_2} = \frac{1}{n} \, \biggl( \,\begin{matrix} t_1  & 0 \\ 0  & t_2 \end{matrix}\, \biggr) \qquad\qquad\text{and}\qquad\qquad \gamma^{(\rmO)}_t = \frac{1}{n} \, \biggl( \,\begin{matrix} 0 & t \\ t & 0 \end{matrix} \,\biggr) \;.
\ee
Here non-degeneracy requires $t_1$, $t_2$, $t$ to be invertible elements of $\bZ_n$.%
\footnote{For $n$ prime, they are just non-vanishing. However these two bicharacters will be used also for $n$ non prime, hence $t_1$, $t_2$, $t$ must be coprime with $n$.}
Note that $\bB = \{0\}$ cannot lead to duality-invariant algebras because it is smaller than Lagrangian.

Consider the case of $\gamma_t^{(\rmO)}$. First we look at Lagrangian algebras associated with subgroups $\bB\cong \bZ_n$ which, according to our general analysis, need to be Lagrangian with respect to $\gamma_t^{(\rmO)}$ because $[\nu] = 0$. The two subgroups $\bB = \bigl\langle (1,0) \bigr\rangle$, $\bigl\langle (0,1) \bigr\rangle$ are always Lagrangian, while $\bB = \bigl\langle (s,1) \bigr\rangle$ is Lagrangian only if 
\be
    2st=0 \mod n
\ee
which can never be satisfied if $n$ is odd.
Then we look at the cases with $\bB=\bA$. In order to satisfy $\phi \bigl( \Rad(\nu) \bigr) = N(\bB) = \{0\}$ in (\ref{eq:firstob1}) we need a discrete torsion \eqref{eq:discrete torsion 3d} with $r\neq 0$. From (\ref{eq:def sigma}) we find
\be
    \sigma = \biggl( \,\begin{matrix} t^{-1} r & 0 \\ 0 & -t^{-1} r \end{matrix}\, \biggr) \;.
\ee
The duality-invariant condition $\sigma^2=1$ reads $(t^{-1}r)^2 = 1 \text{ mod } n$, which can always be satisfied by the values $r = \pm t$.

Consider now the case of $\gamma_{t_1,t_2}^{(\rmD)}$. The subgroups $\bB\cong \bZ_n$ are Lagrangian with respect to $\gamma_{t_1,t_2}^{(\rmD)}$ only when $\bB = \bigl\langle (s,1) \bigr\rangle$ with $t_1 \, s^2 + t_2 = 0 \text{ mod } n$. For $\bB = \bA$, instead, we need a non-vanishing discrete torsion \eqref{eq:discrete torsion 3d}, and since
\be
   \sigma = \biggl( \,\begin{matrix} 0 & - t_1^{-1} r \\ t_2^{-1} r & 0 \end{matrix}\, \biggr) \;,
\ee
the duality-invariance condition reads $r^2 = - t_1 t_2 \text{ mod } n$. This equation and the previous one for $s$ do not always have solutions. For instance, if $t_1=t_2=1$, then $r$ (or $s$) must be a square root of $-1$ which exists for $n = 2,5,13,\dots$ but not for $n=3,7,11, \dots$

In summary, while $\text{TY}(\bZ_n\times \bZ_n)_{\gamma,\epsilon}$ with off-diagonal bicharacter $\gamma$ always trivializes the first obstruction, when the bicharacter is diagonal the category is necessarily anomalous for certain values of $n$ for which the first obstruction forbids the gauging. We also notice that in all of these examples, when there is a duality-invariant Lagrangian algebra associated with $\bB \cong \bZ_n$ we have $n_\nu=1$, while for $\bB \cong \bA$ we have $n_\nu=n$.

\paragraph{3.} We conclude with a more complicated example which is representative of the general case $\bB \subsetneq \bA$ but $[\nu] \neq 0$, hence $\bB$ is non-Lagrangian. Take $\bA = \bZ_4 \times \bZ_4$ which, besides the subgroups we already considered, also has the subgroup
\be
\bB = \bigl\{ (x,2y) \bigm| x \in \bZ_4 \,,\; y \in \bZ_2 \bigr\} \cong \bZ_4\times \bZ_2
\ee
(as well as the similar one with the two factors swapped) hence realizing
\be
N(\bB) = \bigl\{ (0, 2\tilde{y}) \bigm| \tilde{y} \in\bZ_2 \bigr\} \cong \bZ_2 \subset \bA^\vee \;.
\ee
The most general alternating bicharacter on $\bB$ is
\be
    \chi_\nu = \frac{1}{4} \, \biggl( \,\begin{matrix} 0 & a \\ b & 0 \end{matrix}\, \biggr) \qquad\text{with}\qquad 2(a+b) = 0 \mod 4 \;,
\ee
hence $a,b \in \bZ_4$ must be either both even or both odd. If $a,b$ are both even then $\text{Rad}(\nu)= \bB$
and duality invariance cannot be satisfied. If $a,b$ are odd, instead,
\be
    \text{Rad}(\nu) = \bigl\{ (2z,0) \bigm| z \in \bZ_2 \bigr\} \cong \bZ_2 \subset \bZ_4 \times \bZ_2 \;.
\ee
The condition $\phi\bigl( \text{Rad}(\nu) \bigr) = N(\bB)$ cannot be satisfied with the diagonal bicharacter $\gamma_{t_1 ,\, t_2}^{(\rmD)}$, while with the off-diagonal one $\gamma^{(\rmO)}_t$ the condition is met (for both the invertible elements $t=1,3$). The second condition for duality invariance involves
\be
\sigma = \phi^{-1} \circ \psi_\nu = \biggl( \,\begin{matrix} tb & 0 \\ 0 & ta \end{matrix}\, \biggr) \;.
\ee
The condition $\sigma^2=1$ is equivalent to $(tb)^2=(ta)^2=1$ which is automatically satisfied. In this case we get $n_{\nu} = 2$.

\subsection{Second obstruction and equivariantization}
\label{sec: Equivariantization and second obstruction}

In the previous section we rephrased the first obstruction to the gauging of a 2d duality symmetry in terms of the existence of a duality-invariant Lagrangian algebra $\linv$ in the 3d TQFT $\DWA$. 
Gauging $\linv$ leads to a bulk SPT phase $Y \in H^3 \bigl(G, U(1) \bigr)$ for the duality symmetry $G \cong \bZ_2$, which determines the DW twist of the corresponding $G$ gauge theory as explained in \eqref{eq: introtwist}. The total twist $\epsilon_{\text{tot}} = \epsilon \, Y$ in turn determines whether a Neumann boundary condition is allowed (and $\cN$ is anomaly-free).

\begin{figure}[t]
\centering
\begin{tikzpicture}
    \coordinate (in) at (2.5, 1.25); \coordinate (inu) at (2.5, 1.5); \coordinate (ind) at (2.5, 1); \coordinate (out) at (-0.5, 2.25) ; \coordinate (def) at (1, 1.75);
    \draw[color=black] (def) to (out) node[left] {\small $g(z)$};
    \filldraw[color=white!90!green, opacity=0.75] (0, 0) -- (2, 0.5) -- (2, 3.5) node[below left, color=black] {$U_g$} -- (0,3) -- cycle;
    \draw[color=black] (ind) node[right] {\small $x$} to[out=170, in=-100] ($0.5*(in) + 0.5*(def)$) to[out=80, in=160] (inu) node[right] {\small $y$};
    \draw[color=black] ($0.5*(in) + 0.5*(def)$) node[above=0.2cm] {\small $v_{x,y}^z$} to (def) node[circle, fill, inner sep=1.2pt] {};
    \node[right] at (3.5, 1.75) {$= \quad\;\; [\cU_g]_{x ,\, y}^z$};
    \begin{scope}[shift={(7.2, 0)}]
    \coordinate (in) at (2.5, 1.25); \coordinate (inu) at (2.5, 1.5); \coordinate (ind) at (2.5, 1); \coordinate (out) at (-0.5, 2.25) ; \coordinate (def) at (1, 1.75);
    \draw[color=black] ($(def) - (0,0.25)$) to[out=160, in=-100] ($0.5*(def)+0.5*(out)$)  to[out=80, in=160] ($(def) + (0,0.25)$);
    \draw[color=black] ($0.5*(def)+0.5*(out)$) to (out) node[left] {\small $g(z)$};
    \filldraw[color= white!90!green, opacity=0.75] (0,0) -- (2,0.5) -- (2,3.5) node[below left, color=black] {$U_g$} -- (0,3) -- cycle;    
    \draw[color=black] (ind) node[right] {\small $x$} to ($(def) - (0,0.25)$) node[circle, fill, inner sep=1.2pt] {}; 
    \draw[color=black] (inu) node[right] {\small $y$} to ($(def) + (0,0.25)$) node[circle, fill, inner sep=1.2pt] {}; 
    \node[above=0.3cm] at ($0.5*(def)+0.5*(out)$)  {\small $v_{g(x) ,\, g(y)}^{g(z)}$};
    \end{scope}
\end{tikzpicture}
\\[1em]
\begin{tikzpicture}
    \coordinate (in) at (2.5,1.25); \coordinate (out) at (-0.5,2.25) ; \coordinate (def) at (1,1.75);
    \draw[color=red, line width =1] (def) to (out) node[left] {$\linv$};
    \filldraw[color= white!90!green, opacity=0.75] (0,0) -- (2,0.5) -- (2,3.5) node[below left, color=black] {$U_g$} -- (0,3) -- cycle;
    \draw[color=black] (in) node[right] {\small $x$} to ($0.5*(in) + 0.5*(def)$) ;
    \draw[color=red, line width=1, >-] ($0.5*(in) + 0.5*(def)$) node[below=0.25cm] {$\pi_x$} to (def) node[circle, fill, inner sep=1.2pt] {};
    \node[right] at (3.5, 1.75) {$= \quad\;\; \teta_g(x)$};
    \begin{scope}[shift={(7.2,0)}]
    \coordinate (in) at (2.5,1.25); \coordinate (out) at (-0.5,2.25) ; \coordinate (def) at (1,1.75);
    \draw[color=red, line width =1, >-] ($0.5*(def) + 0.5*(out)$) to (out) node[left] {$\linv$};
    \draw[color=black] (def) to ($0.5*(def) + 0.5*(out)$);
    \filldraw[color= white!90!green, opacity=0.75] (0,0) -- (2,0.5) -- (2,3.5) node[below left, color=black] {$U_g$} -- (0,3) -- cycle;
    \draw[color=black] (in) node[right] {\small $x$} to (def) node[circle, fill, inner sep=1.2pt] {};   \node[below=0.25cm, color=red] at ($0.5*(def) + 0.5*(out)$) {$\pi_{g(x)}$};
    \node[above=0.3cm] at ($0.75*(def) + 0.25*(out)$) {\small $g(x)$};
    \end{scope}
\end{tikzpicture}
\caption{Graphical representation of the action of a symmetry defect $U_g$ on the junction spaces $V_{x ,\, y}^z$ (above) and on the projectors $\pi_x$ (below).
\label{fig: Ug action}}
\end{figure}
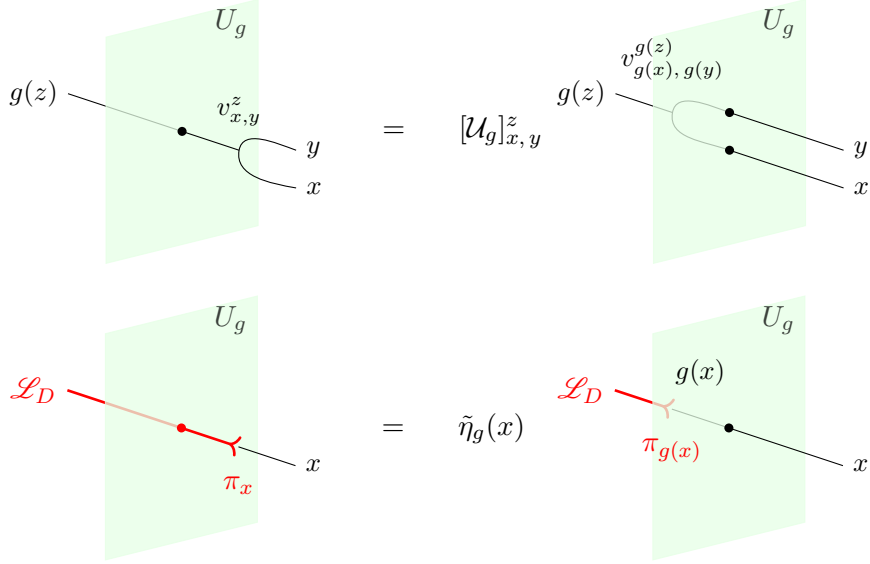

In order to understand the origin of $Y$ we must describe in detail how to make the gauging of $\linv$ consistent with the presence of a 0-form symmetry. Naively this should amount to the requirement that $\linv$ be $G$-invariant as an object: $\Phi(\linv) = \linv$ as stressed in (\ref{Phi-invariance of L_D}). This is however not sufficient, as the algebra $\linv$ comes with a specific choice of morphism \mbox{$m: \linv \times \linv \to \linv$} that is associative and commutative (see Appendix~\ref{app:Frobenius} for the definitions) and a set of projections $\pi_x \in \text{Hom}(\linv, x)$. An \emph{equivariantization} of $\linv$ is the definition of a consistent action of the 0-form symmetry on the projections that leaves $m$ invariant (for more details we refer the reader to Appendix~\ref{app:Frobenius} and \cite{Bischoff:2018juy} for a thorough treatment). To define this structure the proper context is that of $G$-crossed MTCs \cite{Barkeshli:2014cna}. In this framework a symmetry defect $U_g$ acts on the junction spaces $V_{x ,\, y}^z$, where $x,y,z \in \bA \times \bA^{\vee}$ label simple objects\footnote{Throughout this section we leave implicit that all simple objects are invertible and hence all junction spaces are one-dimensional.} , by a unitary automorphism $[\cU_g]_{x ,\, y}^z: V_{x ,\, y}^z \to V_{g(x) ,\, g(y)}^{g(z)}$ as
\be
U_g\bigl( v_{x ,\, y}^z \bigr) = [\cU_g]_{x ,\, y}^z \cdot v_{g(x) ,\, g(y)}^{g(z)} \;
\ee
where $v$ is a chosen basis vector of $V_{x ,\, y}^z$ (see Figure~\ref{fig: Ug action}). The phases $[\cU_g]_{x ,\, y}^z$ have to satisfy several compatibility conditions with the data of the underlying category, in particular consistency with the braiding requires
\be
[\cU_g]_{x ,\, y}^z \; R_{x ,\, y}^z = R_{g(x) ,\, g(y)}^{g(z)} \; [\cU_g]_{y ,\, x}^z \;.
\ee
Using the R-matrices \eqref{choice of R-matrices} and the $G\cong\bZ_2$ action on elements of $\DWA$ one easily sees that this equation admits a simple solution
\be \label{eq:solU}
[\cU_g]_{(a, \alpha),\, (b, \beta)}^{(a+b, \alpha+ \beta)} = \alpha(b)\, ,
\ee
for $g = \underline1$ the generator of $\bZ_2$. 

Now let us come to the equivariantization. For the algebras discussed in Section~\ref{sec: invalgebra}, a consistent\footnote{Commutativity of the algebra requires $m_{x,\, x'}^{z} = m_{x', \, x}^{z} R_{x, \, x'}^{z}$, which, in our case, becomes $m_{x,\, x'}^{z}/m_{x',\,x}^{z} = \chi_\nu(b',b)$.} choice of $m$ is
\be
\label{choice of m-matrix}
m_{x ,\, x'}^{x + x'} = \nu(b' ,\, b) \qquad\text{where}\qquad x = \bigl( b ,\, \beta \psi_\nu(b) \bigr) \in \linv \;.
\ee
In the following we will use $x,y,z, \dots$ to denote elements of $\linv$ in order to lighten the notation. Working in components we expand
\be
m = \bigoplus_{x, \, y} \, m_{x, \, y}^{z}  \qquad\text{and}\qquad m_{x, \, y}^{z} \in V_{x ,\, y}^z 
\ee
where $z = x+y$.
The defects $U_g$ act on the projectors $\pi_x: \linv \to x$ by an automorphism $\teta_g(x): \pi_x \to \pi_{g(x)}$ as follows (see Figure~\ref{fig: Ug action})
\be
U_g(\pi_x) = \teta_g(x) \cdot \pi_{g(x)}
\ee
Using these transformations, $m$ is invariant if  \footnote{Here we use that all objects in the algebra $\linv$ are invertible and appear with multiplicity one in the $\DWA$ theory.}
\be
\label{eq: equivariantizationm}
m_{g(x) ,\, g(y)}^{g(z)} = m_{x ,\, y}^z \; [\cU_g]_{x ,\, y}^z \; \frac{\teta_g(z) }{ \teta_g(x) \, \teta_g(y)} \;.
\ee

The equivariantization datum $\teta$ can be neatly interpreted in cohomology. First acting with gauge transformations $\pi_x \to \mu(x) \, \pi_x$ on the vector spaces associated to $\pi_x$ and $\pi_{g(x)}$ we can identify
\be
\label{eq: gaugefreedommain}
\teta_g(x) \,\sim\, \teta_g(x) \; \frac{\mu\bigl( g(x) \bigr)}{\mu(x)} \;.
\ee
Second, consistency with the group composition law demands that
\be
\label{eq: closedness}
\teta_g(x) \; \teta_h \bigl( g(x) \bigr) = \teta_{gh}(x) \;.
\ee
Interpreting $\teta_g$ as a cochain in $C^1\bigl( \linv, U(1) \bigr)$, so that $\teta \in C^1\bigl( G,\, C^1\bigl( \linv, U(1) \bigr) \bigr)$, we can rewrite \eqref{eq: closedness} and \eqref{eq: gaugefreedommain} in terms of a differential. Using now additive notation, for the sake of clarity and for later convenience, they look, respectively, as
\be
d_\rho \teta = 0 \;, \qquad\qquad\qquad \teta \,\sim\, \teta + d_\rho \mu \;,
\ee
for any $\mu \in C^0\bigl( G,\, C^1\bigl( \linv, U(1) \bigr) \bigr) \cong C^1\bigl( \linv, U(1) \bigr)$. Here $d_\rho$ is a twisted differential, while $\rho$ is the $G$-action on anyons.
We obtain that $\teta$ is naturally an object in twisted group cohomology (see \eg{} \cite{Benini:2018reh} and Appendix \ref{sec: twisted cohomology} for a review):
\be
\teta \, \in \, H^1_\rho \Bigl( G,\, C^1\bigl(\linv, U(1) \bigr) \Bigr) \;.
\ee
Restricting the solution (\eqref{eq:solU}) to elements of $\linv$ we find
\be 
[\cU_{g}]_{x ,\, x'}^{x + x'} = \chi_{\nu}\bigl( b , b' \bigr)=\chi_\nu \bigl( \sigma(b) , \sigma(b') \bigr)^{-1}
\ee
where in the second step we used the relations between the symmetric and antisymmetric bicharacters \eqref{eq:relsymmantisymm}.
Since $m_{g(x) ,\, g(x')}^{g(x + x')} = \nu\bigl( \sigma(b') , \sigma(b) \bigr)$ from (\ref{choice of m-matrix}), then \eqref{eq: equivariantizationm} becomes
\be
\label{eq:condition on the class equiv}
\frac{\nu(b , b') }{ \nu \bigl( \sigma(b') , \sigma(b) \bigr)} = d \teta_g
\ee
which we recognize as the first equation in \eqref{eq:firstob4} with a caveat. The set of solutions for $\teta_g$, with $g= \underline1$, form a torsor over $\linv^{\vee}$ while the solutions of \eqref{eq:firstob4} are related by elements of $\bigl(\bB / \text{Rad}(\nu)\bigr)^{\vee}$, therefore, strictly speaking, the solutions sets of the two equations differ. However we will see below that the two sets of equations give rise to the same second obstruction.\footnote{Physically the extra solutions correpond to symmetry fractionalization patterns between $\bZ_2$ and $\cS$ for which there is no mixed 't Hooft anomaly.}

For later convenience we also notice that the set of solutions to \eqref{eq:condition on the class equiv} for $\tilde\eta$ forms a torsor over
\be
\label{cohomology group of fractionalization}
H^1_\rho \bigl( G ,\, \linv^\vee) = H^1_\rho(G, \cS) \;,
\ee
whose elements we denote by $\eta$. This will be useful for the upcoming reinterpretation of the second obstruction in terms of symmetry fractionalization in Section~\ref{sec: frac3d}. All in all we found that an equivariantization of a duality-invariant Lagrangian algebra $\linv$ is specified by the choice of an element $\tilde\eta \in H^1_\rho \bigl( G,\, C^1 \bigl( \linv, U(1) \bigr) \bigr)$ satisfying \eqref{eq:condition on the class equiv}, and that any two choices differ by an element $\eta$ of $H^1_\rho \bigl( G ,\, \linv^\vee)$.

Given an equivariantization $\teta$ of $\linv$, we ask what is the SPT phase $Y \in H^3 \bigl( G, U(1) \bigr)$ for $G$ that we obtain after gauging $(\linv, \teta)$. Indeed, the theory after gauging has a single genuine line $\unit$ (and thus is an invertible TQFT) but also a single non-genuine topological twist line $M_g$ for each $g \in G$. The spins $\theta_{M_g}$ of such objects are gauge dependent by a $G$-character \cite{Barkeshli:2014cna}. In the presence of a discrete torsion $Y$, the $\theta_{M_g}$'s do not form a $G$-character: their deviation from being a character is physical and is induced by the SPT phase $Y$. In the present case that $G \cong \bZ_2$,%
\footnote{In order not to clutter we will suppress the label $g$ in what follows, as there is only one nontrivial $G$ defect anyway.}
$\theta_M$ does not square to 1 but instead
\be
\theta_M = \sqrt{Y(\underline1, \underline1, \underline1)} \;, 
\ee
the sign of the square root being pure gauge. We can thus detect the $\bZ_2$ SPT phase through the gauge-invariant quantity $\theta_M^2 = Y$. 
We now show how to reproduce \eqref{eq: secondob}. A key fact is that, given a choice of equivariantization $\tilde\eta$ for $\linv$, there is a unique non-genuine twist line $M$ after gauging $(\linv, \teta)$. It is then possible to show, using the defining equation \eqref{eq: localmodules} for a twisted local module $M$ that
\be
\label{eq:misteriouskey}
f_a(b)^{-1} = \teta(b) \qquad\qquad\text{for}\qquad b \in \bB/\text{Rad}(\nu) \qquad\text{and}\qquad \sigma(b) = b \;,
\ee
where $f_a$ is the function introduced in (\ref{def torsor f_a}). The equation holds for all the values of $a$ for which $\text{Hom}(\sigma_a, \, M) \neq 0$.\footnote{To show that the result holds consider $\eqref{eq: localmodules}$ and set $g(x_i)=x_i$. The matrix $r_L$ can then be eliminated on the two sides. Decomposing the module $M_g$ in its components and using the formulas \eqref{choice of R-matrices} for the $R$ matrix gives the desired result.} 
We will use the notation $M^{(a)}$ to account for the different choices one has for the equivariantization $\teta$: upon gauging, each choice leads to a theory with a unique non-genuine operator, however different choices lead to different SPT phases $Y$ and the label $a$ (whose possible values depend on $\linv$ in a complicated way) keeps track of the equivariantization chosen.

Since $\theta_M^2$ must be well defined, the spins squared of the components of $M$ must coincide. Since, as an object, $M^{(a)}$ can be described as the orbit of the twist defect $\sigma_a$ under fusion with the lines of $\linv$, using the fusions in \eqref{eq: fusiontft} we get:%
\footnote{Besides identifying twist defects related by fusion with the lines of $\linv$, one also has to impose locality conditions, that depend on $\tilde\eta$ (see Appendix~\ref{app:Frobenius}). Together these constraints single out a unique twist defect for each choice of equivariantization.}
\be
M^{(a)} = \bigoplus_{u} \sigma_{a + u} \qquad\text{where}\qquad u = b + \phi^{-1} \bigl( \beta \psi_\nu(b) \bigr) \qquad\text{for all}\qquad \bigl( b ,\, \beta \psi_\nu(b) \bigr) \in \linv \; .
\ee
Consistency with the existence of a unique local module requires that $\theta_{\sigma_a}^2 = \theta_{\sigma_{a+u}}^2$, \ie
\be
\theta_{M^{(a)}}^2 = \frac{1}{\sqrt{|\bA|}} \, \sum_{c \, \in \, \bA } f_a(c)^{-1} \stackrel{!}{=} \frac{1}{\sqrt{|\bA|}} \, \sum_{c \, \in \, \bA } f_{a+u}(c)^{-1} = \frac{1}{\sqrt{|\bA|}} \, \sum_{c \, \in \, \bA } f_a(c)^{-1} \, \gamma(u, c)^{-1} \;,
\ee
from which we can extract some consequences. For our purposes it will be enough to consider $u = b + \phi^{-1}(\psi_{\nu}(b))$ with $b \in \bB$, we then impose 
\be
\begin{split}
\theta_{M^{(a)}}^2 & = \frac{1}{|\bB|}\sum_{b \in \bB} \theta_{M^{(a)}}^2 = \frac{1}{|\bB| \sqrt{|\bA|}}\sum_{\substack{b \in \bB \\ c \in \bA}}f_a(c)^{-1} \, \gamma(b + \phi^{-1}(\psi_{\nu}(b)), c)^{-1}\\ &= \frac{1}{|\bB| \sqrt{|\bA|}}\sum_{\substack{b \in \bB \\ c \in \bA}}f_a(c)^{-1} \, \gamma(b, c)^{-1}\, \gamma(\phi^{-1}(\psi_{\nu}(b)), c)^{-1}\,.
\end{split}
\ee
Any $b \in \bB$ can be split as
\begin{equation}\label{eq:splitb}
    b = \iota(\phi^{-1}(\beta)) + s(x)
\end{equation}
with $\beta \in N(\bB)$ and $x \in \bB/\text{Rad}(\nu)$. Here $\iota$ is the inclusion of $\text{Rad}(\nu)$ in $\bB$ and $s: \bB/\text{Rad}(\nu) \rightarrow \bB$ is a section. Using linearity of $\gamma$ and that $\psi_{\nu}(\phi^{-1}(\beta))=0$ we see that the only $\beta$-dependent factor in the summand is $\beta(c)$, so that the sum over $\beta$ constraints $c \in \bB$. We then have
\be
\begin{split}
\theta_{M^{(a)}}^2 & = \frac{|\text{Rad}(\nu)|}{|\bB| \sqrt{|\bA|}}\sum_{\substack{b' \in \bB \\ x \in \bB/\text{Rad}(\nu)}}f_a(b')^{-1} \, \gamma(s(x), b')^{-1}\, \gamma(\sigma(s(x)), b')^{-1}\,.
\end{split}
\ee
We now split also $b'$ as \eqref{eq:splitb} obtaining
\be
\theta_{M^{(a)}}^2  = \frac{|\text{Rad}(\nu)|}{|\bB| \sqrt{|\bA|}}\sum_{\substack{\beta' \in N(\bB) \\ x,x' \in \bB/\text{Rad}(\nu)}}f_a(\phi^{-1}(\beta'))^{-1}f_a(s(x'))^{-1} \, \gamma(s(x), s(x'))^{-1}\, \gamma(\sigma(s(x)), s(x'))^{-1}\,
\ee
where we noticed that $f_{a}(\phi^{-1}(\beta) + b) = f_{a}(\phi^{-1}(\beta)) f_{a}(b)$ for any $\beta \in N(\bB)$ and $b \in \bB$. Because of this $f_a$ restricted on $\text{Rad}(\nu)$ is a character, hence the sum over $\beta'$ yields $\theta_{M^{(a)}}^2=0$ unless $f_a(\phi^{-1}(\beta'))=1$ for any $\beta' \in N(\bB)$, $\ie$ $f_a$ must restrict to the trivial character on $\text{Rad}(\nu)$ to avoid an inconsistent answer. Plugging this in we get
\be
\theta_{M^{(a)}}^2  = \frac{|\text{Rad}(\nu)|^{2}}{|\bB| \sqrt{|\bA|}}\sum_{\substack{\beta' \in N(\bB) \\ x,x' \in \bB/\text{Rad}(\nu)}}f_a(s(x'))^{-1} \, \gamma(s(x)+ \sigma(s(x)), s(x'))^{-1}\,
\ee
notice that, due to the property of $f_a$ mentioned above, this expression is independent of the sections chosen hence we shall drop them in the following. Using \eqref{eq:relsymmantisymm} we rewrite
\be
\gamma(\sigma(x), x') = \gamma(x,\sigma(x'))^{-1}
\ee
so that summing over $x$ constraints $x'$ to be fixed by $\sigma$. All in all the spin of the twist defect is
\be\label{eq:spin arf}
\theta_{M^{(a)}}^2  = \frac{|\text{Rad}(\nu)|}{\sqrt{|\bA|}}\sum_{\substack{b \in \bB /\text{Rad}(\nu) \\ \sigma(b) = b}}f_{a}(b)^{-1} 
\ee
hence, due to \eqref{eq:misteriouskey}, confirming that \be \theta_{M^{(a)}}^2 = \text{Arf}(\teta) = Y \, .\ee Notice that this computation automatically provides with the proper normalization to ensure that  $\Arf(\teta) = \pm 1$.

\paragraph{Example.}
Consider $\abA= \bZ_n \times \bZ_n$ with off-diagonal bicharacter $\gamma_1^{(\rmO)}$. The invariant algebra is
\be
\linv = \bigl\{ \bigl( (a_1, a_2); (- a_2, a_1) \bigr) \bigm| a_1, a_2 \in \bZ_n \bigr\} \;.
\ee
Our choice for the functions $f_a$ in (\ref{def torsor f_a}) is
\be
f_{(a_1 ,\, a_2)}(b_1, b_2) = \exp \biggl(-\frac{2 \pi i}{n} \, b_1 b_2 \biggr) \, \gamma(a, b) \;.
\ee
From this it is simple to show that 
\be
\theta^2_{\sigma_a} = \exp\biggl( -\frac{2 \pi i}{n} \, a_1 a_2 \biggr) \;.
\ee
A module $M^{(a)}$ is given, as an object, by
\be
M^{(a)} = \begin{cases}
\bigoplus_{b \,\in\, \bZ_n} \sigma_{b ,\, a_2} &\text{for $n$ odd} \,, \\
\bigoplus_{b \,\in\, \bZ_n} \sigma_{a_1 + 2 b ,\, a_2} &\text{for $n$ even} \,.
\end{cases}
\ee
Imposing the spin $\theta_{\sigma_a}^2$ to be constant on the orbit $M^{(a)}$ strongly constrains the possible local module candidates.
One finds that for $n$ odd there is only one consistent choice of module $M$, namely $M^{(0,0)}$ while for $n$ even there are four, corresponding to $(a_1, a_2) = \bigl( s_1 ,\, \frac n2 s_2)$ and $s_{1,2} \in \{0,1\}$. Their spins squared are:
\be
\label{eq: moduletable}
\begin{array}{c||c|c|c|c}
   M^{(a)}  & M^{(0, \, 0)} & M^{(1 , \, 0)} & M^{(0,\, n/2)} & M^{(1,\, n/2)}  \\[0.3em]
   \hline \rule{0pt}{1.3em}
   \theta_M^2  & 1 & 1 & 1 & -1  
\end{array}
\ee
It is possible to check that all four satisfy the locality condition \eqref{eq: localmodules} for the four inequivalent choices of $\teta$, parametrized by $H^1_\rho(\bZ_2, \, \bZ_n \times \bZ_n) = \bZ_2 \times \bZ_2$. We will see in the next section how the same result can be obtained in terms of symmetry fractionalization.

\subsection{Second obstruction and symmetry fractionalization}
\label{sec: frac3d}

The discussion in the previous section gave us a description of the second obstruction from a purely bulk perspective. It however requires precise knowledge of the full categorical data of the 3d MTC, hence it is hard to generalize it to higher-dimensional cases. Moreover it leaves one conceptual problem to address: what is the physical interpretation of the different choices of equivariantization from the point of view of the boundary? We make here a proposal that solves both issues: different choices of equivariantization in the bulk lead to different ways to couple the symmetry to backgrounds fields. This goes by the name of \emph{symmetry fractionalization}.%
\footnote{This is a slight abuse of terminology since the term ``symmetry fractionalization'' is commonly used to indicate the decoration of defect junctions by higher-codimension defects, while here we mix two 0-form symmetries. Yet, we use the term in order to better uniformize the 2d discussion here with the 4d one.}

Even though we do not know how to turn on background fields for the non-invertible symmetry directly, we can use the vanishing of the first obstruction to reduce the problem to the discussion of inequivalent couplings to standard $\bZ_2$ background fields on the invertible boundary. There we also have the 0-form symmetry $\cS=\cZ(\abA) / \linv$, which crucially has a mixed anomaly with $G$. It is known \cite{Delmastro:2022pfo, Brennan:2022tyl} that in such cases the cubic $G$ anomaly might not have an intrinsic value: it can be changed by choosing different symmetry fractionalization classes. Analyzing this phenomenon will lead to the required condition for the vanishing of the second obstruction.

Let us start by determining the mixed anomaly between $G$ and $\cS$. The duality action $\Phi$, which leaves $\linv$ invariant, descends to an action on the quotient $\cS=(\bA \times \bA ^\vee )/\linv$, which we already described in detail in Section \ref{sec: invalgebra}. For simplicity we consider here the case $\bB=\bA$, so that $\cS = \bA^\vee$. The general case is qualitatively analogous and we report it in Appendix~\ref{sec:mix anomaly general}. We use the duality isomorphism $\phi$ to write the background for $\abA^\vee$ as $\phi(B)$ with $B\in H^1(X,\bA)$. The partition function of the invertible boundary theory coupled to a background $B$ can be easily expressed in terms of the reference electric boundary:
\be
\label{eq: electric/invariant boundaries}
Z_\text{inv} \bigl[ \phi(B) \bigr] = \sum _{b \,\in\, H^1(X,\bA)} \exp \biggl[2\pi i \int_X b^* \nu + 2\pi i 
 \int_X b \cup \phi(B) \biggr] \; Z_\rme[b] \;.
\ee
Here $b^*\nu \in H^2 \bigl(X , U(1) \bigr)$ is the pull-back of $\nu \in H^2 \bigl( \bA, U(1) \bigr)$, understood in additive notation (see footnote~\ref{foo: pull back}).
The duality maps $Z_\rme$ to the partition function of the magnetic theory $Z_\rmm$, which in turn can be expressed as a discrete gauging of the electric theory:
\be
\label{eq: electric magnetic boundaries}
\Phi \cdot Z_\rme[b] = Z_\rmm \bigl[ \phi(b) \bigr] = \sum _{a \,\in\, H^1(X,\bA)} \exp \biggl[ 2\pi i \int_X \phi(a) \cup b \biggr] \; Z_\rme[a] \;.
\ee
The action of $\Phi$ on the invertible boundary can be derived combining \eqref{eq: electric/invariant boundaries} with \eqref{eq: electric magnetic boundaries}, using that $\Phi$ only acts on the partition functions $Z$, and it reads
\be
\label{eq:anomalous transoformation}
\Phi \cdot Z_\text{inv} \bigl[ \phi(B) \bigr] = \exp \biggl[ 2\pi i \int_X B^* \nu \biggr] \; Z_\text{inv} \bigl[ \phi(\sigma B) \bigr] \;.
\ee
The overall phase stems from a mixed 't~Hooft anomaly between $G$ and $\cS$. 
Crucially, from \eqref{eq:anomalous transoformation} we find that $G \cong \bZ_2$ acts non trivially on $\cS$ through an automorphism $\rho : G \rightarrow \text{Aut}(\cS)$ such that
\be
\rho_{\underline1}(B) = \sigma B \;,
\ee
so that the total symmetry of the invertible boundary is a semidirect product $\cS\rtimes _\rho G$. Thanks to 
\be
\exp \biggl[ 2 \pi i \int_X B^* \bigl( \nu \circ \sigma \bigr) \biggr] = \exp \biggl[ -2 \pi i \int_X B^* \nu \biggr] \;,
\ee
which is the integrated additive version of \eqref{eq:G cond on nu}, the aforementioned anomaly is consistent with the $\bZ_2$ symmetry.
Let us write the inflow action for the anomalous phase, introducing a background filed $A \in H^1(X, \bZ_2)$ for $G$. The general construction is detailed in Appendix~\ref{sec:mixed anomaly 2d}. The bottom line of that discussion is that such anomalies are classified by $\mu \in H^1_\rho \bigl(\bZ_2, H^2 \bigl(\bA, U(1) \bigr) \bigr)$ in terms of which the 3d inflow action is
\be
\label{eq:inflow action semidirect 3d}
    S_\mu = 2\pi i \int_{X_3} \mu(A)\cup B\cup B \;.
\ee
In components this is defined as
\be
\label{eq:anomaly inflow}
\Bigl( \mu(A) \cup B \cup B \Bigr)_{ijkl} = \mu(A_{ij}) \bigl( \rho_{A_{ij}} B_{jk} \,,\, \rho_{A_{ik}} B_{kl} \bigr) \;.
\ee
A gauge variation $A\rightarrow A+d\lambda$ produces a boundary term
\be
\label{eq:inflow variation}
    S_\mu \rightarrow S_\mu + 2\pi i \int _{\partial X_3} \mu(\lambda) \cup B\cup B \;.
\ee
The class $\mu$ can be thought of as a function $\mu: \bZ_2 \rightarrow H^2\bigl( \bA, U(1) \bigr)$ satisfying the twisted cocycle condition (using addivite notation)
\be
\label{eq: twisted homomorphism mu}
\rho_g \, \mu(h) + \mu(g) = \mu(g+h) \;,
\ee
and subject to the the identification
\be
\label{eq: twisted exactness mu}
\mu(g) \,\cong\, \mu(g) + \rho_g \, \xi - \xi \qquad\qquad\text{for any}\qquad \xi \in H^2\bigl( \bA, U(1) \bigr) \;.
\ee
Because of the relation \eqref{eq:G cond on nu}, which in additive notation reads $\sigma \cdot \nu =-\nu$, we can consistently choose
\be
\mu(\underline0) = 0 \;,\qquad\qquad \mu(\underline1) = \nu \;.
\ee
Notice that this makes sense because $\Phi^2$ leaves $Z_{\text{inv}}$ invariant. With this choice, taking a background such that the pull-back of $A$ to the boundary $\partial X_3$ is $\underline0$ and performing a gauge transformation $A \rightarrow A + d\lambda$ with $\lambda\big|_{\partial X_3} = \underline1$, one reproduces the anomalous phase \eqref{eq:anomalous transoformation}. 
This construction also provides a convenient way to determine whether the anomalous phase \eqref{eq:anomalous transoformation} corresponds to a true anomaly or can be cancelled by a local counterterm. Indeed the latter situation occurs if and only if $\mu$ is cohomologically trivial, namely
\be
\label{eq:condition counterterm}
    \nu = \sigma \cdot \xi -\xi
\ee
for some $\xi \in H^2 \bigl( \bA, U(1) \bigr)$. In this case the anomalous phase is eliminated by modifying the action coupled to $B\in H^1( X_2, \bA)$ by the addition of the local counterterm
\be
    S_\text{c.t.} = 2\pi i \int _{X_2} B^*\xi \;.
\ee
If there exists no $\xi$ satisfying \eqref{eq:condition counterterm} then the anomalous phase cannot be cancelled and there is an anomaly. To show the power of this method, let us discuss the example of $\bA = \bZ_n \times \bZ_n$ with diagonal symmetric bicharacter $\gamma_{1,1}^{(\rmD)}$ and 
\be
    \nu \bigl( (x_1,x_2),\, (y_1,y_2) \bigr) = \frac{r}{n} \, x_1 y_2 \qquad\text{with}\qquad r^2 = -1 \mod n \;.
\ee
Then $\sigma$ acts on $\bA$ as $\sigma(x_1,x_2)=(rx_2,-rx_1)$, and since the most general $\xi \in H^2 \bigl( \bA, U(1) \bigr)$ is represented as $\xi \bigl( (x_1,x_2),(y_1,y_2) \bigr) = \frac{s}{n}x_1y_2$ or equivalently as $\xi \bigl( (x_1,x_2), (y_1,y_2) \bigr) = - \frac{s}{n}x_2y_1$ then
\be
    \bigl( \sigma \cdot \xi -\xi \bigr) \bigl( (x_1,x_2), (y_1,y_2) \bigr) = - \frac{2s}{n} \, x_1 y_2 \;.
\ee
For $n$ odd, we can always choose $s = - 2^{-1}r$ hence the anomalous phase can be cancelled by a local counterterm and it is not an anomaly. On the other hand, for $n$ even, $r$ is necessarily odd and thus no choice of $s$ can cancel the anomalous phase: in this case this is a genuine anomaly.

As already argued, the question of what is the value of the pure $G \cong \bZ_2$ anomaly on the invertible boundary is not well-posed until we specify how $G$ couples to a background field $A \in H^1(X_2, G)$. In the boundary global variant where the full symmetry category is invertible, the presence of another 0-form symmetry $\cS$ allows one to make discrete choices for that coupling labelled by a class
\be
    \eta \in H^1_\rho(G,\cS) \;,
\ee
which satisfies the twisted cocycle condition $\rho_g \, \eta(h) + \eta(g) = \eta(g+h)$ and is subject to the identification $\eta(g) \cong \eta(g) + \rho_g \, c - c$ for any $c \in \cS$, similarly to (\ref{eq: twisted homomorphism mu})--(\ref{eq: twisted exactness mu}). So $\eta$ specifies a (twisted) homomorphism from $G$ to $\cS$ which allows one to modify the minimal coupling prescription for $A$, declaring that the latter effectively couples to the diagonal subgroup of $G$ and the image $\eta(G)\subset \cS$. The anomaly cannot be unambiguously determined until we specify $\eta$ because different choices correspond to different $\bZ_2$ subgroups of the full 0-form symmetry group and, due to the mixed anomaly \eqref{eq:inflow action semidirect 3d}, they can have different anomalies.

This phenomenon is sometimes called \emph{symmetry fractionalization}, even though the term is more often used for the mixing of a 0-form symmetry with higher-form symmetries \cite{Delmastro:2022pfo, Brennan:2022tyl}%
\footnote{See, \eg, \cite{PhysRevB.62.7850, Barkeshli:2014cna, PhysRevX.5.041013, Tarantino_2016, PhysRevB.105.125114} and references therein for discussions of symmetry fractionalization in the condensed matter literature.}
(which will be relevant for the 4d/5d case), but we will use the same terminology to emphasize a unified description. The crucial point is that in general there is no canonical choice and we can only talk about differences of anomalies induced by a certain class $\eta$. This is easy to implement at the level of background fields. When $A\in H^1(X_2,\bZ_2)$ is activated, the symmetry fractionalization class changes the background $B\in H^1(X_2,\cS)$ to
\be
    B' = B + A^* \eta = B + \eta(A) \;.
\ee
By plugging this expression into the mixed anomaly \eqref{eq:inflow action semidirect 3d} we change the pure $\bZ_2$ anomaly, classified by $H^3 \bigl( \bZ_2, U(1) \bigr)$, by an extra piece
\be
    S_\text{pure} = 2\pi i \int _{X_3} \mu(A) \cup \eta(A) \cup \eta(A) \,\equiv\, 2 \pi i \int _{X_3} A^*y
\ee
that can be written in terms of a class $y \in H^3 \bigl( \bZ_2, U(1) \bigr)$. An explicit expression for $y(g_1,g_2,g_3)$ can be derived by recasting $\mu(A)\cup \eta(A)\cup \eta(A)$ as
\be
\Bigl( \mu(A) \cup \eta(A) \cup \eta (A) \Bigr)_{ijkl}=  -\mu(-A_{ij}) \Bigl( \eta(A_{jk}) \,,\, \rho_{A_{jk}} \eta(A_{kl}) \Bigr) \;.
\ee
This is useful because $A$ appears with only three different pairs of indices, and we conclude that
\be
    y(g_1,g_2,g_3) = - \mu(-g_1) \Bigl( \eta(g_2) \,,\, \rho_{g_2} \eta(g_3) \Bigr) \;.
\ee
The possible non-triviality of this 3-cocycle is determined by its value at $g_1 = g_2 = g_3 = \underline{1}$, and we will denote simply by $\mu$ and $\eta$ their values at $g=\underline{1}$. Since $\mu =\nu$ and $\rho \, \eta = - \eta$ we obtain
\be
\label{new formula for Y 2d}
y \,\equiv\, y(\underline{1},\underline{1},\underline{1}) = \nu(\eta,\eta) \;.
\ee
Going back to multiplicative notation, we obtain that
\be
    Y = \nu(\eta, \, \eta) \;,
\ee
we will see in examples that coincides with the SPT phase obtained by the equivariantization procedure in Section~\ref{sec: Equivariantization and second obstruction}.

\subsubsection*{Examples}

Let us apply the general discussion to the previously discussed examples.

\paragraph{1.}
The first example is $\abA=\bZ_n$ where a duality-invariant lattice is present only for $n=p^2$. The choice of discrete torsion $\nu$ is trivial, so there is no way to shift the ``bare" Frobenius-Schur indicator $\epsilon$ and the second obstruction vanishes if and only if $\epsilon = 1$.

\paragraph{2.} Next we consider TY($\bZ_n \times \bZ_n$).
Choosing the diagonal bicharacter $\gamma^{(\rmD)}_{1,1}$ in (\ref{eq:duality 3d ZnxZn}), the duality-invariant boundaries are obtained by gauging the full $\bA$ with discrete torsion $\nu$ such that (see \eqref{eq:discrete torsion 3d})
\be
\chi_\nu = \frac{1}{n} \, \biggl( \,\begin{matrix} 0 & r \\ -r & 0 \end{matrix}\, \biggr) \qquad\qquad\text{with}\qquad r^2 = -1 \mod n \;.
\ee
Thus the action $\rho : \bZ_2 \rightarrow \text{Aut}(\bA^\vee)$ is $\rho_{\underline{1}}(a_1,a_2) = (ra_2, -ra_1)$.
We look for all possible symmetry fractionalization classes $\eta \in H^1_\rho(\bZ_2,\bZ_n\times \bZ_n)$, which are determined by $\eta \equiv \eta(\underline{1}) = (x_1,x_2)$ constrained by $x_2 = r x_1$.
Taking into account the identification
\be
(x_1,rx_1) \sim (x_1,rx_1)+(rc_2-c_1,-rc_1-c_2)
\ee
and setting $c_2=0$, $c_1=-1$ we realize that $x_1 \sim x_1+1$ and hence all cocycles are exact:
\be
    H^1_\rho(\bZ_2,\bZ_n\times \bZ_n)=0 \;.
\ee
Thus the phenomenon of symmetry fractionalization is absent in this case and there is only a single equivariantization for $\linv$. The second obstruction again vanishes if and only if $\epsilon = 1$.

Choosing instead the off-diagonal bicharacter $\gamma^{(\rmO)}_1$ is more interesting. As already discussed, the duality-invariant boundaries are associated with the alternating bicharacters
\be
\chi_\nu = \frac{1}{n} \, \biggl( \,\begin{matrix} 0 & r \\ -r & 0 \end{matrix}\, \biggr) \qquad\qquad\text{with}\qquad r^2 = 1 \mod n \;.
\ee
Then $\rho_{\underline1} (a_1,a_2)=(-ra_1,ra_2)$ and the most general cocycle $\eta \in H^1_\rho (\bZ_2, \bZ_n \times \bZ_n)$ has $\eta=(x_1,x_2)$ with
\be
    (r-1) \, x_1 = 0 \mod n \;,\qquad\qquad (r+1) \, x_2 = 0 \mod n \;,
\ee
and is subject to the identifications $x_1\sim x_1-(r+1)c_1$, $x_2\sim x_2-(r-1)c_2$.
Without loss of generality we can take $r=1$, so that $2x_2=0$ and $x_1\sim x_1+2$. Hence for $n$ odd there is no symmetry fractionalization while for $n$ even:
\be
H^1_\rho(\bZ_2,\bZ_n\times \bZ_n) = \bZ_2\times \bZ_2 \qquad\qquad\text{($n$ even)}
\ee
generated by $\eta_{s_1,\, s_2} = \bigl( s_1, \frac n2 s_2 \bigr)$ with $s_{1,2} \in \{0,1\}$.
A representative for $\nu $ is 
\be
    \nu(a,b) = \exp \biggl( \frac{2\pi i}{n} \, a_1 b_2 \biggr)
\ee
and therefore
\be
    Y = \nu(\eta,\eta)= \exp \bigl( \pi i \, s_1 s_2 \bigr) = \operatorname{Arf}(\teta) \;.
\ee
Thus the second obstruction vanishes if and only if
\be
\epsilon = 1 \quad\text{and}\quad s_1 s_2= 0 \;,\qquad\quad\text{or}\qquad\quad \epsilon = -1 \quad\text{and}\quad s_1 s_2 = 1 \;,
\ee
in agreement with the discussion in \cite{Thorngren:2019iar} for the case $\bA= \bZ_2 \times \bZ_2$ and with our computations using the equivariantization of $\linv$ around \eqref{eq: moduletable}.

\section{Anomalies of duality symmetries in 3+1 dimensions}
\label{sec: 5d}

We now extend the classification of anomalies for non-invertible duality defects to the four-dimensional case. As in 2d, we find that there are two obstructions to gauging a non-invertible duality symmetry. The first obstruction again hinges upon the absence of a duality-invariant bulk Lagrangian algebra $\linv$. This maps to the fact that the 4d theory $\cT$ coupled to the Symmetry TFT must admit a duality-invariant global variant. The second obstruction is the presence of a cubic anomaly:
\be
\epsilon_\text{tot} \in \Omega_5^\text{spin} (BG) \;,
\ee
which can be contaminated by a mixed anomaly involving the 0-form symmetry $G$ and a 1-form symmetry $\cS$ through a symmetry fractionalization mechanism similar to the 2d case, now encoded in a class $\eta \in H^2_\rho(G, \cS)$.

Well known examples of 4d theories with self-duality symmetries are the free Maxwell theory, super-Yang-Mills theories with $\cN=4$ supersymmetry and whose gauge algebra is invariant under Langlands duality (\ie, ADEFG as well as $B_2 \cong C_2$) \cite{Choi:2021kmx, Kaidi:2022uux, Choi:2022zal, Antinucci:2022vyk} and various theories of class $\cS$ \cite{Bashmakov:2022jtl, Antinucci:2022cdi}. Understanding the anomalies in these symmetries has immediate interesting consequences. For example, it has been recently observed \cite{Damia:2023ses} that the $\cN=1^*$ massive deformation of $\cN=4$ SYM preserves a self-duality symmetry. The well-known results about vacuum degeneracy in $\cN=1^*$ can then be reinterpreted as anomaly matching conditions. A second natural application is constrain which $\cN=3$ theories can be described through a discrete gauging of $\cN=4$, which we comment about in the conclusions.

\subsection{Duality defects}

Much of our analysis in Section \ref{sec: 3d} can be generalized to self-duality defects in four-dimensional theories that are self-dual under the gauging of a 1-form symmetry $\bA$ \cite{Choi:2021kmx, Choi:2022zal, Kaidi:2022uux}. 
Again the self-duality must be supplied with a choice of isomorphism $\phi: \bA \rightarrow \bA^\vee$. While a complete description of the underlying fusion 3-category $\cC$ is still out of reach, some of the relevant data can be spelled out explicitly.%
\footnote{The Symmetry TFT analysis offers a complementary viewpoint on the data constituting the duality category which might be easier to handle. We explain how the data we describe here are matched in Section \ref{sec: 5dsymtft}} 
As stated in the introduction, this is a graded category with the grading being implemented by the duality group $G$. The fusion rules take the form
\be
\label{eq: duality4dfusion}
a \times \cN_g = \cN_g \times a = \cN_g \;,\qquad\qquad \cN_g(\Sigma) \times \overline{\rule{0pt}{0.9em}\cN}_{\!g}(\Sigma) = \!\! \sum_{\gamma \,\in\, H^1(\Sigma,\, \bA)} \!\! a(\gamma) = C_\bA(\Sigma) \;,
\ee
where $\cN$ is the duality interface, $\Sigma$ the 3-manifold where it lives, $a$ a 1-form symmetry surface, and $C_\bA$ the condensate of $\bA$. The fusion of $\cN_g \times \cN_h$ is also known, and is group-like at the level of connected components, $\ie$  forgetting the appearance of condensates (see footnote \ref{footnote: connected components}).
It was analyzed in \cite{Antinucci:2022vyk, Antinucci:2022cdi}.

\begin{figure}[t]
\centering
\begin{tikzpicture}[scale=1.75]          
%blue defect
    \fill[left color=white!50!blue, right color= white] (1.44, 0.88) arc (88: 270: 0.3 and 0.31) -- (2.5,-0.05) -- (2.5,0.55) -- cycle;  
    \draw[color= blue, line width = 0.8] (1.55, 0.85) -- (2.5, 0.54);
    \draw[color= blue, line width = 0.8] (1.4, 0.27) -- (2.5, -0.06);                          
%red defect
    \filldraw[color=red!50!white, fill=white!70!red, opacity=0.8] (0.2, 0.2) -- (2.2, -0.3) -- (2.2, 1.2) -- (0.2, 1.7) -- cycle;
%endline of blue defect, small piece
    \draw[line width=1, color=black] (1.244, 0.787) arc (490: 557: 0.3);
%endline of green defect, left piece
    \draw[line width=1, color=black] (1.2, 0.52) arc(310: 645: 0.3);
%green defect             
    \fill[color=white, right color=white!50!green, left color=white, opacity=0.5] (0.9, 1.03) arc (110: -80: 0.3 and 0.3) -- (0, -0.05) -- (0, 0.55) -- cycle;
    \draw[color=green!90!black, line width = 0.83] (0.905, 1.037) -- (0, 0.55);
    \draw[color=green!90!black, line width = 0.8] (1.04, 0.45) -- (0, -0.05);
%endline of the green defect, right piece
    \draw[line width=1, color=black] (1.2, 0.52) arc(310: 470: 0.3);                        
%endline of the blue defect
    \draw[line width=1.] (1.15,0.47) arc (200:470:0.3);
%nodes with text
    \node[below] at (0.5,1.5) {$\cN$};
    \node[left] at (-0. , 0.55) {$a$};
    \node[below] at (2.7,0.) {$\phi(b)$};
    \node[right] at (3.0, 0.55) {$ = \quad \gamma(a, b)$};
\begin{scope}[shift={(4.7, 0)}]
%blue defect
\begin{scope}[shift={(0.2, -0.3)}]
    \fill[left color=white!50!blue, right color= white] (1.44, 0.88) arc (88: 270: 0.3 and 0.31) -- (2.5,-0.05) -- (2.5,0.55) -- cycle;  
    \draw[color= blue, line width = 0.8] (1.55, 0.85) -- (2.5, 0.54);
    \draw[color= blue, line width = 0.8] (1.4, 0.27) -- (2.5, -0.06);                          
\end{scope}                      
%red defect
    \filldraw[color=red!50!white, fill=white!70!red, opacity=0.8] (0.2, 0.2) -- (2.2, -0.3) -- (2.2, 1.2) -- (0.2, 1.7) -- cycle;
%endline of the green defect #1
    \draw[line width=1, color=black] (1.2, 0.52) arc(310: 680: 0.3); 
%green defect             
    \fill[color=white, right color=white!50!green, left color=white, opacity=0.5] (0.9, 1.03) arc (110: -70: 0.3 and 0.3) -- (0, -0.05) -- (0, 0.55) -- cycle;
    \draw[color=green!90!black, line width = 0.8] (0.889, 1.029) -- (0, 0.55);
    \draw[color=green!90!black, line width = 0.8] (1.145, 0.48) -- (0, -0.05);
%endline of the green defect #2                    
    \draw[line width=1] (1.145, 0.482) arc(297: 473: 0.3);                        
%endline of the blue defect
\begin{scope}[shift={(0.2, -0.3)}]
    \draw[line width=1] (1.15, 0.47) arc (200: 560: 0.3);
\end{scope}        
    \node[below] at (0.5, 1.5) {$\cN$};
    \node[left] at (0, 0.55) {$a$};
    \node[below] at (3, 0.6) {$\phi(b)$};          
\end{scope}
\end{tikzpicture}
\caption{\label{fig: brading on surface} Braiding of lines $W^L_a$ and $W^R_b$ on the duality defect. Unlinking the line configuration gives rise to the symmetric bicharacter $\gamma(a,b)$.}
\end{figure}
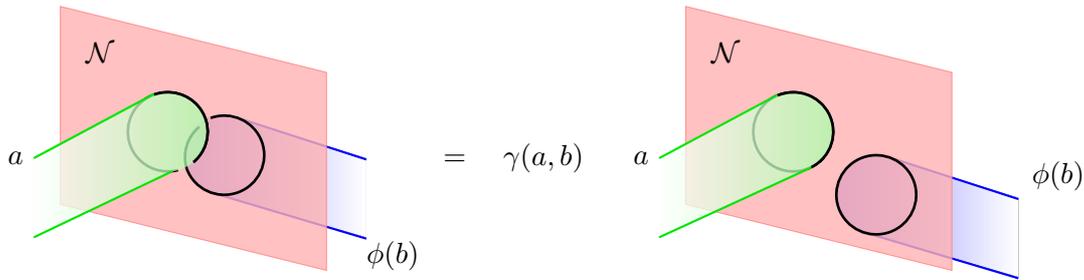

A first piece of categorical data can be obtained noticing that the 1-form symmetry surfaces $a$ can end topologically on $\cN$ thus defining topological line operators $W_a^{L,R}$, where $L$/$R$ encode the side (Left or Right) on which the 1-form symmetry surface ends.%
\footnote{One could think of those as 2-morphisms $W^R_a: a \times \unit_\cN \to \unit_\cN$ and $W^L_a : \unit_\cN \times a \to \unit_\cN$, where $\unit_\cN$ is the identity endomorphism of $\cN$.}
These line defects must compose according to the $\abA$ group law, modulo undetectable decoupled objects:%
\footnote{The importance of modding out such decoupled TQFTs has been recently emphasized in \cite{Copetti:2023mcq} in a related context.} 
\be
W^{L,R}_{a} \times W^{L,R}_b = W^{L,R}_{a + b} \;.
\ee
Following the same logic as in the Tambara-Yamagami case, we consider the braiding between endlines of 1-form symmetry surfaces $a$ and $\alpha$ ending on the two sides of the duality defect $\cN$.
Using electric-magnetic duality we can map $W^L_b$ to an 't~Hooft line $T^L_{\phi^{-1}(b)}$, which braids canonically with $W_a^L$. We conclude that the braiding between $W_a^R$ and $W_b^L$ is given by a symmetric bicharacter $\gamma$:
\be
\cB_{W_a^R ,\, W_b^L} = \gamma(a, b) \;,
\ee
where the symmetry of $\gamma$ follows from the fact that we should get the same result if we worked in the magnetic frame instead. The configuration is depicted in Figure~\ref{fig: brading on surface}.

The lines $W_a^R$ and $W_a^L$ form a $3d$ TQFT $\cA$, but such description is clearly non-minimal: lines of the form 
\be
\cK_a = W_a^R \times W_{-a}^{L}
\ee 
are decoupled from the bulk 1-form symmetry and constitute an undetectable sector $\cA_0$. Quotienting this out gives the minimal description $\cA_{\text{min}}$ of the category of lines living on the defect.%
\footnote{Formally this is obtained by stacking the orientation reversal of $\cA_0$ and gauging the diagonal $\bA$ symmetry: $\cA_\text{min} = (\cA \times \overline{\cA}_0)/ \bA$.}
This produces, in general, a set of lines $L_a$ forming a minimal $\abA$ TQFT $\cA^{\abA, \, q}$ \cite{Hsin:2018vcg}, with $q$ the quadratic refinement of the symmetric bilinear form determined by $\gamma$. This resonates with previous results obtained from the Symmetry TFT perspective \cite{Antinucci:2022vyk}.

Finally, as in the two-dimensional case, we can associate to $\cN$ a pure $G$ anomaly $\epsilon$. This is as an higher analogue of the Frobenius-Schur indicator. For $G=\bZ_2$ such analogy is exact, as $\epsilon$ is reflected in a time reversal anomaly on the duality defect \cite{Hason:2020yqf, Cordova:2019wpi}. 
Alternatively, $\epsilon$ can be understood as a standard $G$ 't~Hooft anomaly on four-manifolds with trivial $H_2(X, \bA)$. On these manifolds, $\cN_g$ behaves as a standard invertible symmetry according to the fusion rules \eqref{eq: duality4dfusion}.
At the level of the Symmetry TFT, the presence of a nontrivial $\epsilon$ gives a DW twist for the theory $\cZ(\cC)$.
All in all, we find that the known data defining a self-duality category in 4d, or at least a subset of them, are given by a pure anomaly $\epsilon$ for the self-duality group and a symmetric bicharacter $\gamma : \bA \times \bA \rightarrow U(1)$. 

In the ensuing analysis we will make two simplifying assumptions. First, we will consider duality defects on spin manifolds, $w_2(TX)=0$. The classification of discrete gauging operations (global variants of a gauge theory) is different on non-spin manifolds, as the set of discrete theta angles is larger.%
\footnote{As an illuminating example, consider $\bA=\bZ_n$ with $n$ even. On generic manifolds, discrete torsion terms are classified by $H^4 \bigl( B^2\bZ_n, U(1) \bigr) = \bZ_{2n}$, while on spin manifolds the order-two element of this group vanishes due to the Wu formula $B \cup B = B \cup w_2(TX) \text{ mod } 2$. This discussion generalizes to arbitrary $\bA$ in a straightforward manner.}  
Physically this amounts to the possibility of assigning a well-defined spin to lines as this cannot be screened by heavy neutral fermions \cite{Ang:2019txy}. This restriction has physical consequences on the obstruction theory outlined above: some duality defects can be anomaly free on spin manifolds, but anomalous in the presence of a nontrivial $w_2$.\footnote{Loosely speaking, this is some kind of mixed anomaly with gravity, due to the dependence on $w_2(TX)$.}. 
As a prototypical example, consider the $\mathfrak{su}(2)$ $\cN=4$ SYM theory. This admits an $S$-invariant global variant $SO(3)_-$ on spin manifolds. On non-spin (but orientable) manifolds this variant splits into $SO(3)_-^\text{b}$ and $SO(3)_-^\text{f}$, where b/f (bosonic/fermionic) refer to the spin of the generator of the lattice of genuine lines. According to \cite{Ang:2019txy} (Appendix~C) the two objects are interchanged by $S$. Thus, although the duality symmetry in $SU(2)$ $\cN=4$ SYM might be non-anomalous on spin manifolds, it is anomalous on generic orientable manifolds.%
\footnote{We will briefly comment on the interpretation of this fact from the point of view of gapped phases in Appendix~\ref{app: invTFT}.}

Our second assumption is to consider duality defects for which $G$ does not contain fermion parity. This for example excludes the vanilla $S$-duality of the $\cN=4$ SYM theory, for which $S^4 = (-1)^F$, but includes the situation where $S$ is twisted by a discrete R-symmetry \cite{Damia:2023ses}. At the practical level, this implies that the relevant cobordism classification for cubic $G$ anomalies is given by $\Omega_5^\text{spin}(BG)$ as opposed to $\Omega_5^{\text{spin}-G}(\text{pt})$. Both groups have been computed \eg{} in \cite{Hsieh:2020jpj, Debray:2023yrs}.

\subsection{Symmetry TFT and Lagrangian algebras}
\label{sec: 5dsymtft}

The Symmetry TFT for 4d duality defects can be described in close analogy with the 2d case \cite{Kaidi:2022cpf, Antinucci:2022vyk}. We start from a 5d Dijkgraaf-Witten theory for a 1-form symmetry $\bA$ with trivial twist. This has topological surface operators labelled by pairs $(a, \alpha) \in \bA \times \bA^\vee$ with antisymmetric canonical braiding
\be\label{eq: braiding 5d}
\cB_{(a_1, \alpha_1) , (a_2, \alpha_2)} = \alpha_1(a_2) \; \alpha_2(a_1)^{-1} \quad \in\, U(1) \;.
\ee
As in three dimensions, the 5d pure 2-form gauge theory for $\bA$ enjoys \emph{electric-magnetic} duality, corresponding to a choice of isomorphism $\phi$. There is an important difference, though, with respect to the 3d case. The most general ansatz for a duality is 
\be
\begin{array}{cccc}
S: & \bA \times \bA^\vee & \rightarrow & \bA \times \bA^\vee \\
 & (a,\alpha) & \mapsto & \bigl( I \circ \phi^{-1}(\alpha) ,\, \phi(a) \bigr)
\end{array}
\ee
for some automorphism $I: \bA \rightarrow \bA$ to be determined. Let $\gamma : \bA \times \bA \rightarrow U(1)$ be the bicharacter associated with $\phi$, namely $\gamma(a,b) = \phi(a)b$, then $S$ preserves the braiding if and only if
\be
\gamma\bigl( a,\, I\circ \phi^{-1}(\alpha) \bigr) \; \gamma\bigl( \phi^{-1}(\alpha),\, a \bigr) = 1 \;.
\ee
This equation may have multiple solutions, depending on the Abelian group $\bA$, but here we limit ourselves to EM dualities that can be defined universally. If $I$ is the identity then $\gamma$ is an antisymmetric non-degenerate bicharacter, which however does not exist for all Abelian groups%
\footnote{For instance $\bA = \bZ_n$, $n\neq 2$ does not admit any.}
and thus we will not study this case any further. On the other hand, if $I$ is the inversion
\be
I(a)=-a
\ee
then $\gamma$ must be a symmetric non-degenerate bicharacter, which always exists. We will thus consider this case in the following.%
\footnote{Other automorphisms $I$ may exist and lead to EM dualities for certain Abelian groups $\bA$.}
With this choice of definition, $S$ is an order-four automorphism:
\be
S^2(a,\alpha)=(-a,-\alpha) \qquad\Longrightarrow\qquad S^2=C \;,
\ee
where we defined the charge-conjugation operator $C: \bA \times \bA^\vee \rightarrow \bA \times \bA^\vee$. The 5d $\DWA$ theory enjoys a larger set of 0-form symmetries, for any group $\bA$. Indeed we can define another generator 
\be
T:(a ,\, \alpha) \mapsto \bigl( a + \phi^{-1}(\alpha) ,\, \alpha \bigr) \;, 
\ee
and in this way construct an order-three automorphism of $\bA \times \bA^\vee$:
\be
CST: (a ,\, \alpha) \mapsto \bigl( \phi^{-1}(\alpha) ,\, -\alpha -\phi(a) \bigr) \qquad\text{such that}\qquad (CST)^3 = \unit \;.
\ee
The Symmetry TFT for the duality or triality defects is then defined by gauging the group $G$ generated by $S$ or $CST$, respectively. This gauging admits a choice of discrete torsion, which on spin manifolds is classified by
\be
\epsilon \,\in\, \Omega_5^\text{spin}(BG) \;,
\ee
and can be thought of as the higher analogue of the Frobenius-Schur indicator we introduced before.

The same argument for the first obstruction corresponding to the absence of $G$-invariant Lagrangian algebras in the $\DWA$ theory carries over to the 5d case. We are thus led to study the properties of gapped boundaries of the pure 2-form gauge theory for $\bA$. These are labelled by two discrete choices, as in 2d:
\begin{itemize}
    \item a subgroup $\bB \subset \bA$ to be gauged;
    \item a class $[\nu] \in H^4 \bigl( B^2\bB, U(1) \bigr)$ specifying the discrete torsion.
\end{itemize}
Recall that in 2d the discrete-torsion classes are classified by alternating bicharacters. The analog here is the identification of $H^4 \bigl( B^2\bB, U(1) \bigr)$ with the dual of the universal quadratic group $\Gamma(\bB)$ (see \cite{Kapustin:2013qsa, Benini:2018reh} for  details):
\be\label{eq: universal quadratic group}
    H^4 \bigl( B^2\bB, U(1) \bigr) \,\cong\, \Gamma(\bB)^\vee \;.
\ee
This means that any discrete torsion class $[\nu]$ can be represented by a quadratic function $   q_\nu: \bB \rightarrow U(1)$.
The group $\Gamma(\bB)$ is equipped with a quadratic function $\cQ: \bB \rightarrow \Gamma(\bB)$ such that for any Abelian group $V$, any quadratic function $q: \bB \rightarrow V$ factorizes as $q=\tilde{q} \circ \cQ$ with $\tilde{q}: \Gamma(\bB) \rightarrow V$ a group homomorphism. The topological term implementing the discrete torsion is
\be
    S_\text{torsion} = \int_{X_4} B ^*\nu = \int_{X_4} \tilde{q} \bigl( \fP(B) \bigr) \;,
\ee
where $\fP(B) = B^*\fP \in H^4\bigl( X_4, \Gamma(\bB) \bigr)$ is the pull-back of the universal Pontryagin square class $\fP \in H^4 \bigl( B^2\bB, \Gamma(\bB) \bigr)$ whilst $\tilde{q}$ is the homomorphism associated with the quadratic function $q_\nu$.

Crucially, if $X_4$ is a four-dimensional spin manifold, then two discrete torsions $\nu$, $\nu'$ leading to two quadratic functions $q_\nu$, $q_{\nu'}$ which are different quadratic refinements of the same bicharacter lead to the same topological term \cite{Hsin:2018vcg, Ang:2019txy}: $\int_{X_4} B^*\nu = \int_{X_4} B^*\nu'$.
Thus by working on spin manifolds we can safely label topological manipulations of the boundary theory in terms of a choice of subgroup $\bB \subset \bA$ and of symmetric bicharacter $\chi_\nu : \bB \times \bB \rightarrow U(1)$. Then most of the results will be closely analogous to the 2d/3d case, just replacing antisymmetric with symmetric bicharacters.

As explained, on spin manifolds we can label the Lagrangian algebras $\calL_{\bB, [\nu]}$ in terms of the data $(\bB, \chi_\nu)$. The corresponding gapped boundary has a 1-form symmetry 
\be
    \cS = \bigl( \bA \times \bA^\vee \bigr) / \calL_{\bB, [\nu]} \;.
\ee
One can easily adapt the 3d discussion in order to explicitly write the form of the general Lagrangian algebra. The symmetric bicharacter $\chi_\nu: \bB \times \bB \rightarrow U(1)$ induces a group homomorphism $\psi_\nu: \bB \rightarrow \bB^\vee$ as in the 3d case.
Given a pair $(\bB,\chi_\nu)$ we construct the Lagrangian algebra $\calL_{\bB, [\nu]} \subset \bA \times \bA^\vee$ as 
\be
\label{Lagrangian algebras in 5d}
\calL_{\bB, [\nu]} = \Bigl\{ \bigl(b,\, \beta \psi_\nu(b) \bigr) \in \bA \times \bA^\vee \Bigm| b \in \bB \,,\quad \beta \in N(\bB) \Bigr\} \;.
\ee
This has cardinality $|\bA|$ and is Lagrangian since $\cB_{(b_1,\beta _1) , (b_2,\beta _2)} = \chi_\nu (b_2,b_1) \: \chi_\nu (b_1,b_2)^{-1} = 1$, where $(b,\beta)$ is a shorthand for $\bigl( b, \beta\psi_\nu(b) \bigr)$ and we used the symmetry of $\chi_\nu$. 
As in the 3d case (see Appendix~\ref{sec:app B}) one can show that all Lagrangian algebras of the 5d $\DWA$ theory are of this form.

\subsection{First obstruction}

After fixing a choice of electric-magnetic duality, we ask what are the conditions for a duality-invariant Lagrangian algebras $\linv = \Phi(\linv)$ to exist. We will study two cases: $\Phi= S$ (duality) and $\Phi = CST$ (triality). Other cyclic 0-form symmetry groups, when they exist, can be treated similarly. As we previously showed, all Lagrangian algebras are of the form (\ref{Lagrangian algebras in 5d}).
To verify whether a lattice is $\Phi$-invariant, as in 3d we impose that the pairing between $\calL$ and $\Phi(\calL)$ be trivial. The analysis is analogous to the 3d case. For both choices of $\Phi$, we find the necessary condition
\be
\phi\bigl( \text{Rad}(\nu) \bigr) = N(\bB) \;,
\ee
where $\text{Rad}(\nu)$ is the kernel of $\psi_\nu$. As in the 3d case this implies that $|\bB|^2=k|\bA|$ for some positive integer $k\in \bN$ and still $\bB$ cannot be smaller than Lagrangian. Notice however since $\chi_\nu$ is now symmetric rather than antisymmetric we cannot conclude that $|\bA|$ (and in particular $k$) is a perfect square. Indeed we will see explicit counterexamples, hence showing that in higher categories the obstruction from non-integer quantum dimensions of \cite{Chang:2018iay} is not true.

The remaining conditions depend on $\Phi$ and are listed below.

\paragraph{Duality.} The automorphism $\sigma = \phi^{-1} \psi_\nu$ of $\bB/\text{Rad}(\nu)$ must satisfy
\be
\sigma^2 = -1 \; .
\ee
In particular $\sigma$ allows to relate the two symmetric bicharacters as
\be \label{eq: fob4d}
\gamma\bigl( \sigma(a),\, b \bigr) = \chi_\nu(a, b) \ .
\ee
From the two equations above it follows that $\sigma$ is an order-two automorphism of the group of symmetric bilinear forms on $\bB/\text{Rad}(\nu)$:
\be
\chi_\nu \bigl( \sigma(a) , \sigma (b) \bigr) \; \chi_\nu(a, b) = 1 \;.
\ee

\paragraph{Triality.} The automorphism $\tau = \phi^{-1} \psi_\nu$ must satisfy
\be
1 + \tau + \tau^2 = 0 \;.
\ee
It is simple to show that the above implies that $\tau$ is an order-three operation: $\tau^3 = 1$.
Also in this case, the restriction to $\bB/\text{Rad}(\nu)$ of
\be
\gamma \bigl( \tau(a) ,\, b \bigr) = \chi_\nu(a, b)
\ee
 holds.
Using the two above equations it follows that $\tau$ is an order-three automorphism of the group of symmetric bilinear forms on $\bB/\text{Rad}(\nu)$:
\be
\label{triality constraint on chi_nu}
\chi_\nu \bigl( \tau^2(a) , \tau^2(b) \bigr) \; \chi_\nu \bigl( \tau(a), \tau(b) \bigr) \; \chi_\nu(a,b) = 1 \;.
\ee

\subsubsection*{Examples}

\paragraph{1.} Let us study the case of $\bA = \bZ_n$ with the standard symmetric bicharacter $\gamma(a_1, a_2)= \exp \bigl( \frac{2 \pi i}{n} \, a_1 a_2 \bigr)$. Consider a factorization $n = p q$ and a subgroups 
\begin{equation}
\bB =\left\{ b q \ | \ b=0,...,p-1\right\}\cong \bZ_p    
\end{equation}
so that $N(\bB) \cong \bZ_q$. Since duality invariance requires $\bB$ to contain $\phi^{-1}( N(\bB))$ as a subgroup, $q$ must divide $p$ and we set $p = \ell q$.
A choice of $\psi_\nu$ is associated with another symmetric bicharacter defined on $\bB$:
\begin{equation}
\chi_\nu(b_1, b_2) = \exp \left( \frac{2 \pi i r}{p} \, b_1 b_2 \right) \ ,    
\end{equation}
where $r \in \{0, \dots, p-1\}$.
Notice that $\text{Rad}(\nu) \cong \bZ_{\gcd(r,p)}$ hence imposing $\phi\bigl( \text{Rad}(\nu) \bigr) = N(\bB)$ forces $\text{gcd}(r,p)=q$, namely $r = s q$ with $\gcd(\ell,s)=1$. Furthermore, since $\gamma(qb_1,qb_2)=\exp{\left(\frac{2\pi iq}{p}b_1b_2\right)}$ over $\bB/\text{Rad}(\nu)\cong \bZ_p/\bZ_q\cong \bZ_l$ we have
\begin{equation}
    \sigma (b)=\phi^{-1}\psi _\nu (b)=sb \ \text{mod}(\ell)
\end{equation}
Thus we find that:
\begin{enumerate}
\item On spin manifolds, there is a duality-invariant $\linv$ for $\bA = \bZ_n$ if $n = \ell q^2$ and $-1$ is a quadratic residue $\text{mod} (\ell)$:
    \be \label{eq: Zn duality inv}
    s^2 = - 1 \  \text{mod}(\ell)  \, .
    \ee
\item On spin manifolds, there is a triality-invariant $\linv$ for $\bA = \bZ_n$ if $n = \ell q^2$ and the equation
    \be
    s^2 + s + 1 = 0 \  \text{mod}(\ell) \, ,
    \ee
    admits solutions.
\end{enumerate}
These results coincide with the recent classification \cite{Apte:2022xtu} of 4d topological $\bZ_n$ gauge theories that are duality or triality invariant on spin manifolds. As in the 3d case, we provide a precise connection between the two approaches in Appendix~\ref{app: invTFT}.

We notice a simple number theoretic consequence of this result. The duality invariance condition can never be satisfied with $\bB=\bA=\bZ_n$ if $n$ is a multiple of 4. Indeed it is simple to see that if $n=2m$ then $s\in \bZ_n$ such that $s^2=-1\ \text{mod}(2m)$ must be $1\ \text{mod}(2)$. But then if $m$ is even $s=2x+1$ cannot satisfy $s^2=-1 \ \text{mod}(2m)$. Analogously triality invariance cannot be satisfied if $n$ is multiple of $9$.

\paragraph{2.} Another interesting case to consider is $\abA = \bZ_2 \times \bZ_2$ which is the one-form symmetry group of $\text{Spin}(4k)$ gauge theory.  On $\bZ_2 \times \bZ_2$ there are four symmetric non-degenerate quadratic forms 
\be
\gamma^{(D)} = \frac{1}{2}\mat{1 & 0 \\ 0 & 1} \, , \ \ \gamma^{(O)}=\frac{1}{2}\mat{0 & 1 \\ 1 & 0} \, , \ \ \gamma^+ = \frac{1}{2}\mat{1 & 1 \\ 1 & 0} \, , \ \ \gamma^- = \frac{1}{2}\mat{0 & 1 \\ 1 & 1}.
\ee
In this case $-1$ acts as the identity on $\abA$ and duality is an involution. Thus given any choice of $\gamma$ the first obstruction is cancelled by choosing $\bB = \bA$, $\sigma=1$ and $\chi_\nu = \gamma$.
The case of triality is slightly more involved. Let us consider $\abB=\abA$. It is simple to show that the only two $\bZ_2 \times \bZ_2$ isomorphisms $\tau$ satisfying $\tau^2 + \tau + 1 = 0$ are  $\tau^\pm = \smat{1 & 1 \\ 1 & 0}$, $\smat{0 & 1 \\ 1 & 1}$ which are inverses to each other. If $\gamma= \gamma^{(D)}$ we can solve the triality obstruction by taking  $\chi_\nu = \gamma^\pm$ and $\tau =\tau^\pm $, similarly if $\gamma = \gamma^{\pm}$ we can take $\chi_\nu = \gamma^{(D)}$ and $\tau=\tau^\mp$. On the other hand, if $\gamma = \gamma^{(O)}$, $\gamma(\tau^\pm(a), \, b)$ is not symmetric and the obstruction is present for $\abB=\abA$. Let us then consider $\gamma = \gamma^{(O)}$ and $\abB= \bZ_2$. Since $N(\abB)$ is also $\bZ_2$ we must have that $\abB = \phi(N(\abB))$. It is simple to verify that taking $\abB$ to be the diagonal $\bZ_2$ this is indeed satisfied. We conclude that the first obstruction for $\abA = \bZ_2\times \bZ_2$ vanishes for both duality and triality.

This example, combined with the previous one, allows to discuss the first obstruction for $\cN=4$ $\text{Spin}(2m)$ SYM (and its global variants). Recall that the one-form symmetry group is:
\be
\abA = \begin{cases}
 \bZ_4 , \, \ \ \ &\text{if} \ m = 2k +1 \, , \\
 \bZ_2 \times \bZ_2 , \, \ \ \ &\text{if} \ m = 2k \, .
\end{cases}
\ee
We thus find that the first obstruction identically vanishes in all of the relevant cases.

\subsection{Second obstruction and symmetry fractionalization}

While in absence of duality- (or triality-) invariant Lagrangian algebras the non-invertible self-duality symmetry is anomalous, when such invariant algebra does exist the anomalies are determined by those on the invariant boundary, where the symmetry is invertible. The philosophy is the same as in the 2d/3d case: due to a mixed anomaly between the 1-form symmetry $\cS = \abA \times \abA^\vee/\linv$ and the invertible duality symmetry $G$ we can shift the value of the pure $G$ anomaly by changing the symmetry fractionalization class $\eta \in H^2_\rho(G, \, \cS)$. We now determine the mixed anomaly in the simpler case $\bB=\bA$, the generalization to proper subgroups being straightforward but technically tedious.

\paragraph{Duality.} In the case of $\Phi = S$ and so $G = \bZ_4$ the invariant partition function is given by:%
\footnote{For simplicity we omit the normalization factors due to gauging.}
\be
\label{expression for Z_inv}
Z_\text{inv} \bigl[ \phi(B) \bigr] = \sum_{b \,\in\, H^2(X,\,\bA)} \exp \biggl( 2\pi i \int_{X_4} b^*\nu + 2 \pi i \int_{X_4} \phi(B) \cup b \biggr) \; Z_\rme[b] \;,
\ee
where $Z_\rme$ is the partition function corresponding to the reference electric boundary condition, while $\nu$ is defined through a bicharacter $\chi_\nu$ such that
\be
\gamma \bigl( \sigma(a),\, b \bigr) = \chi_\nu(a, b) \qquad\text{and}\qquad \sigma^2 = -1 \;.
\ee
The action of $S$-duality on $Z_\text{inv}$ is easily determined using the action of $S$-duality on the electric theory: 
\be
S \cdot Z_\rme[B] = \sum_{a \,\in\, H^2(X,\, \bA)} \exp \biggl( 2\pi i \int_{X_4} \phi(B) \cup a \biggr) \; Z_\rme[a] \;.
\ee
We find
\be
\label{eq: 4dinvtrafo}
S \cdot Z_\text{inv} \bigl[ \phi(B) \bigr] = G_{\nu} \exp \biggl( 2 \pi i \int_{X_4} B^*\nu \biggr) \; Z_\text{inv} \bigl[ \phi(\sigma B) \bigr] \;,
\ee
where $G_\nu \equiv \sum_{b \in H^2(X,\bB)} \exp \bigl( 2 \pi i \int_X b^*\nu \bigr)$.
Here, assuming that $X_4$ is spin, we used the simplifying relation
\be
\exp \biggl( 2 \pi i \int_{X_4} B^* \bigl( \nu \circ \sigma \bigr) \biggr) = \exp \biggl( - 2 \pi i \int_{X_4} B^* \nu \biggr) \;.
\ee
Assuming that $X_4$ is simply connected (and thus $H^2(X_4, \bZ)$ has no torsion classes) and spin, one can show that the Gauss sum $G_\nu$ is equal to 1 \cite{Apte:2022xtu}.
In a similar way we can verify that
\be
S^2 \cdot Z_\text{inv} \bigl[ \phi(B) \bigr] = Z_\text{inv} \bigl[ -\phi(B) \bigr] = C \cdot Z_\text{inv} \bigl[ \phi(B) \bigr] \;.
\ee
Eqn.~\eqref{eq: 4dinvtrafo} implies that the $\bZ_4$ symmetry generated by $S$ acts on the 1-form symmetry of the theory through $\sigma$, \ie, the symmetry is a split 2-group with nontrivial action $\rho: G \to \text{Aut}(\bA)$ \cite{Baez2004, Kapustin:2013qsa, Benini:2018reh} given by $\rho_{\underline1}(a) = \sigma\, a$.
Furthermore, the overall phase $\exp \bigl( 2 \pi i \int_X B^* \nu \bigr)$ should be thought of as encoding a mixed anomaly
\be
\mu \,\in\, H^1_\rho \Bigl( \bZ_4 ,\, H^4 \bigl( B^2\bA ,\, U(1) \bigr) \Bigr) \qquad\qquad\text{where}\qquad\qquad \mu(\underline1) = \nu \ee
and $\underline1$ is the generator of $G = \bZ_4$, much as in the 2d case.

\paragraph{Triality.} For $\Phi = CST$ and so $G = \bZ_3$ we have the same expression (\ref{expression for Z_inv}) for $Z_\text{inv}\bigl[ \phi(B) \bigr]$, but with the class $\nu$ now satisfying (\ref{triality constraint on chi_nu}) in terms of a $\tau$ such that $\tau^2 + \tau + 1 = 0$. T-duality acts on the electric boundary as
\be
T \cdot Z_e[B] \equiv \exp\biggl(- 2 \pi i \int_{X_4} B^* \gamma \biggr) Z_e[B]\,.
\ee
Then
\be
(CST) \cdot Z_\rme[B] = \exp\biggl( 2 \pi i \int_{X_4} B^* \gamma \biggr) \sum_{a \,\in\, H^2(X, \, \bA)} \exp\biggl( 2 \pi i \int_{X_4} \phi(B) \cup a \biggr) \; Z_\rme[a]
\ee
with $B^*\gamma$ any class stemming from a quadratic refinement of $\gamma$, \ie{} the Pontryagin square induced by $\gamma$, and we find 
\be
(CST) \cdot Z_\text{inv} \bigl[ \phi(B) \bigr] =G_{\gamma + \nu} \, \exp\biggl(2 \pi i \int_{X_4}  B^*\nu \biggr) \; Z_\text{inv} \bigl[ \phi(\tau B) \bigr]
\ee
Here we used that, on spin manifolds, $
\exp\bigl[ 2 \pi i \int B^* \bigl( \nu + \nu \circ \tau + \nu \circ \tau^2 \bigr) \bigr] = 1$. It also holds that
\be
(CST)^3 \cdot Z_\text{inv} \bigl[ \phi(B) \bigr] = Z_\text{inv} \bigl[\phi(B) \bigr] \;.
\ee
As before, the result is interpreted by saying that the split 2-group is twisted by the $\bZ_3$ symmetry and the overall phase comes from a mixed anomaly
\be
\mu \,\in\, H^1_\rho \Bigl( \bZ_3 ,\, H^4 \bigl( B^2\bA ,\, U(1) \bigr) \Bigr) \qquad\qquad\text{where}\qquad\qquad \mu(\underline1) = \nu \;.
\ee

\bigbreak

We thus conclude that, similarly to the $3$d case, the $5$d mixed anomaly is determined by a class 
\begin{equation}
    \mu \in H^1_\rho \Bigl( G ,\, H^4 \bigl( B^2\bA, U(1) \bigr) \Bigr) \cong H^1_\rho(G,\Gamma(\bA)^\vee)
\end{equation}
namely a function from $G$ to the group of quadratic functions over $\bA$ satisfying
\be
\label{eq:consistency}
    \rho_g \, \mu(h) + \mu(g) = \mu(g+h) \;
\ee
and subject to the the identification
\be
\mu(g) \,\cong\, \mu(g) + \rho_g \, \xi - \xi \qquad\qquad\text{for any}\qquad \xi \in H^4\bigl( B^2\bA, U(1) \bigr) \;.
\ee
The full detailed derivation of the anomaly inflow is given in Appendix~\ref{sec: anomaly 4d} and we find 
\be
\label{eq:5d inflow}
    S_\mu = 2\pi i \int_{X_5} \mu(A)\cup \fP_\rho(B) \;.
\ee
To reproduce the anomalous phase arising in the boundary theory we have to compare this phase with the boundary term arising in $S_\mu$ from $A+d\lambda$ when we set the pull-back of $A$ to the boundary to zero, as well as the boundary value of $\lambda$ equal to the element of the group $G$ for which we compute the variation. This determines all the values of $\mu(g)$ for $g\in G$. We can check that the consistency (\ref{eq:consistency}) of these values is satisfied.
In the case of duality $G=\bZ_4$, since $\nu$ satisfies $\nu \bigl( \sigma(a), \sigma(b) \bigr) = -\nu(a,b)$, we deduce that
\be
    \mu(\underline1) = \mu(\underline3) = \nu \;,\qquad\qquad \mu(\underline0) = \mu(\underline2) = 0 \;.
\ee
It is obvious that \eqref{eq:consistency} is satisfied.

For triality $G=\bZ_3$ the crucial relation is 
\be
\label{eq:triality condition}
    \gamma(a,b) \,+\, \gamma \bigl( \tau(a), \tau(b) \bigr) \,+\, \gamma \bigl( \tau^2(a), \tau^2(b) \bigr) = 0 \;.
\ee
By looking at the anomalous phases that we got this implies that
\be
\mu(\underline0) = 0 \;,\qquad\quad
\mu(\underline1)(a,b) = \gamma \bigl( \tau(a) , \tau(b) \bigr) \;,\qquad\quad
\mu(\underline2)(a,b) = \gamma \bigl( \tau(a) , \tau(b) \bigr) + \gamma(a,b) \;.
\ee
Among the consistency relations \eqref{eq:consistency}, the only non-trivial (and independent) ones to check are:
$\tau \mu(\underline1) + \mu(\underline1) = \mu(\underline2)$, $\tau\mu(\underline2) + \mu(\underline1) = 0$ and $\tau^2 \mu(\underline2) + \mu(\underline2) = \mu(\underline1)$, which are indeed satisfied thanks to \eqref{eq:triality condition}.

Given such a mixed anomaly, we are now able to discuss the pure $G$ anomaly. The philosophy is the same as in the 2d/3d case: combining the choice of symmetry fractionalization with the mixed anomaly we can induce an extra contribution to the pure anomaly for the invertible duality symmetry. The details are however slightly different.

In 4d symmetry fractionalization is classified by $\eta \in H^2_\rho(G, \bA)$, which, as opposed to the 2d case where it corresponds to the choice of a $G$ subgroup of the full symmetry, here it corresponds to the choice of a 1-form symmetry defect $\eta(g,h)\in \bA$ inserted along the junction of the intersection of $g,h$ and $gh$ defects. This amounts to redefine the coupling of the 0-form symmetry to a background, prescribing that $B$ is shifted to
\be
    B' = B + A^*\eta  \in H^2_\rho (X,\bA) \;.
\ee
By plugging  this expression into the mixed anomaly \eqref{eq:5d inflow} we shift the pure $G$ anomaly by an extra piece
\be
\label{eq: 5d pure anomaly}
S_\text{pure}  =  2\pi i \int_{X_5}  \mu(A) \cup \fP_\rho(A^*\eta) \,\equiv\, 2 \pi i \int_{X_5} A^*y
\ee
that can be written in terms of a class $y \in H^5 \bigl( G, U(1) \bigr)$. In order to work out an explicit expression for this class we rely on a working assumption. We note that the Pontryagin square operation, when the homology group $H_2(X_5,\bZ)$ is torsion-free, can be written as a cup product%
\footnote{\label{ft:product}The expression (\ref{expr for Pontryagin sq}) should be interpreted as follows. One writes $\bA = \oplus_i \bZ_{n_i}$ and lift $A^* \eta$ to $\oplus_i\bZ$, which is always possible for finite Abelian groups. In $\oplus_i\bZ$ we can take the product among the various components of the lift, then (\ref{expr for Pontryagin sq}) is obtained restricting the result to $\Gamma\left(\oplus_i \bZ_{n_i}\right)=\bigoplus_i \Gamma\left(\bZ_{n_i}\right) \oplus \bigoplus_{i<j} \bZ_{n_i} \otimes \bZ_{n_j}$. If $X_5$ has torsion 1-cycles the Pontryagin square is not a cup product and in order to write it in components we need Steenrod's cup products (see \eg{} \cite{Benini:2018reh}).} \cite{Kapustin:2013qsa}:
\be
\label{expr for Pontryagin sq}
\fP_\rho(A^*\eta) = A^*\eta \cup A^*\eta \; .
\ee
On the other hand the pure $G$ anomaly is non-trivial when the homology group $H_1(X_5,\bZ)$ contains torsion \cite{Debray:2023yrs}. Therefore, in order to do the computation, we pick a bulk spin manifold $X^*_5$ with torsion 1-cycles but with torsion-free 2-cycles so as to write \eqref{eq: 5d pure anomaly} as
\be
S_\text{pure}  =  2\pi i \int _{X^*_5}  \mu(A)\cup A^*\eta \cup A^*\eta \;.
\ee
Then it is easy to conclude that
\be
y \bigl( g_1,g_2,g_3,g_4,g_5 \bigr) = \langle -\mu(-g_1) \, ,\eta(g_2,g_3) \, \rho_{g_2 + g_3} \, \eta(g_4,g_5)\rangle \;.
\ee
where the product in the second entry should be interpreted as in footnote \ref{ft:product}. When the second entry is the image of a quadratic function $\gamma : \bA \rightarrow \Gamma(\bA)$ the above expression can be rewritten in a simpler form using the universal property defining $\Gamma(\bA)$ (see the discussion around \eqref{eq: universal quadratic group}). In particular if we can find a representative for $\eta$ that is invariant under the $\rho$ action, setting $g_2=g_4$ and $g_3 = g_5$, we have
\be
y \bigl( g_1,g_2,g_3,g_2,g_3 \bigr) = \langle -\mu(-g_1) \, ,\eta(g_2,g_3) \,\eta(g_2,g_3)\rangle = -\mu(-g_1)(\eta(g_2, g_3)) \;.
\ee

\subsubsection*{Examples}
We now discuss how this general story applies to examples where $\bA=\bZ_n$ and $G$ is either $\bZ_4$ or $\bZ_3$, namely duality and triality respectively. This has some consequence for the anomaly structure of $\cN=4$ SYM theories with gauge group $\SU(n)$ at $\tau=i, e^{\frac{2\pi i}{3}}$ respectively.

Several technical details on the computations of the twisted cohomology groups are based on the following known result (see \eg \cite{HatcherBook}). If $G\cong \bZ_k$, denoting $f=\rho _{\underline{1}}\in \text{Aut}(\bA)$ (note that $f^k=1$), then
\begin{equation}\label{eq:rwisted cohomology forumla}
    H^n_\rho(G,\bA)\cong \left\{ \begin{array}{cc}
     \displaystyle    \frac{\text{Ker}(1-f)}{\text{Im}(1+f+f^2+...+f^{k-1})} & \  \ \ \ \ \text{if $n$ is even}  \\ \\ 
    \displaystyle   \frac{\text{Ker}(1+f+f^2+...+f^{k-1})}{\text{Im}(1-f)}   & \ \ \ \ \ \text{if $n$ is odd}
    \end{array} \right.
\end{equation}
The symmetry fractionalization classes are classified by $H^2_\rho(G,\bA)$, and we notice that in both the duality and triality examples we have
\begin{equation}
    1+f+f^2+...+f^{k-1}=0
\end{equation}
by virtue of the relations $\sigma ^2=-1$, $\tau ^2+\tau +\tau =1$. Hence for us
\begin{equation}\label{eq:twisted H^2}
    H^2_\rho(G,\bA)=\text{Ker}(1-f)=\left\{ a\in \bA \ | \ \rho_{\underline{1}}(a)=a \right\}=\text{Fix}_{\rho_{\underline{1}}}(\bA) \ .
\end{equation}
This also gives a hint for the form of the explicit representatives of the non-trivial twisted cocycles as
\begin{equation}
    \eta_x(\underline{1},\underline{1})=x \ , \ \ \ x\in \text{Fix}_{\rho_{\underline{1}}}(\bA)
\end{equation}
\paragraph{Duality.}
For the case of duality $G\cong \bZ_4$ we have
\begin{equation}
    \rho_{\underline{1}}(a)=ta \ , \ \ \ t^2=-1 \ \text{mod}(n) \ .
\end{equation}
Using \eqref{eq:twisted H^2} we get
\begin{equation}
    H^2_\rho(\bZ_4, \bZ_n)\cong \left\{ \begin{array}{cc}
        \bZ_2 & \ \ \ \ \text{if $n$ is even}  \\
        0 & \ \ \ \ \text{if $n$ is odd}
    \end{array}\right.
\end{equation}
and in the even case the cocycles can be represented
\begin{equation}
    \eta _s (\underline{1},\underline{1})= \eta _s(\underline{3},\underline{3})= \eta _s(\underline{1},\underline{3})= \eta _s(\underline{3},\underline{1})=\frac{n}{2}s \ , \ \ \ \ \ s=0,1
\end{equation}
with all the other values vanishing. By setting $n=2m$, the pure anomaly is determined by the value of the 5-cocycle $Y\in H^5(\bZ_4, U(1))$ in $g_1=...=g_5=\underline{1}$ and we get
\begin{equation}
    Y=q _{\nu}(    \eta _s(\underline{1},\underline{1}))=e^{2 \pi i ts^2\frac{m}{4}}\,.
\end{equation}
We conclude that for $n$ odd the pure duality anomaly on the invertible boundary is the bare one, given by $\epsilon \in H^5(\bZ_4,U(1))\cong \bZ_4$, while for $n$ even the cancellation depends on the possible values of $Y$. Recall that the first obstruction never vanishes when $m$ is even. Therefore the possible values of $Y$ are 
\be
Y = \exp\left( \dfrac{\pi i}{2} t (2k+1) \right) \ \ \ \ \ \  \text{for }n=2(2k+1)\,.
\ee

In the $\cN=4$ theory with gauge group $\SU(n)$ at $\tau=i$ the non-invertible duality symmetry is anomalous whenever it is intrinsically non-invertible, on spin manifolds we have given the relevant condition for $\abA= \bZ_n$ around equation \eqref{eq: Zn duality inv}.
If the defect is non-intrinsically non-invertible the anomaly automatically vanishes provided we combine the duality with an appropriate R-symmetry rotation in order to have a $\bZ_4$ operation (see \eg{} \cite{Argyres:2016yzz ,Bourton:2018jwb}). Indeed following \cite{Hsieh:2019iba} and using that $\Omega^{spin}_5(B\bZ_4) \cong \bZ_4$ one gets \cite{Damia:2023ses}
\begin{equation}
    \epsilon=60(n-1)-24(n^2-1) \ \text{mod}(4) = 0\, ,
\end{equation}
therefore one should choose the trivial fractionalization class to cancel the second obstruction. One could also consider other definitions of S-duality which do not involve the R-symmetry, in such cases the relevant bordism group $\Omega^{\text{spin}-\bZ_8}(pt) = \bZ_{32} \oplus \bZ_2$ is larger and our techniques would need to be refined in order to appropriately account for the cubic anomaly.

A similar conclusion applies to Maxwell theory, for which $S^4=1$ and the anomaly $56 \ \text{mod}(4) = 0$ also identically vanishes.

\paragraph{Triality.}
In the triality case $G\cong \bZ_3$,
\begin{equation}
    \rho_{\underline{1}}(a)=ta \ , \ \ \ \ t^2+t+1=0
\end{equation}
for which we get
\begin{equation}
    H^2_\rho(\bZ_3, \bZ_n)\cong \left\{ \begin{array}{cc}
        \bZ_3 & \ \ \ \ \text{if $n=0 \ $mod(3) }  \\
        0 & \ \ \ \ \text{otherwise}
    \end{array}\right.
\end{equation}
and the (non)trivial cocycles are 
\begin{equation}
    \eta _s (\underline{1},\underline{1})= \eta _s(\underline{2},\underline{2})= \eta _s(\underline{1},\underline{2})= \eta _s(\underline{2},\underline{1})=\frac{n}{3}s \ , \ \ \ \ \ s=0,1,2
\end{equation}
with all the other values vanishing. Setting $n=3m$, the class $Y\in H^5(\bZ_3,U(1))\cong \bZ_3$ is determined by\footnote{One can easily check that, when $n=3m$ is also even, so $m = 0 \text{ mod } (2)$, the choice of quadratic refinement for $q_\gamma$ is immaterial. }
\begin{equation}
    Y=\left[\mu(\underline{2}) \Bigl( \eta_s(\underline{1},\underline{1}) ,\, \rho_{\underline{2}} \, \eta _s(\underline{1},\underline{1}) \Bigr)\right]^{-1}=q_{\gamma}\left( \eta _s (\underline{1},\underline{1}) \right)=  \left\{ \begin{array}{ll}
         \exp{\left(2 \pi i \dfrac{k}{3} \right)}   & \text{if $m =2k$}  \\ 
         & \\
         \exp{\left( 2 \pi i \dfrac{4 k + 2}{3}\right)}   & \text{if $m =2k +1 $}
    \end{array} \right.
     \, . 
\end{equation}
Again we can apply these results to the case of triality symmetry appearing in $\cN=4$ SYM at $\tau = e^{2 i \pi/3}$. The triality defect is non-intrinsic when there exist $t \in \bZ_n$ such that $1+t+t^2=0 \  \text{mod} (n)$. When this is the case we can ask about the second obstruction. To apply our methods we are not forced to combine the naive $CST$ operation with an R-symmetry rotation to eliminate fermion parity, since $(CST)^3 = \unit$. Then, by the same token as the duality case and knowning that $\Omega^{spin}_5(B\bZ_3) \cong \bZ_9$, we have 
\be
\epsilon = 60(n-1) \ \text{mod}(9) = -3(n-1) \ \text{mod}(9) \,.
\ee
Notice that $Y$ is valued in the $\bZ_3$ subgroup of the $\bZ_9$ anomaly group, then to compare $Y$ to the $\epsilon$ above we need to multiply by $3$. When $n=1 \ \text{mod}(3)$ then $\epsilon = 0$ and there is no choice of fractionalization, therefore the second obstruction vanishes. For $n=2 \ \text{mod}(3)$ we find $\epsilon = 6$ and the triality defect is always anomalous. Finally when $n= 0$ mod $3$ we have $\epsilon = 3$ and a simple computation shows that the second obstruction can be cancelled only when $n= 3m $  with $m=1 \text{ mod } (3)$.

In Maxwell theory instead the anomaly is $56 \ \text{mod}(9) = 2$ and cannot be cancelled by any choice of symmetry fractionalization. We conclude that the triality symmetry in Maxwell theory is always anomalous due to the second obstruction.

\section{A check from dimensional reduction}
\label{sec: dimred}

As a check of our results, we show that the obstruction theory of Section~\ref{sec: 5d} is consistent with the one for Tambara-Yamagami categories upon dimensional reduction on an orientable 2-manifold $W$.
We treat explicitly the case that $W$ is a torus $T^2$, but the generalization to any Riemann surface $\Sigma_\fg$ is straightforward. Physically this should be expected, indeed the simplest example of a 4d theory enjoying self-duality is Maxwell theory, which upon compactification on $T^2$ reduces to the theory of two compact bosons.%
\footnote{Plus a decoupled 2d Maxwell sector that we ignore. Such a sector has a 1-form and a $(-1)$-form symmetry (associated to a $2\pi$ shift of the theta angle), associated to the 0-form and 2-form symmetries of the Symmetry TFT.}
In this example the complexified gauge coupling $\tau$ is mapped to the position of the 2d CFT on the Narain moduli space. Such a theory is well known to enjoy Tambara-Yamagami-type symmetries if the point on the conformal manifold is chosen appropriately \cite{Thorngren:2021yso}.

Compactifying the 5d Dijkgraaf-Witten theory for $\bA$ on the torus is a simple exercise. The resulting 3d TQFT has a 1-form symmetry $\widetilde{\bA}\times \widetilde{\bA}^\vee$ where
\be
\widetilde{\bA} = \bA \times \bA \;,
\ee
together with a 0-form and a 2-form symmetry, both being $\bA \times \bA^\vee$, which we neglect in the following discussion. 
Given a choice $\phi$ for the isomorphism that enters into the 5d duality symmetry, the defect $\Phi$ also implements a $\bZ_4$ symmetry in 3d:
\be
\Phi (a_1 ,\, a_2 ;\, \alpha_1 ,\, \alpha_2) = \bigl( -\phi^{-1}(\alpha_2) ,\, \phi^{-1}(\alpha_1) ;\, -\phi(a_2) ,\, \phi(a_1) \bigr) \;,
\ee
where $(a_1,a_2) \in \widetilde{\bA}$ and $(\alpha_1,\alpha_2) \in \widetilde{\bA}^\vee$.
To get a $\bZ_2$ symmetry we compose this transformation with the internal S-duality of the torus, which also squares to charge conjugation and sends $(a_1, \, a_2; \, \alpha_1, \, \alpha_2) \to (a_2, \, - a_1; \, \alpha_2, \, - \alpha_1)$. The resulting $\bZ_2$ symmetry, which we dub $\widetilde{\Phi}$ acts as:
\be
\widetilde{\Phi} (a_1, \, a_2 ; \, \alpha_1, \, \alpha_2) = (\phi^{-1}(\alpha_1), \, \phi^{-1}(\alpha_2); \, \phi(a_1), \, \phi(a_2)) \, ,
\ee
or, using the $\widetilde{\abA}$
\be
\begin{array}{ccc}
\widetilde{\Phi}: \widetilde{\abA}\times \widetilde{\abA}^{\vee} & \longrightarrow & \widetilde{\abA}\times \widetilde{\abA}^{\vee}  \\
     (\widetilde{a},\widetilde{\alpha}) & \longrightarrow & ( \widetilde{\phi} ^{-1}(\widetilde{\alpha}), \widetilde{\phi}(\widetilde{a})) 
\end{array}
\ee
with $\widetilde{\phi}:\abA\times \abA\rightarrow \abA^{\vee}\times \abA^{\vee}$ given by $\widetilde{\phi}(a_1,a_2)=(\phi(a_1),\phi(a_2))$.

\subsubsection*{First and second obstruction upon dimensional reduction}

We now discuss how the first obstruction in 5d is mapped to the first obstruction in 3d language after compactification. Clearly not all Lagrangian algebras $\calL$ in the 3d description can descend from a 5d description, so we must first characterize them. 
Recall that, in 5d, algebras where described by a choice of subgroup $\bB$ of $\bA$ together with a discrete torsion $[\nu] \in H^4 \bigl( B^2\bB ,\, U(1) \bigr)$. Upon reduction on $T^2$ this should map to a specific class $[\tilde\nu] \in H^2 \bigl( B \widetilde{\bB} ,\, U(1) \bigr)$, where $\widetilde{\bB} = \bB \times \bB$.
Expanding the 5d background $B = B_1 \theta_1 + B_2 \theta_2$ with $\theta_i$ a basis of $H^1(T^2,\bZ)$ (we neglect the 0-form and 2-form symmetries), we find:
\be
\int_{T^2} B^* \nu = B_1 \cup_\nu B_2 - B_2 \cup_\nu B_1 \;,
\ee
where $\cup_\nu$ is the cup product induced by the symmetric bilinear form $\chi_\nu$. The bicharacter corresponding to $\tilde{\nu}$ is then, in matrix and additive notation,
\be
\label{eq: chardimred}
\chi_{\tilde{\nu}} = \biggl( \begin{matrix} 0  & \chi_\nu \\ -\chi_\nu & 0 \end{matrix} \biggr) \;.
\ee
A 3d Lagrangian algebra $\widetilde{\calL}$ induced from 5d then is of the form
\be
\widetilde{\calL} = \Bigl\{ \bigl( \tilde{b} ,\, \tilde{\beta} \psi_{\tilde\nu} (\tilde{b}) \bigr) \Bigm| \tilde{b} \in \widetilde\bB \,,\quad \tilde{\beta} \in N(\widetilde\bB) \Bigr\} \;, 
\ee
where $\psi_{\tilde\nu}: \widetilde\bB \to \widetilde\bB^\vee$ is the homomorphism associated with the antisymmetric bicharacter \eqref{eq: chardimred}.
Since $\text{Rad}(\tilde\nu) = \text{Rad}(\nu) \times \text{Rad}(\nu) $ the 5d condition $\phi(N(\bB)) = \text{Rad}(\nu)$ implies $\phi(N(\widetilde\bB)) = \text{Rad}(\widetilde\nu)$ in 3d.
On the other hand, the map $\tilde\sigma = \tilde\phi{}^{-1} \psi_{\tilde\nu}$ is given by:
\be
\tilde\sigma = \biggl( \begin{matrix} 0 & \sigma \\ -\sigma & 0 \end{matrix} \biggr) \;, 
\ee
which is an involution $\tilde\sigma^2=1$. We have thus shown that solutions to the first obstruction in 5d always descend to solutions to the first obstruction in 3d.

Let us now discuss the second obstruction. We notice that the 5d discrete torsion $\epsilon$, when reduced on $T^2$, trivializes. This is because the torus (as well as any Riemann surface) does not have torsion 1-cycles. Thus it is not possible to detect the 5d second obstruction in 3d after compactification on a Riemann surface. Indeed, from the point of view of symmetry fractionalization, we have $G \cong \bZ_n$ and for any Abelian group $\bA$ we get
\be
H^1_\rho( \bZ_n ,\, \bA ) = \text{Ker}(1 + f) / \text{Im}(1-f) \;,
\ee
with $f = \rho_{\underline{1}}$.
Applying this to the case $\bA = \widetilde\bB / N(\widetilde\bB)$ and $f = \tilde\sigma$ it is simple to prove that the twisted cohomology group is trivial for any choice of $\bB$.%
\footnote{Using that $\sigma^2=-1$ we find that $\text{Ker}(1 + \tilde\sigma)$ is spanned by elements $(b_1,b_2) \in  \widetilde\bB/N(\widetilde\bB)$ such that $b_2 = \sigma(b_1)$. An element of $\text{Im}(1-\tilde\sigma)$ instead is of the form $(b_1, b_2)= \bigl( x - \sigma(y) ,\, \sigma(x) + y \bigr)$. A simple manipulation shows that this is equivalent to $b_2 = \sigma(b_1)$.}
Thus there are no fractionalization classes and therefore the second obstruction always trivializes.

\section{Conclusions and applications}
\label{sec: conclusions}

Let us conclude by mentioning some immediate applications of our results, as well as some interesting open problems.

\paragraph{4d $\boldsymbol{\cN=3}$ theories.}
It has been appreciated in the past that a class of 4d $\cN=3$ theories may be obtained from a discrete gauging of the $\cN=4$ duality symmetry for special values of $\tau$ \cite{Argyres:2016yzz, Bourton:2018jwb}. More precisely, given a $\bZ_k$ subgroup of $SL(2, \, \bZ)$ and a fixed coupling $\tau_k$, where $k=2,3,4,6$,%
\footnote{To be precise, since the duality group is $\text{Mp}(2, \, \bZ)$ the discrete groups are actually $\bZ_3, \, \bZ_4, \, \bZ_8, \, \bZ_{12}$ as charge conjugation squares to fermion number $C^2 = (-)^F$. The combined duality - R symmetry transformation however lies is $\bZ_k$ with $k$ as in the main text.}
we can combine this transformation with a $\bZ_k$ R-symmetry rotation in the Cartan of $SU(4)$ so that the combined action preserves $\cN=3$ supersymmetry. As the gauge coupling $\tau=\tau_k$ must be fixed to its self-dual value, these theories have no exactly marginal deformation and are inherently strongly coupled.
The case of $k=2$ is special, as the symmetry is charge conjugation, hence it preserves the full $\cN=4$ supersymmetry, and is invertible.
We will thus concentrate on the cases $k=4$ (corresponding to the $S$ transformation) and $k=3$ (corresponding to the $CST$ transformation) and gauge group $SU(n)$. 
As the duality symmetry is non-invertible, it must be gauged together with (a subgroup of) the $\bZ_n$ 1-form symmetry and our results imply that this is only consistent if the first obstruction vanishes. Thus there is a severe constraint on the possible $\cN=3$ theories which can be obtained in this way. For example our results show that there is no such theory for $n=3$ and $k=4$. 
We must also check the vanshing of the second obstruction.
The joint duality/R-symmetry anomaly is given by \cite{Damia:2023ses}:
\be
60 (n - 1) - 24 (n^2-1) \ \begin{cases}
    &\text{mod }4 \, , \ \ \ \ \text{if} \ k = 4 \\
    &\text{mod }9 \, , \ \ \ \ \text{if} \ k=3 \, 
\end{cases} \,.
\ee
For the duality case the cubic anomaly is identically trivial, thus the vanishing of the first obstruction is a sufficient condition for the gauging to be consistent. For triality instead it is given by $6 \text{ mod }9$ when $n = 3 m + 2$ and is zero otherwise. It has been checked in \cite{Damia:2023ses} that this anomaly identically trivializes when the first obstruction vanishes. Therefore also in the triality case the gauging is consistent if the first obstruction vanishes. This also implies that, when $n=3m$, we must choose the trivial fractionalization class $\eta \in H^2_\rho(\bZ_3, \, \bZ_{3 m})$. 

In some special cases the S-fold construction of \cite{Garcia-Etxebarria:2015wns} gives rise to discrete gaugings of $\cN=4$ SYM \cite{Aharony:2016kai}. These are engineered by 2 D3-branes probing a $k=3 , \, 4 , \, 6$ S-fold and lead to a discrete gauging of $SU(3), \, SO(5)$ and $G_2$ $\cN=4$ SYM respectively. Our analysis can be applied to the first two cases which, following the discussed examples, indeed are free of anomalies for triality and duality respectively. It would certainly be interesting to understand whether our methods can give some insight also on $\cN=3$ theories which cannot be obtained by a discrete gauging procedure from $\cN=4$ and, in particular, if they enlarge the list of generalized symmetries of S-folds described recently in \cite{Etheredge:2023ler, Amariti:2023hev}.

\paragraph{A mixed anomaly.}
We have mentioned in Section \ref{sec: invalgebra} that the space of duality-invariant Lagrangian algebras is larger on spin manifolds. Similarly one can argue, for example following \cite{Apte:2022xtu}, that the first obstruction in the 4d case has less solutions if the spacetime $X$ is not spin. 
This should be rephrased as the presence of a mixed 't Hooft anomaly between the non-invertible symmetry $\cN$ and gravity, sourced by a nontrivial second Stiefel-Whitney class $w_2(X)$. A well known example is the symmetry $\text{TY}(\bZ_2)_{1,1}$ of the Ising CFT. As a bosonic symmetry this is anomalous as the first obstruction cannot be cancelled. However, if we consider it on spin manifolds $X$ only, the obstruction is absent since the bulk algebra $\linv = \left\{ (0,0) , \, (1,1) \right\}$ is manifestly duality invariant. Such an algebra can only be condensed on spin manifolds as $\theta_{(1,1)}=-1$. On the field theory side it is well known \cite{Chang:2018iay, Thorngren:2021yso, Seiberg:2023cdc} that fermionizing the Ising CFT into a Majorana fermion the duality symmetry $\cN$ becomes the invertible $(-1)^{F_L}$ which is anomaly free. A similar example in 4d, as already stated before, if the $\cN=4$ $SU(2)$ SYM theory, whose duality symmetry is anomaly-free on spin manifolds (after combining it with an R-symmetry rotation) it is anomalous by the first obstruction when $X$ is non-spin.
It would be nice to make this idea more precise.

\paragraph{Duality-invariant RG flows.}
In both 2d and 4d, duality-symmetric theories allow for a plethora of interesting RG flows which preserve the non-invertible symmetry. In the former case they have been studied in \cite{Thorngren:2021yso}, while in the latter an initial study has appeared recently \cite{Damia:2023ses}. As in the 2d case, the anomalies for the duality symmetry can lead to strong constraints on the possible low energy phases. A simple example is the $\cN=1^*$ \cite{Donagi:1995cf, Dorey:1999sj, Dorey:2000fc, Dorey:2001qj, Aharony:2000nt} deformation of $\cN=4$ SYM at $\tau=i$, which, in the presence of the first obstruction, necessarily leads in the IR either to spontaneous symmetry breaking of the non-invertible symmetry, or to a self-dual Coulomb phase \cite{Damia:2023ses}. A related problem deserving further study in the light of our results is the deformation of the $SU(2), \, SU(3), \, SU(4)$ $\cN=4$ theory by the Konishi operator. This must lead in the IR either to an $\cN=0$ CFT or to chiral symmetry breaking in order to match the cubic $\SU(4)$ anomaly. Consistency of these scenarios with the intricate pattern of non-invertible symmetries and their anomalies might allow to put stringent constraints on the possible IR phases. This problem is currently under investigation.

\paragraph{Intrinsic versus anomalous.} In our work we have seen that, in the context of duality symmetries, the concept of ``instrinsic'' \cite{Kaidi:2022cpf} implies that the duality symmetry is anomalous. 
Such concept is not unique to duality symmetries, and can be rephrased as the statement that the symmetry category $\cC$ is not Morita equivalent to any category of the form $n\text{Vec}_G$ for some (higher) group $G$. 
It would be interesting to understand how far the relationship between 't Hooft anomalies and intrinsic defects extends.

\paragraph{Duality-invariant boundary conditions.}
It is known \cite{Jensen:2017eof,Thorngren:2020yht} that the presence of an 't Hooft anomaly for a symmetry $\cC$ forbids the existence of a $\cC$-invariant boundary condition.\footnote{Strictly speaking the argument of \cite{Thorngren:2020yht} only applies to 2d. However, given a representation of the higher dimensional 't Hooft anomalies in terms of defect configurations, it should be possible to extend it to general symmetry categories.} Our results can in principle be used to constrain the existence of duality-invariant conformal boundary conditions, building on the results of \cite{Gaiotto:2008ak,Gaiotto:2008sa,DiPietro:2019hqe}\footnote{See also \cite{Cordova:2023ent} for more details on the action of the non-invertible duality symmetry on boundary conditions in Maxwell theory.} for $\cN=4$ SYM and free Maxwell theory, respectively. This problem is currently under investigation \cite{CTizz}.

\section*{Acknowledgements}
We thank Michele del Zotto, Kantaro Ohmori and Yifan Wang for discussions. CC whishes to thank especially Yifan Wang for clarifications on Tambara Yamagami symmetries and their anomalies. We are grateful to the NORDITA institute for the hospitality  and to organizers of the conference \emph{Categorical Aspects of Symmetry} where this work has been completed. AA, FB, CC and GR are supported by the ERC-COG grant NP-QFT
No. 864583 “Non-perturbative dynamics of quantum fields: from new deconfined phases of
matter to quantum black holes”. The authors are partially supported by the INFN “Iniziativa Specifica ST\&FI”.

\appendix

\section{Gauging in fusion categories}
\label{app:Frobenius}

In this appendix we briefly review well known material about gauging in Fusion categories and Modular tensor categories (possibly extended by zero-form symmetry). A complete review of the underlying formalism can be found in \cite{Fuchs:2002cm, Barkeshli:2014cna} respectively.

\paragraph{Gauging and algebras.} Gauging a generalized symmetry in two dimensions corresponds to the definition of a special symmetric Frobenius algebra $\cA \subset \cC$. This is described by a triplet:
\be
\cA \equiv (\cA, \, m \, , \eta) \, , \ \ \ m \in \Hom(\cA \times \cA , \, \cA) \, , \ \ \ \eta \in \Hom(\unit, \, \cA) \, ,
\ee
where $\cA = \bigoplus_{x_i} Z_i(\cA) \, x_i $ is an object in $\cC$, and we define $Z_i(\cA) = \dim(\Hom(\cA, \, x_i))$. We use $x_i$ to denote the simple objects in $\cC$.
The maps $\pi_i$ are projectors $\pi_i : \cA \to x_i$ onto the simple components of $\cA$ and can be used to recast the commuting diagrams below in tensor-valued expressions.
The algebra morphism $m$ satisfies \be F_{\cA ,\, \cA ,\, \cA}^m = 1 \ee ($m$ trivializes the associator). Furthermore $\eta \circ m = \text{id}_\cA$. We will henceforth suppress $\eta$ for simplicity. The algebra also has a dual structure 
\be
(\Delta, \, \bar{\eta}) \, , \ \ \ \Delta \in \Hom(\cA, \, \cA \times \cA) \, , \ \ \ \bar{\eta} \in \Hom(\cA , \, \unit) \, .
\ee
Satisfying $\Delta \circ m = \text{id}_{\cA}$. Furthermore $\Delta$ and $m$ satisfy the so called Frobenius condition
\bea
\begin{tikzcd}
 \cA \times \cA \arrow[dr, "m"] \arrow[dd, "\unit \times \Delta"] \arrow[rr, "\Delta \times \unit"] & & \cA \times \cA \times \cA \arrow[dd, "\unit \times m"] \\
  & \cA \arrow[dr, "\Delta"] & \\
  \cA \times \cA \times \cA \arrow[rr, "m \times \unit"] & & \cA \times \cA
\end{tikzcd}
\eea
Which ensures that crossing moves from any direction can be performed safely. In three dimensions an algebra must satisfy an additional condition which ensures that it is compatible with the braided structure:
\bea
\begin{tikzcd}
    \cA \times \cA \arrow[r, "b"] \arrow[dr, "m"] & \cA \times \cA \arrow[d, "m"] \\
     & \cA
\end{tikzcd}
\eea
such an algebra is called \emph{commutative}.
The gauging of a symmetry $\cA$ is implemented by a thick enough network of $\cA$ defects, with morphisms $m$ and $\Delta$ at three-valent junctions. 
To understand the symmetry of the theory after gauging we must introduce the concept of modules. First let us introduce the category of (left) $\cA$-modules $\text{Mod}_\cA$. 
Its elements are doublets $(M, r_L)$ with $M$ an object an $r_L \in \Hom(\cA \times M, \, M)$ a morphism allowing the algebra to end on $M$. $r_L$ must satisfy a natural compatibility condition:
\bea
\begin{tikzcd}
    \cA \times \cA \times M \arrow[r, " r_L"]  \arrow[d, "m"] & \cA \times M \arrow[d, "r_L"] \\
    \cA \times M \arrow[r, "r_L"] & M 
\end{tikzcd}
\eea
This equation allows to interpret $r_L$ as a sort of representation of the algebra $\cA$ on $r_L$. Physically the category $\text{Mod}_\cA$ describes an $\cA$-invariant boundary condition. In two dimensions the category describing the symmetry after gauging $\cA$ is the bimodule category $\text{Bimod}_{\cA- \cA}$ of $\cA$ bimodules. A bimodule $(B, \, r_L, \, r_R)$ is both a left and a right module for $\cA$ such that the left and right actions commute:
\bea
\begin{tikzcd}
    \cA \times B \times \cA \arrow[r, "r_L"] \arrow[d,"r_R"] & B \times \cA \arrow[d, "r_R"] \\
    \cA \times M \arrow[r, "r_L"] & M
\end{tikzcd}
\eea
Finally, the symmetry after gauging $\cA$ in 3d is described by local modules $\text{Mod}^{loc}_\cA$ of the commutative algebra $\cA$. These are modules which are compatible with the braiding with $\cA$. In particular given a left module morphism $r_L$ we define the right morphism $r_R$ by \be r_R = r_L \circ b\ee
With the consistency condition:
\bea
\begin{tikzcd}
    \cA \times M \arrow[r, "b \circ b"] \arrow[dr, "r_L"]& \cA \times M \arrow[d, "r_L"] \\
    & M
\end{tikzcd}
\eea
This implements the intuition that objects after the gauging must braid trivially with $\cA$. It is known that the dimension of the category of local modules is:
\be
\dim(\text{M}^{loc}_\cA) = \frac{\dim(\cC)}{\dim(\cA)} \, , \ \ \ \dim(\cA) = \sum_{i \, \in \, \text{simple}} Z_i(\cA) \ \dim(x_i) \, .
\ee
Since the dimension of a fusion category must be $\geq 1$ there is a notion of maximality in gauging commutative algebras, which implies that:
\be
\dim(\cA) \leq \dim(\cC) \, ,
\ee
when the inequality is saturated the algebra $\cA$ is called \emph{Lagrangian} and is denoted by the letter $\calL$. There are more or less standard techniques to construct the category of modules, which employ the fact that the formal tensor product:
\be
\text{Ind}_{\cA}(x_i) = \cA \times x_i \, ,
\ee
gives a (reducible) left $\cA$ module. These modules are called ``induced" and the construction of $\text{Mod}_\cA$ boils down to the decomposition of induced modules. The interested reader can consult \cite{Fuchs:2002cm, Fuchs:2004dz} for a review of these techniques.

\paragraph{Theories with 0-form symmetry.} We will also need some material about 3d theories enriched with a zero-form symmetry $G$. These are the so-called G-crossed extensions and we refer to \cite{Barkeshli:2014cna} for a complete review. 
A G-crossed extension is described by a graded tensor category 
\be\cC = \bigoplus_{g \, \in \, G} \cC_g  \ee
with $\cC_g$ the $g$-twisted sector of the zero-form symmetry. This can be thought of as a two-category $\Sigma\cC$ with a single connected component $|\pi_0(\Sigma \cC)|=1$, the twist defects giving a base of homomorphisms $\sigma_g \in \Hom(U_g, \, \unit)$. We will use $x_i$ to denote objects in the untwisted sector, while $\sigma_{g, \, i}$ denote (simple) twist defects.
The fusion product on the twist defects is graded:
\be
\cC_g \times \cC_h \subset \cC_{g h} \, .
\ee
The zero-form symmetry naturally acts on the defects in $\cC$ via an automorphism of the fusion algebra:
\be
U_g\left[ X_h \right] = g(X)_{h^{-1} g h} \, \in \cC_{h^{-1} g h} \, , 
\ee
we will henceforth restrict to an abelian zero-form symmetry $G$. The symmetry also acts on the junction spaces $V_{ (g,i) , \, (h, j) }{}^{(gh, k)}$ by (unitary) isomorphisms
\be
\cU_{g} : V_{ (g_1, i), \, (g_2, \, j)}{}^{(g_1g_2, k)} \to V_{(g_1, g(i)) , \, (g_2, g(j))}{}^{(g_1g_2, g(k) )} \, ,
\ee
while the $G$ composition law is encoded in a morphism 
\be
\lambda_{x_i}(g,h): \, g(h(x_i)) \to gh(x_i) \, .
\ee
The category comes with a graded associator $\alpha$ and a graded braiding isomorphism:
\be
b : X_g \times Y_h \to g(Y)_h \times X_g \, ,
\ee
also satisfying $G$-crossed versions of the pentagon and hexagon equations.
The number of simple objects $\sigma_{g, \, i}$ in the $g$-twisted sector is equal to the number of $g$-invariant local lines $x_i \in \cC_0$, $g(z_i)= x_i$. This follows from modularity of the Hilbert space on $T^2$ with $G$ backgrounds.
The dimension of each graded category $\cC_g$ is $G$ invariant, thus:
\be
\dim(\cC_g) = \dim(\cC_0) \, , \ \ \ \dim(\cC) = |G| \dim(\cC_0) \, .
\ee

\paragraph{Gauging and equivariantization.}
There are two natural operations which can be introduced in this setting. The first is gauging (A subgroup of) the zero-form symmetry $G$. This leads to a larger modular tensor category $\cC/G$ which has dimension:
\be
\dim(\cC/G) = |G| \dim(\cC) \;,
\ee
$\cC/G$ has an anomaly free one-form symmetry $\text{Rep}(G) = \pd{G}$ which gives charge to the liberated $g$-twisted sectors. The category after gauging hence is still graded by this charge:
\be
\cC/G = \bigoplus_{g \in G} \cD_g \, .
\ee
The way in which simple objects of $\cC/G$ are constructed is familiar from the theory of orbifolds. A simple object $\sigma_{g, \, i}$ is equivariantized into an orbit:
\be
\Sigma_{g, \, i} = \bigoplus_{h \in G/\text{Stab}(\sigma_{g, i})} h\left( \sigma_{g, i} \right) \, ,
\ee
where
\be
\text{Stab}(X) = \left\{ g \in G : g(X) = X \right\} \, ,
\ee
is the stabilizer subgroup. This object can furthermore be dressed by dual symmetry lines which carry a representation $\pi$ of $\text{Stab}(\sigma_{g , i})$. We thus get:
\be
|\text{Stab}(\sigma_{g, i})| \ \ \text{objects} \ \ \ \Sigma_{g, i}^\pi \, .
\ee
The second is to gauge an algebra $\cA_0 \subset \cC_0$. Let $H \subset G$ be the stabilizer of $\cA_0$:
\be
H = \left\{ g \in G : g(\cA_0)= \cA_0 \right\}
\ee
we say that $\cA_0$ preserves a subgroup $H$ of the zero form symmetry.
To specify an $H$-invariant algebra we must also associate a consistent $H$-action to the data $(m, \, \eta)$. This constitutes an \emph{equivariantization} of $\cA_0$ and is generally not unique nor it is guaranteed to exist. The required conditions are simple to summarize. First we require the stabilizer $H$ to leave the algebra morphism fixed
\be \label{eq: mequiv}
m_{g(x) ,\, g(y)}^{g(z)} = m_{x ,\, y}^z \; [\cU_g]_{x ,\, y}^z \; \frac{\teta_g(z) }{ \teta_g(x) \, \teta_g(y)} \;.
\ee
requires the definition of maps $\teta_g(x)$ on the projectors $\pi_x : \cA_0 \to x$:
\be
\pi_{x_i} \to \teta_{g}(x_i) \, \pi_{g(x_i)} \, ,
\ee
furthermore the maps $\teta_g(x_i)$ must compose nicely under the $H$ action, namely
\be \label{eq: Gcomp}
\teta_g(x_i) \, \teta_{h}(g(x_i)) = \lambda_{x_i}(g,h) \, \teta_{gh}(x_i) \, .
\ee

A solution to equation \eqref{eq: mequiv}, \eqref{eq: Gcomp} is not guaranteed to exist, and its existence is tied to the splitting of a certain short exact sequence \cite{Bischoff:2018juy}.  
Even if a solution exists it must be modded out by the appropriate gauge transformations. Suppose that $\Hom(\cA_0, \, x_i)$ is at most one dimensional, then $\eta_g$ is a one-cochain and we can redefine:
\be \label{eq: gaugefreedom}
\pi_{x_i} \to \mu(x_i) \, \pi_{x_i} \, , \ \ \ \teta_{g}(x_i) \to \teta_{g}(x_i) \frac{\mu(x_i)}{\mu(g(x_i))} \, .
\ee
With this settled gauging $\cA_0$ preserves an $H$ subgroup of the zero-form symmetry. The resulting category is:
\be
\cC/\cA_0 = \bigoplus_{h \in H} \cC_h/\cA_0 \, ,
\ee
each entry having dimension:
\be
\dim(\cC_h/\cA_0) = \dim(\cC_0)/\dim(\cA_0) \, ,
\ee
by consistency. We now describe in some detail the objects of the twisted category $\cC/\cA_0$. Clearly since $\cA_0$ has trivial grading it is possible to define twisted module categories:
\be
\text{Mod}_\cA^g 
\ee
by doublets $(M_g, \, r_L)$ with $M_g \in \cC_g$ and $r_L$ a left map $r_L: \cA_0 \times M_g \to M_g$. The interesting part of the construction involves making these modules local. In particular the braiding map:
\be
b: M_g \times \cA_0 \to \cA_0 \times M_g \, ,
\ee
induces a nontrivial action of $g$ onto the module morphism $r_L$ sending 
\be
r_L(x_i) \to \teta_g(x_i) \, r_L(g(x_i)) \, ,
\ee
\bea
\begin{tikzpicture}[scale=0.75]
      \draw[color=red, line width=1] (1,0.75) -- (0.2,1.8*0.75);
     \filldraw[color= white!70!green, opacity=0.5] (0,0) -- (2,0.5) -- (2,3.5) -- (0,3) -- cycle;
    \draw[color=red, line width=1] (2,0) -- (1,0.75);
        \draw[color= black, line width=1] (0,0) -- (0,3);
        \draw[color=red, line width=1] (-0.2, 2.2*0.75) -- (-0.5, 2.5*0.75) -- (0, 3*0.75);
        \draw[fill=blue] (0, 3*0.75) circle (0.05);
        \draw[fill=black] (1,0.75) circle (0.025);
        \node[below] at (0,0) {$M_g$}; \node at (2.25,3.75) {$g$};
        \node at (2.25,-0.25) {$\cA_0$};
        \node at (-1.25, 2.5*0.75) {$g(\cA_0)$};
        
        \begin{scope}[shift={(7,0)}]
        \draw[color=red, line width=1] (1,0.75) -- (0.2,1.8*0.75);
     \filldraw[color= white!70!green, opacity=0.5] (0,0) -- (2,0.5) -- (2,3.5) -- (0,3) -- cycle;
    \draw[color=red, line width=1, >-] (1.25,0.75*0.75) -- (1,0.75);
    \draw[color=black] (1.25,0.75*0.75) -- (2,0);
        \draw[color= black, line width=1] (0,0) -- (0,3);
        \draw[color=red, line width=1] (-0.2, 2.2*0.75) -- (-0.5, 2.5*0.75) -- (0, 3*0.75);
        \draw[fill=blue] (0, 3*0.75) circle (0.05);
        \draw[fill=black] (1,0.75) circle (0.025);    
         \node[below] at (0,0) {$M_g$};
           \node at (2.25,-0.25) {$x_i$};
         \node at (3.75,1.5) {$= \ \ \teta_g(x_i)$};
         \begin{scope}[shift={(7,0)}]
              \draw[color=black] (1,0.75) -- (0.2,1.8*0.75);
     \filldraw[color= white!70!green, opacity=0.5] (0,0) -- (2,0.5) -- (2,3.5) -- (0,3) -- cycle;
    \draw[color=black] (1.25,0.75*0.75) -- (1,0.75);
    \draw[color=black] (1.25,0.75*0.75) -- (2,0);
        \draw[color= black, line width=1] (0,0) -- (0,3);
        \draw[color=black] (-0.2, 2.2*0.75) -- (-0.5, 2.5*0.75) -- (-0.25, 2.75*0.75); \draw[color=red, line width=1, >-] (-0.25, 2.75*0.75) -- (0, 3*0.75);
        \draw[fill=blue] (0, 3*0.75) circle (0.05);
        \draw[fill=black] (1,0.75) circle (0.025);  
          \node[below] at (0,0) {$M_g$};
           \node at (2.25,-0.25) {$x_i$};
              \node at (-1.25, 2.5*0.75) {$g(x_i)$};
         \end{scope}
        \end{scope}
\end{tikzpicture}
\eea
The local bimodule condition is encoded in the commutative diagram:
\bea
\begin{tikzcd}
    \cA_0 \times M_g \arrow[r, "b"] \arrow[drr, "r_L"] & M_g \times \cA_0 \arrow[r, "b"] & g(\cA_0) \times M_g \arrow[d, "g(r_L)"] \\
     & & M_g
\end{tikzcd}
\eea
Or, in components:
\be \label{eq: localmodules}
r_L(x_i) = \teta_g(x_i) R_{x_i , \, M_g} \cdot R_{M_g, \, x_i} \cdot r_L(g(x_i)) \, .
\ee
Thus the specification of $\teta$ influences the structure of the $H$-twisted sectors after gauging $\cA_0$.

\section{Lagrangian algebras for DW theories}\label{sec:app B}
In this appendix we prove that any Lagrangian algebra of $\DWA$ is of the form
\begin{equation}
    \calL_{\abB,[\nu]}=\left\{(b,\beta \psi _\nu(b) \ | \ b\in \abB \ , \ \beta \in N(\abB) \ \right\}
\end{equation}
for some subgroup $\abB\subset \abA$ and a class $[\nu]\in H^2(\abB, U(1))$, and that their associated boundary conditions correspond to a theory obtained from the electric boundary by gauging $\abB$ with discrete torsion $[\nu]$.

We denote by $\pi _{\abA} : \abA \times \abA^\vee \rightarrow \abA$ and $\pi _{\abA^{\vee}}: \abA\times \abA^\vee \rightarrow \abA^\vee$ the projections on the two factors. Let $\calL \subset \abA\times \abA^\vee$ be Lagrangian. We define a subgroup of $\abA$
\begin{equation}
    \abB=\pi_{\abA} (\calL)\subset \abA \ .
\end{equation}
Notice that this not necessarily a subgroup of $\calL$. On the other hand any element of the form $(0,\beta) $ with $\beta \in N(\abB)$ has trivial braiding with any element of $\calL$, hence since $\calL$ is maximal, $N(\abB) \subset \pi _{\abA^\vee}(\calL)$ must be a subgroup of $\calL$. Using the short exact sequence
\begin{equation}
    1\longrightarrow N(\abB) \longrightarrow \abA^\vee \longrightarrow \abB^\vee \longrightarrow 1
\end{equation}
we realize any element of $\abA^\vee$, and in particular of $\pi_{\abA^\vee}(\calL)$, as a pair $\beta \omega$, $\beta \in N(\abB)$, $\omega \in \abB^\vee$. All the elements of $\calL$ are then of the form $(b,\beta \omega)$, $b\in \abB$, $\beta \in N(\abB)$, $\omega \in \abB^\vee$, but since $|\calL|=|\abA|=|\abB| |N(\abB)|$ there must exists a homomorphism $\psi :\abB\rightarrow \abB^\vee$ such that
\begin{equation}
    \omega=\psi(b) \ .
\end{equation}
Finally $\psi$ cannot be arbitrary in order $\calL$ to be Lagrangian, since the elements must have vanishing spin. Defining a bicharacter $\chi :\abB\times \abB\rightarrow U(1)$ as $\chi(b_1,b_2)=\psi(b_1)b_2$, then imposing that $(b,\beta \psi(b))$ has trivial spin we get
\begin{equation}
    1=\theta_{(b,\beta \psi(b))}=\chi(b,b) \ .
\end{equation}
Thus $\chi$ is alternating and defines a class $[\nu]\in H^2(\abB,\bR/\bZ)$, hence $\calL=\calL_{\abB,[\nu]}$.

Now we aim to prove that the boundary defined by $\calL_{\abB,[\nu]}$, where the symmetry is 
\begin{equation}
    \cS=\frac{\abA\times \abA^\vee}{\calL_{\abB,[\nu]}}\cong \calL_{\abB,[\nu]}^\vee
\end{equation}
is obtained from the electric boundary by gauging $\abB$ with discrete torsion $[\nu]$. First we notice that $\calL_{\abB,[\nu]}$ is an extension of $\abB$ by $N(\abB)$ determined as follows. Let $\widehat{c} \in H^2(\abB^\vee,N(\abB))$ be the class associated with 
\begin{equation}
    1\longrightarrow N(\abB)\longrightarrow \abA^\vee \longrightarrow \abB^\vee\longrightarrow 1 \ .
\end{equation}
This is determined by Pontryagin duality from $1\rightarrow \abB \rightarrow \abA \rightarrow \abA/\abB\rightarrow 1$, which is associated with a class $c\in H^2(\abA/\abB, \abB)$. $\widehat{c}$ enters in the composition rules of elements of $\abA^\vee$ when they are represented as pairs $\beta \eta $, $\eta \in \abB^\vee, \beta \in N(\abB)$:
\begin{equation}
    \beta _1\eta _1+\beta _2\eta _2=( \beta_1+\beta_2+\widehat{c}(\eta _1,\eta_2))(\eta_1+\eta _2) \ .
\end{equation}
The elements of $\calL_{\abB,[\nu]}$ can be realized as pairs of $b\in \abB$ and $\beta \in N(\abB)$
\begin{equation}
    [b,\beta]\equiv (b,\beta \psi _\nu(b))
\end{equation}
and their composition is
\begin{equation}
\begin{array}{rl}
    [b_1,\beta _1]+[b_2,\beta _2] & = \Bigl[b_1+b_2,\beta _1+\beta _2+\widehat{c}\left(\psi _\nu(b_1),\psi _\nu(b_2)\right) \Bigr]
\end{array} \ .   
\end{equation}
We conclude that $\calL_{\abB,[\nu]}$ is an extension 
\begin{equation}\label{eq:L as extension}
1    \longrightarrow N(\abB) \longrightarrow \calL_{\abB,[\nu]}\longrightarrow \abB \longrightarrow 1
\end{equation}
determined by
\begin{equation}
    \psi_\nu^*(\, \widehat{c}\, )\in H^2(\abB, N(\abB)) \ .
\end{equation}
Taking the Pontryagin dual of \eqref{eq:L as extension} we get
\begin{equation}
    1\longrightarrow \abB^\vee \overset{\iota}{\longrightarrow} \cS \overset{\pi}{\longrightarrow} \abA/\abB\longrightarrow 1 
\end{equation}
whose class is
\begin{equation}
 \widetilde{c}=   \psi _\nu \circ c \in H^2(\abA/\abB, \abB^\vee) \ .
\end{equation} 

To show that this is the correct symmetry structure on the boundary theory obtained by gauging $\abB$ with discrete torsion $[\nu]$, we consider its partition function coupled to a background
\begin{equation}
    B=\iota(B_1)+s(B_2) \in H^1(X, \cS)
\end{equation}
where $s: \abA/\abB\rightarrow \cS$ is a section of $\pi$ and $B_1, B_2$ are gauge fields valued in $\abB^\vee$ and $\abA/\abB$ respectively. $B_2$ is closed but $B_1$ it is not: its differential is equal to the pull-back through $B_2$ of the extension class in $H^2(\abA/\abB, \abB^\vee)$.
On the other hand the dynamical gauge field $B'$ valued in $\abB$ satisfies
\begin{equation}
    dB'=B_2^*(\, c \, )
\end{equation}
and the partition function is
\begin{equation}
    Z_{\abB,[\nu]}=\sum _{B'}\exp{\left(\int _X\left( B'^* (\, \nu \, )+B_1\cup B'\right) \right)} Z_e[B', B_2] \ .
\end{equation}
The exponent is not gauge invariant under $B'\rightarrow d\rho$ unless $B_1$ satisfies 
\begin{equation}
    \psi _\nu (dB')-dB_1=0 \ .
\end{equation}
This determines the modified cocycle condition for $B_1$ as
\begin{equation}
    dB_1=B_2^*(\, \widetilde{c}\, )
\end{equation}
hence proving that $\cS$ is the right symmetry after gauging $\abB$ with discrete torsion $[\nu]$.

\section{General duality-invariant Lagrangian algebras}

In this appendix we report technical details regarding Lagrangian algebras $\calL_{\abB,[\nu]}$ for general $\abB$. In particular we give the proof of their duality invariance and compute the mixed 't Hooft anomaly with the invertible duality symmetry in that cases, extending the discussion for $\abB=\abA$ given in the main text. For concreteness we look at the 2d/3d case, but the 4d/5d one is analogous.

\subsection{Proof of duality invariance}
\label{sec:general duality invariance}

In  Section \ref{sec: invalgebra} we claimed the result for the general classification of duality invariant Lagrangian algebras of $\cZ(\abA)=\abA\times \abA^\vee$. The most general Lagrangian algebra is in the form
\begin{equation}
    \calL_{\abB, [\nu]}=\left\{(b,\beta \psi _\nu(b)) \ , | \ b\in \abB \ , \beta \in N(\abB)  \right\}
\end{equation}
and this is duality invariant if and only if
\begin{enumerate}
    \item $\phi \left(\text{Rad} (\nu) \right)=N (\abB)$. 
\item The injective homomorphism
\begin{equation}
    \sigma : \abB/\text{Rad} (\nu) \rightarrow \abA/\text{Rad} (\nu) \ .
\end{equation}
has image equal to $\sigma \left(\abB/\text{Rad} (\nu)\right)=\abB/\text{Rad} (\nu)$, and 
\begin{equation}
    \sigma ^2=1 \ .
\end{equation}
\end{enumerate}
To prove it we first notice that since $\calL$ and $\Phi (\calL) $ are both Lagrangian, they are equal if and only if all their lines are mutually transparent. In other words for all $b,b'\in \abB$, $\beta ,\beta' \in \cN(\abB)$ we must have
\begin{equation}\label{eq:duality ivariant condition'}
    \left[\phi (b') b\right] \, \beta \left[\phi ^{-1} (\beta '\psi _{\nu}(b'))\right] \, \psi _{\nu}(b)\left[ \phi ^{-1} \left(\beta '\psi _{\nu}(b') \right)\right] =1 \ .
\end{equation}

First we prove that the two conditions above are necessary. If \eqref{eq:duality ivariant condition'} is satisfied, let us specialize it to $\beta =1$. If $b\in \text{Ker}(\psi _{\nu})$, then
\begin{equation}
    \phi (b)b'=1 \, \ \ \ \forall b'\in \abB
\end{equation}
implying that $\phi (b)\in \cN(\abB)$. Thus $\phi(\text{Ker}(\psi _{\nu}))\subset \cN (\abB)$. On the other hand, for generic $b $ but $b'=1$ we have
\begin{equation}
 \psi _{\nu} (b)\phi ^{-1}(\beta ')=\chi_\nu (b,\phi ^{-1}(\beta '))=\left(\psi _{\nu}\left(\phi ^{-1}(\beta ')\right) b\right)^{-1}=1
\end{equation}
namely $\phi ^{-1}(\beta ')\in \text{Ker}(\psi _{\nu})$. This proves that
\begin{equation}
    \phi \left(\text{Ker} (\psi _{\nu}) \right)=\cN (\abB) 
\end{equation}
and thus we can define $\sigma : \abB/\text{Ker}(\psi _{\nu})\rightarrow \abA/\text{Ker}(\psi _{\nu})$ as above. Let us now keep $\beta $ generic but $\beta '=1$. If also $b=1$ we get
\begin{equation}
    \beta\left(\sigma b'\right)=1
\end{equation}
which implies that $\sigma \left(\abB/\text{Ker}(\psi _{\nu})\right)\subset \abB/\text{Ker}(\psi _{\nu})$. On the other hand if $\beta =1$, for any $b,b'$ we have
\begin{equation}
    \phi (b')b=\psi _{\nu}\left(\sigma b'\right)b
\end{equation}
namely $\sigma ^2=1$.

Conversely we can prove that these two conditions are also sufficient. Since $\phi ^{-1}(\cN (\abB))\subset \abB$ and $\sigma \left(\abB/\text{Ker}(\psi _{\nu})\right)\subset \abB/\text{Ker}(\psi _{\nu})$, then $\beta \left(\phi ^{-1}(\beta')\right)=1$ and $\beta (\sigma (b'))=1$. Moreover since $\phi^{-1}(\cN (\abB))=\text{Ker}(\psi _{\nu})$ we have 
\begin{equation}
    \psi _{\nu}(b)\left[\phi ^{-1}(\beta')\right]=\chi_\nu \left[b,\phi ^{-1}(\beta')\right]=\left[\psi _{\nu}\left(\phi ^{-1}(\beta')\right) b\right]^{-1}=1
\end{equation}
and the equation \eqref{eq:duality ivariant condition'} is reduced to
\begin{equation}
    \left[\phi (b')b\right]\left[\psi _{\nu}(b) \left(\sigma b'\right)\right]=\gamma (b',b) \chi_\nu (b, \sigma b')=1
\end{equation}
since the condition $\sigma ^2=1$ is equivalent to
\begin{equation}
    \gamma (b',b)=\chi_\nu (\sigma (b'),b)
\end{equation}
This completes the proof.

For what follows it is important to discuss few consequence of this theorem.
\begin{itemize}
    \item Because of $\phi(\text{Rad}(\nu))=N(\bB)$ it follows that the short exact sequence $1\rightarrow \text{Rad}(\nu)\rightarrow \abA\rightarrow \abA/\text{Rad}(\nu)\rightarrow 1$ is the image under $\phi^{-1}$ of $1\rightarrow N(\bB)\rightarrow \bA^\vee\rightarrow \bB^\vee\rightarrow 1$. In other words there is a commutative diagram
    \bea\label{eq:commutative diagram'}
\begin{tikzcd}
S_1: & 1 \arrow[r]  & N(\abB) \arrow[r] \arrow[d,leftarrow, "\phi"] &  \abA^\vee \arrow[r] \arrow[d,leftarrow, "\phi"] & \abB^\vee  \arrow[r] \arrow[d,leftarrow, "\phi"] & 1 \\
S_2: & 1 \arrow[r] & \text{Rad}(\nu) \arrow[r] & \abA \arrow[r] & \abA/\text{Rad}(\nu) \arrow[r]  & 1     \ .
\end{tikzcd}
\eea
\item Taking the Pontryagin dual of the diagram \eqref{eq:commutative diagram'} and using the symmetry of $\phi$, namely $\phi^\vee=\phi$ we get an other commutative diagram
\bea\label{eq:commutative diagram}
\begin{tikzcd}
S_3=S_1^\vee: & 1 \arrow[r]  & \abB \arrow[r] \arrow[d,rightarrow, "\phi"] &  \abA \arrow[r] \arrow[d,rightarrow, "\phi"] & \abA/\abB  \arrow[r] \arrow[d,rightarrow, "\phi"] & 1 \\
S_4=S_2^\vee: & 1 \arrow[r] & N(\text{Rad}(\nu)) \arrow[r] & \abA^\vee \arrow[r] & \text{Rad}(\nu)^\vee \arrow[r]  & 1     \ .
\end{tikzcd}
\eea
\item It is simple to prove that there is a canonical isomorphism $N\left(\text{Rad}(\nu)\right)/N(\abB) \cong \left(\bB/\text{Rad}(\nu)\right)^\vee $. Then using that $\phi(\abB)=N(\text{Rad}(\nu))$ and $\phi(\text{Rad}(\nu))=N(\abB)$ we found a commutative diagram
\bea\label{eq:commutative diagram''}
\begin{tikzcd}
S_5: & 1 \arrow[r]  & \text{Rad}(\nu) \arrow[r] \arrow[d,rightarrow, "\phi"] &  \abB \arrow[r] \arrow[d,rightarrow, "\phi"] & \abB/\text{Rad}(\nu)  \arrow[r] \arrow[d,rightarrow, "\phi"] & 1 \\
S_6: & 1 \arrow[r] & N(\bB) \arrow[r] & N\left(\text{Rad}(\nu)\right) \arrow[r] &  \left(\bB/\text{Rad}(\nu)\right)^\vee \arrow[r]  & 1     \ .
\end{tikzcd}
\eea
as well as its Pontryagin dual
\begin{equation}  \label{eq:commutative diagram'''}
    \begin{tikzcd}
S_7=S_5^\vee : & 1 \arrow[r]  & \left(\bB/\text{Rad}(\nu)\right)^\vee \arrow[r] \arrow[d,leftarrow, "\phi"] &  \abB^\vee \arrow[r] \arrow[d,leftarrow, "\phi"] & \text{Rad}(\nu)^\vee  \arrow[r] \arrow[d,leftarrow, "\phi"] & 1 \\
S_8=S_6^\vee : & 1 \arrow[r] & \abB/\text{Rad}(\nu) \arrow[r] & \abA/\text{Rad}(\nu)  \arrow[r] & \bA/\bB\arrow[r]  & 1     \ .
\end{tikzcd}
\end{equation}
\end{itemize}

\subsection{Mixed anomaly in the general case}
\label{sec:mix anomaly general}

The discussion of this appendix is quite technical and involves heavy notation. We will use several short exact sequences which we denote as 
\begin{equation}
    S_m \ \ : \ \ \ \   1\longrightarrow \abB_m \overset{\iota_m}{\longrightarrow} \abA _m\overset{\pi _m}{\longrightarrow} \abA_m/\abB_m \longrightarrow 1  \ .
\end{equation}
$\iota _m,\pi _m, s_m$ denote respectively the inclusion, the projection and a section of $\pi _m$. $S_m$ induces an extension class $c_m\in H^2(\abA_m/\abB_m, \abB_m)$. The sequences which will be used are the $S_1,...,S_8$ introduced in appendix \ref{sec:general duality invariance}.

Moreover we will systematically decompose gauge fields valued in $\abA_m$ in terms of gauge fields values in the subgroup and the quotient as
\begin{equation}
    a_m=\iota _m(b_m)+s_m(b_m')\ .
\end{equation}
$b_m'$ is closed while
\begin{equation}\label{eq:extensions on backgrounds}
    db_m=b_m'^*(c_m) \ .
\end{equation}
All the fields have a subscript labelled the short exact sequence, and both the field valued in the subgroup and in the quotient are denoted by the same letter, but the one in the quotient is always primed. An important remark is in order. The relations \eqref{eq:extensions on backgrounds} mean that in presence of a non-trivial extension the background for the subgroup is the sum of an ordinary cohomology class and a particular cochain solving the constraint, which is dependent on the background for the quotient. Hence all path integrals are intended to be done in order: one first integrate the background for the subgroup (the unprimed one), or more precisely its cohomology part, and then the background for the quotient (the primed one).

Let $\linv$ be a duality invariant algebra associated with $(\abB, [\nu])$. We want to compute the mixed anomaly between $\cS=\cZ(\abA)/\linv$ and the duality $G\cong \bZ_2$ on the invertible boundary. This is obtained from the electric boundary by gauging $\abB$ with discrete torsion $[\nu]$. A gauge field $A\in H^1(X,\abA)$ can be decomposed according to the sequence $S_3$ as
\begin{equation}
\label{eq:A decomposition}
    A=\iota_3 (b_3)+s_3(b_3')
\end{equation}
After gauging $\abB$ with torsion the dual symmetry $\cS$ is an extension of $\abA/\abB$ by $\abB^\vee$ with extension class $\widetilde{c}=-\psi _\nu \circ c_1 \in H^2(\abA/\abB,\abB^\vee)$ (see appendix \ref{sec:app B}), then a background field for $\cS$ is given by a pair $B,b'_3$ valued respectively in $\abB^\vee$ and $\abA/\abB$ where 
\begin{equation}
    dB=b_3'^*(\widetilde{c})\ .
\end{equation}
The partition function on the invertible boundary is
\begin{equation}
    Z_{\text{inv}}[B,b'_3]=\sum _{b_3}\exp{\left(2\pi i \int b_3^*(\nu)+2\pi i \int B\cup b_3\right)}Z_e[b_3,b'_3]\ .
\end{equation}
By acting with the duality on the electric boundary we get the magnetic one, corresponding to the gauging of $\abA$ with trivial torsion:
\begin{equation}
    \Phi \cdot Z_e[b_3,b_3']=\sum _{a\in H^1(X,\abA)}\exp{\left(2\pi i \int \phi(a)\cup \left(\iota _3(b_3)+s_3(b'_3)\right)\right)}Z_e[a] \ .
\end{equation}
We decompose the $\abA$ valued field $a$ according to the sequence $S_3$:
\begin{equation}
    a=\iota _{3}(a_3)+s_3(a_3') \ .
\end{equation}
Because of the commutative diagram \eqref{eq:commutative diagram} $\phi(a) \in H^1(X,\abA^\vee)$ has a decomposition using $S_4$,
\begin{equation}
    \phi(a)=\iota _4(x_4)+s_4(x_4') \ , \ \ \ \ x_4=\phi(a_3) \ , \ \ x_4'=\phi(a_3') \ .
\end{equation}
Furthermore it is useful to decompose $b_3$ using $S_5$
\begin{equation}
    b_3=\iota _5(y_5)+s_5(y_5') \, .
\end{equation}
Hence, using that $\nu$ vanishes on $\text{Rad}(\nu)$, we have
\begin{equation}
\begin{array}{rl}
  \Phi \cdot Z_{\text{inv}}[B,b_3']=    &   \displaystyle \sum_{y_5, y_5', x_4, x_4'} \exp \biggl( 2\pi i \int y_5'^* \nu + B\cup (\iota _5(y_5)+s_5(y_5')) + {} \\
     & \displaystyle + \phi(a)\cup \Bigl( \iota _3\iota _5(y_5) + \iota _3s_5(y_5')+s_3(b_3') \Bigr) \biggr) \; Z_e[a_3,a_3']
\end{array} 
\end{equation}
Since $y_5$ appears linearly the sum over it gives a delta function imposing
\begin{equation}\label{eq:constraint from y5}
    \pi_7\left(B+\pi_1(\phi(a))\right)=0 \ \  \longleftrightarrow \  \  B+\pi_1(\phi(a)) \in \iota _7 \left(\left(\abB/\text{Rad}\right)^\vee\right)
\end{equation}
We notice that $\iota _3\iota _5=\iota_2$, and since $\phi(a)=\iota _4(x_4)+s_4(x_4')$ \eqref{eq:constraint from y5} can be rewritten as
\begin{equation}
    x_4'=-\pi_7(B) \ .
\end{equation}
This delta function will be solved by the $x_4'$ sum which however must be performed only after the $x_4$ sum.

Now we can integrate out $y_5'$. It appears quadratically hence the sum over it can be performed by solving its equation of motion 
\begin{equation}
    \psi_\nu(y_5')+B+\pi_1(\phi(a))=0 \ .
\end{equation}
This equation can be inverted in virtue of \eqref{eq:constraint from y5} and using that $\psi_\nu :\bB/\text{Rad}(\nu) \rightarrow \left(\bB/\text{Rad}(\nu)\right)^\vee$ is invertible. Plugging the result back we get
\begin{equation}
    \Phi \cdot Z_{\text{inv}}[B,b_3']=\sum _{a_3,a_3'}\exp{\left(2\pi i \int \phi(a)\cup s_3(b_3')-\Big(\psi _\nu^{-1}(B+\pi_1 \phi(a)) \Big)^*\nu \right)}\delta(x_4'+\pi_7 (B))Z_e[a]
\end{equation}
We decompose $B$ using $S_7$ as
\begin{equation}
    B=\iota _7(B_7)+s_7(B_7')
\end{equation}
and the duality maps
\begin{equation}
    \Phi (B_7,B_7',b_3')=\left( \sigma ^\vee (B_7) ,\phi(b_3'),\phi^{-1}(B_7')\right) \ .
\end{equation}
The sum over $a_3'$ solves the delta function, while the one over $a_3$ reconstructs $Z_{\text{inv}}$ up to a multiplicative factor which gives the anomaly:
\begin{equation}
     \Phi \cdot Z_{\text{inv}}[B_7,B_7',b_3']=\exp{\left(-\int \left(\phi^{-1}(\sigma^\vee B_7) \right)^* \nu\right)}Z_{\text{inv}}\left( \sigma ^\vee (B_7) ,\phi(b_3'),\phi^{-1}(B_7')\right) 
\end{equation}
\section{Twisted cohomology and anomalies}\label{sec: twisted cohomology}
Here we provide details on the topological actions in both 3d and 5d which we used to cancel the mixed anomaly between the duality symmetry and the 0-form (in 2d) or the 1-form (in 4d) symmetry, when we go to the invariant boundary. In the 2d case this is an anomaly for a semi-direct product, while in 4d is an anomaly for a split 2-group. In both cases we do not discuss the full anomaly, but only the one linear in the duality gauge field $A\in H^d(X,G)$ which is relevant for our discussion.
\subsection{Anomaly for a semi-direct product in 2d}\label{sec:mixed anomaly 2d}
We consider a semi-direct product $\abA\rtimes _\rho G$, where $\rho: G\rightarrow \text{Aut}(\abA)$. This is associated with a short exact sequence
\begin{equation}
    1\longrightarrow \abA \overset{\iota}{\longrightarrow }\abA\rtimes _\rho G \overset{\pi}{\longrightarrow} G\longrightarrow 1
\end{equation}
which splits, namely it admits a section $s: G\rightarrow \abA\rtimes _\rho G$ which is a group homomorphism. Any element can be written uniquey as $\iota(a)s(g)$, $a\in \abA$, $g\in G$, with product rule
\begin{equation}
    \iota(a_1)s(g_1)\times \iota(a_2)s(g_2)=\iota(a_1+\rho_{g_1}(a_2))s(g_1+g_2) \ .
\end{equation}
In particular
\begin{equation}
\label{eq:commutation semidirect}
    s(g)\iota(a)s(g^{-1})=\iota(\rho _g(a))  \ .
\end{equation}

Semi-direct products are generically non-abelian, hence we only have 1-form gauge fields. These are classes $\cA\in H^1(X,\abA\rtimes _\rho G )$, namely 
\begin{equation}
    (d\cA)_{ijk}=\cA_{jk}\cA_{ik}^{-1}\cA_{ij}=1 \ , \ \ \ \ \ \cA_{ij}\sim \Lambda_i^{-1}\cA_{ij}\Lambda_j
\end{equation}
where the order of multiplication matters. Since $\cA_{ij}\in \abA\rtimes _\rho G $, we can write
\begin{equation}
    \cA_{ij}=\iota(B_{ij}) s(A_{ij})
\end{equation}
where $B\in C^1(X,\abA)$, $A\in C^1(X,G)$. Using the commutation relation \eqref{eq:commutation semidirect}, the cocycle condition $(d\cA) _{ijk}=1$ is equivalent to
\begin{equation}
    (dA)_{ijk}=A_{jk}-A_{ik}+A_{ij}=0 \ , \ \ \ \ \ \ (d_{\rho(A)} B)_{ijk}=\rho_{A_{ij}}B_{jk}-B_{ik}+B_{ij}=0 \ .
\end{equation}
The identification $\cA_{ij}\sim \Lambda _i^{-1}\cA_{ij}\Lambda _j$, upon decomposing $\Lambda_i=\iota(\theta_i)s(\lambda_i)$ become
\begin{equation}\label{eq:gauge transf semidirect}
    A_{ij}\sim A_{ij}-\lambda_i+\lambda _j \ , \ \ \ \ B_{ij}\sim\rho _{\lambda_i}^{-1}\left(B_{ij}+\rho_{A_{ij}}\theta _j-\theta _i\right) \ .
\end{equation}
Hence $A$ defines a class in cohomology group $H^1(X,G)$, while $B$ a class in the \emph{twisted} cohomology group $H^1_{\rho}(X,\abA)$, also called cohomology with local coefficients.

We are interested in the anomaly for $\abA\rtimes _\rho G$ whose 3d inflow action is quadratic in $B$ and "linear" in $A$. The word linear is in quote since $B$ is a twisted class, and thus $A$ will appear not only linearly, but also in the twisting. This anomaly is characterized by a characteristic class of $\abA\rtimes _\rho G$ bundles taking value in
$H^1_\rho(G,H^2(\abA,\bR/\bZ))$ \cite{Kapustin:2013uxa}. Such a class can be thought of as a function $\mu$ on $G$ with values in the group of alternating bicharacter over $\abA$, satisfying
\begin{equation}
    \rho_g \mu(h)+\mu(g)=\mu(g+h) \ .
\end{equation}
The $G$-action on bicharacters is given by \eqref{eq:G action on bicharacters}. $\mu$ is subject to the identification
\begin{equation}
    \mu(g)\sim \mu(g)+\rho_g\xi-\xi \ , \ \ \ \ \ \xi \in H^2(\abA^\vee,\bR/\bZ)   \ .
\end{equation}
Notice that $\mu(0)=0$, so that $\mu(-g)=-\rho_g^{-1}\mu(g)$.

Given $A\in H^1(X,G)$ we construct $\mu(A)\in C^1(X,H^2(\abA,\bR/\bZ))$, which satisfies the twisted cocycle condition
\begin{equation}\label{eq:cocycle condition mu(A)}
    \left(d_{\rho(A)}\mu(A)\right)_{ijk}=\rho_{A_{ij}}\mu(A_{jk})-\mu(A_{ik})+\mu(A_{ij})=\mu(A_{ij}+A_{jl})-\mu(A_{ik})=0 \ .
\end{equation}
Moreover under a gauge transformation $A\rightarrow A+d\lambda$ is changes by
\begin{equation}\label{eq:gauge variation mu}
\begin{array}{rl}
   \mu(A_{ij})\rightarrow \mu(A_{ij}+\lambda_j-\lambda_i)=  &      \rho_{\lambda_i}^{-1}\mu(A_{ij}+\lambda_j)+\mu(-\lambda_i)= \\
    = & \rho_{\lambda_i}^{-1}\left(\rho_{A_{ij}}(\mu(\lambda_j))-\mu(\lambda_i)+\mu(A_{ij})\right)=\\
    =& \rho_{\lambda_i}^{-1}\left(\left(d_{\rho(A)}\mu(\lambda)\right)_{ij}+\mu(A_{ij})\right)
\end{array} \ ,
\end{equation}
hence $\mu(A)\in H^1_\rho(X,H^2(\abA,\bR/\bZ))$. 

Given $B\in H^1_\rho(X,\abA^\vee)$ we can form the cup product $\mu(A)\cup B\cup B \in H^3(X,\bR/\bZ)$ as
\begin{equation}\label{eq:anomaly semidirect indices}
    \left(\mu(A)\cup B\cup B\right)_{ijkl}=\mu(A_{ij})\left(\rho_{A_{ij}}B_{jk},\rho_{A_{ik}}B_{kl}\right) \ .
\end{equation}
Under a gauge variation $A\rightarrow A+d\lambda$ we have
\begin{equation}\begin{array}{rl}
    \left(\mu(A)\cup B\cup B\right)_{ijkl} \rightarrow  &  \rho_{\lambda_i}^{-1}\left(\mu(A_{ij})+\left(d_{\rho(A)}\mu(\lambda)\right)_{ij}\right)\Big(\rho_{\lambda_i}^{-1} \rho_{A_{ij}}B_{jk}, \rho_{\lambda_i}^{-1}\rho_{A_{ik}}B_{kl}\Big)  \\ \\
    = &   \left(\mu(A)\cup B\cup B\right)_{ijkl}+ \left(d_{\rho(A)}\mu(\lambda)\right)_{ij}\Big(\rho_{A_{ij}}B_{jk},\rho_{A_{ik}}B_{kl} \Big) \ .
\end{array}     
\end{equation}
This means that
\begin{equation}
    \mu(A)\cup B\cup B\rightarrow \mu(A)\cup B\cup B+d_{\rho(A)}(\mu(\lambda))\cup B\cup B=\mu(A)\cup B\cup B+d\left(\mu(\lambda)\cup B\cup B\right) \ .
\end{equation}
We write the inflow action as 
\begin{equation}\label{eq:inflow action semidirect 3d appendix}
    S_{\mu}=2\pi i \int _{X_3}\mu(A)\cup B\cup B \ .
\end{equation}
If $X_3$ this is gauge invariant, while if $\partial X_3=X_2$ we get a boundary term
\begin{equation}\label{eq:anomalous variation inflow}
    S_{\mu}\rightarrow S_{\mu}+2\pi i \int _{X_2}\mu(\lambda)\cup B\cup B \ .
\end{equation}
\subsection{Anomaly for a split 2-group in 4d}\label{sec: anomaly 4d}
In 4d we have an analog story, where $\abA$ is now a 1-form symmetry. The full symmetry structure is a split 2-group, which is a higher categorical version of a semi-direct product. The definitions can be found in \cite{Baez2004} and a more physical discussion is in \cite{Kapustin:2013qsa, Benini:2018reh}. Here we simply use two facts which from our viewpoint can be motivated as being the straightforward generalization of the discussion on semi-direct products. 

First, a background field for the split 2-group is an ordinary cohomology class $A\in H^1(X,G)$ and a twisted cohomology class $B\in H^2_\rho(X,\abA)$. The latter means that
\begin{equation}
    \left(d_{\rho(A)}B\right)_{ijkl}=\rho_{A_{ij}}B_{jkl}-B_{ikl}+B_{ijl}-B_{ijk}=0
\end{equation}
and there is an identification (gauge transformations)
\begin{equation}\label{eq:gauge transf split 2-group}
    A_{ij}\sim A_{ij}+\lambda_j-\lambda _i \ , \ \ \ \  \ \ \ \ B_{ijk}\sim \rho_{\lambda_i}^{-1}\left(B_{ijk}+\rho_{A_{ij}}\theta_{jk}-\theta_{ik}+\theta_{ij} \right)
\end{equation}
which are the obvious generalizations of \eqref{eq:gauge transf semidirect}.

Second, the piece of the anomaly for the split 2-group which is "linear" in $A$ and quadratic in $B$ is labelled by a characteristic class of 2-group gauge bundles
\begin{equation}
    \mu \in H^1_\rho(G,H^4(B^2\abA,\bR/\bZ)) \ .
\end{equation}
One can show \cite{Kapustin:2013qsa} that $H^4(B^2\abA,\bR/\bZ)$ is isomorphic to $\Gamma(\abA)^\vee$, the Pontryagin dual of the universal quadratic group of $\abA$, which can be identified with the group of quadratic functions $q:\abA \rightarrow \bR/\bZ$ (see \cite{Kapustin:2013qsa, Benini:2018reh} for precise definitions and details). The $G$ action on them is naturally given by
\begin{equation}
    (\rho_g q)(a)=q\left(\rho_g^{-1}a\right) \ .
\end{equation}

The construction of the 5d anomaly inflow is very similar to the semi-direct product, thus we skip many details. Given $A\in H^1(X,G)$ we construct $\mu(A)$ which satisfies \eqref{eq:cocycle condition mu(A)} and \eqref{eq:gauge variation mu}, thus defines a class in $H^1_\rho(X,\Gamma(\abA)^\vee)$. Recall that $H^4(B^2\abA,\Gamma(\abA))\cong \text{Hom}(\Gamma(\abA),\Gamma(\abA))$ has a distinguished element $\fP$, the Pontryagin square, such that 
\begin{equation}
    B\in H_\rho^2(X,\abA) \ ,\ \ \ \ B^*\fP=\fP_\rho(B)\in H_\rho^4(X,\Gamma(\bA)) \ .
\end{equation}
The action of $G$ on $\Gamma(\abA)$ is induced by the one on $\Gamma(\abA)^\vee$ in such a way to make the natural pairing $\langle \ , \ \rangle :\Gamma(\abA)\times \Gamma(\abA)^\vee\rightarrow \bR/\bZ$ invariant. Under $A\rightarrow A+d\lambda$ the latter transforms as 
\begin{equation}
    \fP_\rho(B)_{i_0,...,i_4}\rightarrow \rho_{\lambda_{i_0}}^{-1}\fP_\rho(B)_{i_0,...,i_4} \ .
\end{equation}
Using the pairing between $\Gamma(\abA)$ and $\Gamma(\abA)^\vee$ we construct $\mu(A)\cup \fP_\rho(B)\in H^5(X,\bR/\bZ)$ as 
\begin{equation}
    \left(\mu(A)\cup \fP_\rho(B)\right)_{i_0,...,i_5}=\Big\langle\mu(A)_{i_0i_1}, \rho_{A_{i_0i_1}}\fP_\rho(B)_{i_0,...,i_4}\Big\rangle
\end{equation}
Under $A\rightarrow A+d\lambda$ we have
\begin{equation}
      \mu(A)\cup \fP_\rho(B) \rightarrow  \mu(A)\cup \fP_\rho(B) +d\Big(  \mu(\lambda)\cup \fP_\rho(B) \Big) \ .
\end{equation}
We conclude that the 5d inflow action is
\begin{equation}
    S_\mu=2\pi i \int _{X_5}  \mu(A)\cup \fP_\rho(B) \ .
\end{equation}

\section{Equivariantization for 2-algebras}
In this Appendix we show how the symmetry fractionalization datum $\eta$ appears in theory of equivariantization for symmetric 2-algebras \cite{decoppet2023rigid,Decoppet:2022dnz}. 
We refer to there works for a complete introduction to the formalism and we limit ourselves to highlight the important steps for our application. We will assume that all objects are invertible for simplicity. We stress that a full definition of the equivariantization procedure is still an open problem which should be object of future study.

In five dimensions 2-categories are sylleptic \cite{Johnson-Freyd:2021tbq}, meaning that there exists a 2-morphism $\sigma$ such the braiding 1-morphism $b: X \times Y \to Y \times X$ satisfies:
\bea
\begin{tikzcd}
    X \times Y \arrow[rr, "="] \arrow[dr, swap, "b"]& \arrow[d, Rightarrow, "\sigma"]  &  X \times Y \\
    & Y \times X \arrow[ur, swap, "b"] & 
\end{tikzcd}
\eea
Physically $\sigma(X,Y)$ encodes the braiding data between five dimensional surfaces $B_{X, Y} = \sigma(X,Y)/\sigma(Y,X)$. 
A symmetric two-algebra $\cA^{[2]}$ describes the gauging of a 2-categorical symmetry in five dimensions, is it described by:\footnote{We omit the unit morphisms and its various higher morphisms for the sake of simplicity, since they will not play a role in our presentation.}
\begin{itemize}
    \item An object $\cA = \ds \bigoplus_{x_i \, \text{simple}} Z_x \, x \, \in \, \cC$. We will assume $Z_x = 0, \, 1$.
    \item A one-morphisms $m : \cA \times \cA \to \cA$.
    \item Two two-isomorphisms $\mu, \beta$ which uplift the associativity and commutativity relations for $m$ to
    \bea
 \begin{tikzcd}
     \cA \times \cA \times \cA \arrow[rr, "m \times id"] \arrow[d, swap, "id \times m"]& & \cA \times \cA \arrow[d, "m"] \arrow[dll, Rightarrow, "\mu"] \\
     \cA \times \cA \arrow[rr, swap,  "m"] & & \cA 
 \end{tikzcd} 
 \ \ \ 
 \begin{tikzcd}
     & \cA \times \cA \arrow[dr, "m"] \arrow[d, Rightarrow, "\beta"] & \\
    \cA \times \cA \arrow[rr, swap, "m"] \arrow[ur, "b"] & {} & \cA 
 \end{tikzcd}
    \eea
\end{itemize}
$\mu$ and $\beta$ are furthermore subject to various higher algebraic identities which we do not report. 
A zero-form symmetry $G$ acts on the algebra $\cA$, more specifically:
\begin{itemize}
\item To an object $x \in \cA$ we associate a one-isomorphism (an invertible line) $\varphi_g(x): x \to g(x)$ localized on the $U_g$ surface. 
\item To the algebra morphism $m$ we associate a two-morphism $\mu_g(x,y) : \varphi_g(x) \times \varphi_g(y) \to \varphi_g(x \times y)$. Similarly for $\beta$.
\item To an algebra 2-isomorphism we associate an identity for the equivariantization data. 
\end{itemize}
All these data must be compatible with the natural multiplicative structure for $G$ defects. In particular we can consider the action of $ghk \simeq (gh)k \simeq g(hk) \in G$ on the algebra $\cA$. Each three-valent junction $g \times h \to gh$ defines a 2-isomorphism $\eta_{g,h}(x) : \varphi_g(x) \times \varphi_h(g(x)) \to \varphi_{gh}(x)$. Clearly $\eta$ is best though of as a two-cochain $\eta(g,h)$ with values in the Pontryagin dual of $\cA$, $C^2(G, \, \cA^\vee)$.\footnote{As in the case of three-dimensions, the correct statement is that it is a torsor over such group.} Consistency of $\eta$ with the associativity of the $G$ defects implies (we draw below an horizontal section of the configuration)
\bea
&\rho_g \eta(h,k) \, \eta(g,hk) = \eta(g,h) \, \eta(gh,k) \, , \ \ \ \rho_g \eta(h,k)[x] = \eta_{h,k}(g(x)) \, \\
&\begin{tikzpicture}[scale=1.15]
\coordinate (g_hk) at (1.5,2); \coordinate (h_k) at (2.25, 1); \coordinate (g_h) at (0.5,1); \coordinate (gh_k) at (1.5,2); \coordinate (g) at (0,0); \coordinate (h) at (1.5,0); \coordinate (k) at (3,0); \coordinate (ghk) at (1.5,3);
\filldraw[color=white,fill=white!90!blue] (-0.5,0) to (g) to (g_hk) to (ghk) to (-0.5,3) -- cycle;
\filldraw[color=white,fill=white!90!green] (g) to (g_hk) to (h_k) to (h) -- cycle;
\filldraw[color=white,fill=white!90!red] (h) to (h_k) to (k) --cycle;
\filldraw[color=white, fill=white!90!yellow] (3.5,0) to (k) to (g_hk) to (ghk) to (3.5,3) -- cycle;
\draw[line width=1] (g) node[below] {$g$} to (g_hk) node[left] { $\eta_{g, hk}(x)$} to (ghk) node[above] {$ghk$}; \draw[line width=1] (g_hk) to (h_k) node[right] { $\eta_{h,k}(g \cdot x)$} to (h) node[below] {$h$}; \draw[line width=1] (h_k) to (k) node[below] {$k$};
\node[left] at ($0.5*(g) + 0.5*(g_hk)$) {\footnotesize$\varphi_g(x)$}; \node[left] at ($0.25*(h) + 0.75*(h_k)$) {\footnotesize$\varphi_h(g \cdot x )$};  \node[right] at ($0.5*(k) + 0.5*(h_k)$) {\footnotesize$\varphi_k(g h \cdot x)$}; \node[right] at ($0.5*(g_hk) + 0.5*(ghk)$) {\footnotesize$\varphi_{ghk}(x)$};
\draw[fill=black]  (g_hk) circle (0.075); \draw[fill=black]  (h_k) circle (0.075);
\node at (4.75,1.5) {\large$\simeq$};
\begin{scope}[shift={(6,0)}]
\coordinate (g_hk) at (1.5,2); \coordinate (h_k) at (2.25, 1); \coordinate (g_h) at (0.75,1); \coordinate (gh_k) at (1.5,2); \coordinate (g) at (0,0); \coordinate (h) at (1.5,0); \coordinate (k) at (3,0); \coordinate (ghk) at (1.5,3);
\filldraw[color=white,fill=white!90!blue] (-0.5,0) to (g) to (g_hk) to (ghk) to (-0.5,3) -- cycle;
\filldraw[color=white,fill=white!90!green] (g) to (g_h) to (h) -- cycle;
\filldraw[color=white,fill=white!90!red] (h) to (g_h) to (gh_k) to (k) --cycle;
\filldraw[color=white, fill=white!90!yellow] (3.5,0) to (k) to (g_hk) to (ghk) to (3.5,3) -- cycle;
\draw[line width=1] (g) node[below] {$g$} to (g_h) node[left] {\small $\eta_{g, h}(x)$} to (gh_k) node[right] {$\eta_{gh,k}(x)$} to (ghk) node[above] {$ghk$}; \draw[line width=1] (g_hk) to (k) node[below] {$k$}; \draw[line width=1] (g_h) to (h) node[below] {$h$};    
\node[left] at ($0.5*(g) + 0.5*(g_h)$) {\footnotesize$\varphi_g(x)$}; \node[right] at ($0.25*(h) + 0.75*(g_h)$) {\footnotesize$\varphi_h(g  \cdot x )$};  \node[right] at ($0.5*(k) + 0.5*(gh_k)$) {\footnotesize$\varphi_k(gh \cdot x)$}; \node[left] at ($0.5*(g_hk) + 0.5*(ghk)$) {\footnotesize$\varphi_{ghk}(x)$};
\draw[fill=black]  (g_h) circle (0.075); \draw[fill=black]  (gh_k) circle (0.075);
\end{scope}
\end{tikzpicture}
\eea
$\eta$ defines a closed two-cochain with values in the dual of $\cA$.\footnote{As in the 3d case, the corrects statement is that it is a torsor over such group.} This datum matches the symmetry fractionalization data we have described in Section \ref{sec: Equivariantization and second obstruction}.

\section{First obstruction and invariant TQFTs}
\label{app: invTFT}

In this last appendix we review the relationship between the first obstruction and the existence of duality-invariant TQFTs with symmetry $\bA$. This was explained in \cite{Thorngren:2019iar} for the case of the Tambara-Yamagami category $\TYA$ and in \cite{Apte:2022xtu} for 4d theories with $\bZ_n$ 1-form symmetry.

\paragraph{Self-dual TQFTs in 2d.}
A two-dimensional TQFT with symmetry $\bA$ can be characterized by its unbroken subgroup $\bB$ (in the sense of spontaneous symmetry breaking) and an SPT phase for $\bB$ described by $\nu \in H^2 \bigl( B\bB, U(1) \bigr)$. The partition function is as in (\ref{Z of 2d TQFT}):
\be
Z[B] = \begin{cases}
\exp \bigl( 2 \pi i \int\! B^* \nu \bigr) \qquad &\text{if } \pi(B) = 0 \;, \\
0 &\text{otherwise} \;,
\end{cases}
\ee
where $\pi$ is the projection in the short exact sequence $1 \to \bB \xrightarrow{i\,} \bA \xrightarrow{\pi\,} \bA/\bB \to 1$. The duality action on the TQFT is defined as
\be 
\cN \cdot Z[B] =  \frac{1}{\sqrt{ \rvert H^1(X, \bA) \rvert}} \sum_{a \, \in \, H^1(X, \bA)} \exp\biggl( 2\pi i \int_X a \cup_\gamma B \biggr) \, Z[a] \;.
\ee
Let us assume that $\nu$ satisfies \eqref{eq:firstob1}--\eqref{eq:firstob3}. We evaluate the above equation on the torus $T^2$, however this can equivalently be done on any Riemann surface by choosing a decomposition of $H_1(\Sigma_g)$ in A- and B-cycles. We find:
\be
\cN \cdot Z[B_1, \, B_2] = \frac{1}{\lvert\bA\rvert} \sum_{a_1, a_2 \,\in\, \bB} \chi_\nu(a_1, a_2) \, \gamma(a_1, B_2) \, \gamma(a_2, B_1)^{-1}  \;,
\ee
where we used that $Z[a] = 0$ if $\pi(a) \neq 0$ to restrict the sum. We first perform the sum over the subgroup $\text{Rad}(\nu)$ of $\abB$. This is zero unless:
\be
B \in \phi^{-1} \left( N(\text{Rad}(\nu)) \right)  \, ,
\ee
this group coincides with $\abB$ (and thus would give to correct delta function) if and only if $\phi(\text{Rad}(\nu)) = N(\abB)$.\footnote{One uses the fact that, for any subgroup $\abB$, $\phi^{-1} N (\phi^{-1}N(\abB)) = \abB $.} Thus \eqref{eq:firstob1} ensures that preserved subgroup is the same after the action of $\cN$.
Finally we perform the sum over (say) $a_1$ restricted to $\bB/\text{Rad}(\nu)$ which fixes $a_2 =  \sigma B_2$ thus:
\be
\cN \cdot Z[B_1, B_2] = \gamma\bigl( \sigma B_2, B_1)^{-1} \, \delta\bigl( \pi(B_1) \bigr) \, \delta\bigl( \pi(B_2) \bigr) = Z[B_1, B_2] \;,
\ee
where \eqref{eq:firstob2} and \eqref{eq:firstob3} ensure that the SPT phase for $\bB$ remains the same.

\paragraph{Self-dual TQFTs in 4d.}
A similar reasoning is valid in 4d. Without assuming whether the four manifold $X$ is spin we can defined a 4d $\abA^{(1)}$ TQFT by specifying the preserved subgroup $\abB$ and an $\abB$ SPT:
\be
\nu \in H^2(B^2\abB, \, U(1)) \, ,
\ee
and define
\be
Z[B] = \begin{cases}
 \exp\left(2 \pi i \int B^* \nu \right) \, , \ \ \ &\text{if} \ \pi(B)=0 \\
 0 \, , \ \ \ &\text{otherwise} \, .
\end{cases}
\ee
We then have\footnote{This normalization makes sense on manifolds without torsion 2-cycles, on which we can trade it for the common one \cite{Choi:2021kmx} by an Euler number counterterm.}
\be
\cN \cdot Z[B] =  \frac{1}{\sqrt{H^2(X, \abB)}} \sum_{a \, \in \, H^1(X, \, \abA)} \exp\left(2\pi i \int_X a \cup_\gamma B \right) \, \exp\left(2 \pi i \int q_\nu(a)\right) \, .
\ee
First we perform the sum over $\text{Rad}(\nu)$. Since the bicharacter vanishes identically when evaluated on such elements it can be argued that its refinement $q$ also does, independently on the choice of characteristic element. The sum then becomes linear and sets the result to zero unless 
\be
B \in \phi^{-1} N(\text{Rad}(\nu)) = \abB \, .
\ee
The sum over the quotient $\abB/ \text{Rad}(\nu)$ is quadratic and we solve it by shifting $a \to a + \sigma(B)$, which decouples the two fields owning to
\be
\chi_\nu(\sigma(a), \, b) = \gamma(a,\, b)^{-1} \, .
\ee
We are left with:
\bea
&\cN \cdot Z[B] &&= \begin{cases}
G_\nu \, \exp\left(2\pi i \int B^*\left[ 2 \nu + \nu\circ \sigma \right] \right)    \, , \ \ \ &\text{if} \, \pi(B) = 0 \, , \\
0 \, , \ \ \ &\text{otherwise}
\end{cases} \, ,  \\
\ \ &G_\nu &&= \frac{1}{\sqrt{H^2(X, \, \abB/\text{Rad}(\nu))}} \sum_{a \, \in \, H^2(X, \,\abB/\text{Rad}(\nu))} \exp\left(2 \pi i \int a^* \nu \right) \, .
\eea
If $X$ is spin the quadratic refinement does not depend on the choice of characteristic element and we can lift identities for $\chi_\nu$ to identities for $q$. Since 
\be
\chi_\nu(\sigma a, \, \sigma b) = \chi_\nu(a,b)^{-1} \, , 
\ee
on spin manifolds $\nu \circ \sigma = - \nu$ as classes. Furthermore, it is possible \cite{Apte:2022xtu} to prove that the Gauss sum $G_\nu$ is unity. We find that the TQFT $Z[B]$ is duality-invariant on spin manifolds. On the other hand, on non-spin manifolds, we need to impose the stronger condition:
\be
q_\nu(\sigma B) =  q_\nu(B)^{-1} \, , \ \ \  B \in \abB/\text{Rad}(\nu) \, 
\ee
on the quadratic refinement of $\chi_\nu$. A similar story applies to triality defects with minimal modifications.

\bibliographystyle{ytphys}
\baselineskip=0.85\baselineskip
\bibliography{TopGravity}

\end{document}